\documentclass[10pt, runningheads]{llncs}
\usepackage{graphicx}
\usepackage[dvipsnames]{xcolor}
\usepackage{cite}
\usepackage{amsmath,amssymb,amsfonts}
\usepackage{algorithmic}
\usepackage{graphicx}
\usepackage{textcomp}
\usepackage{xcolor}
\usepackage{tikz}
\usepackage{tikz-qtree}
\usetikzlibrary{mindmap,trees}
\usetikzlibrary{arrows}
\usetikzlibrary{calc}
\usepackage{etoolbox}
\usepackage{graphicx}
\usepackage[font={bf,it,small},labelsep=colon]{caption}
\usepackage{subcaption}
\usepackage[hidelinks,colorlinks=false]{hyperref}
\AtBeginEnvironment{quote}{\singlespacing\small}
\usepackage{lipsum} 
\usepackage{caption}
\usepackage{setspace}
\usepackage{makecell}
\AtBeginEnvironment{quote}{\singlespace\vspace{-\topsep}\small}
\AtEndEnvironment{quote}{\vspace{-\topsep}\endsinglespace}
\usepackage{todonotes}
\usepackage{xspace}
\usepackage{color}
\usepackage{enumitem}
\usepackage{flushend}
\usepackage{comment}
\usetikzlibrary{tikzmark}
\usepackage[hang,flushmargin]{footmisc} 

\newcommand{\name}{{\sc BASCPS}\xspace}
\newcommand{\ignore}[1]{}

\begin{document}
\title{\textit{\name}: How does behavioral decision making impact the security of cyber-physical systems?}

% \thanks{This research was supported by grant CNS-1718637 from the National Science Foundation. The opinions expressed in this publication are those of the authors. They do not purport to reflect the opinions or views of sponsor.}
%
%\titlerunning{Abbreviated paper title}
% If the paper title is too long for the running head, you can set
% an abbreviated paper title here
%
\author{Mustafa Abdallah\inst{1} \and
Daniel Woods\inst{1} \and
Parinaz Naghizadeh\inst{2} \and
Issa Khalil \inst{3} \and
Timothy Cason \inst{1} \and
Shreyas Sundaram \inst{1} \and
Saurabh Bagchi \inst{1}
} 
\authorrunning{M. Abdallah et al.}
% First names are abbreviated in the running head.
% If there are more than two authors, 'et al.' is used.
%
\institute{Purdue University, Indiana, USA \\
\email{\{abdalla0,woods104,cason,sundara2,sbagchi\}@purdue.edu}\\ \and
Ohio State University, Ohio , USA \\
\email{naghizadeh.1@osu.edu} \\
%\and Carnegie Melon University, Pittsburgh, USA 
\and Qatar Computing Research Institute, Hamad bin Khalifa University, Doha, Qatar\\
\email{ikhalil@hbku.edu.qa}}
\maketitle              % typeset the header of the contribution
\begin{abstract}
We study the security of large-scale cyber-physical systems (CPS) consisting of multiple interdependent subsystems, each managed by a different defender. Defenders invest their security budgets with the goal of thwarting the spread of cyber attacks to their critical assets. We model the security investment decisions made by the defenders as a security game. While prior work has used security games to analyze such scenarios, we propose \emph{behavioral security games}, in which defenders exhibit characteristics of human decision making that have been
identified in behavioral economics as representing typical human cognitive biases. This is important as many of the critical security decisions in our target class of systems are made by humans. 

We provide empirical evidence for our behavioral model through a controlled subject experiment. We then show that behavioral decision making leads to a suboptimal pattern of resource allocation compared to non-behavioral decision making. We illustrate the effects of behavioral decision making using two representative real-world interdependent CPS. In particular, we identify the effects of the defenders' security budget availability and distribution, the degree of interdependency among defenders, and collaborative defense strategies, on the degree of suboptimality of security outcomes due to behavioral decision making. In this context, the adverse effects of behavioral decision making are most severe with moderate defense budgets. Moreover, the impact of behavioral suboptimal decision making is magnified as the degree of the interdependency between subnetworks belonging to different defenders increases. We also observe that selfish defense decisions together with behavioral decisions significantly increase security risk.
\keywords{Security games \and Cyber-physical systems \and Human behavioral decision making \and NESCOR \and SCADA.}
\end{abstract}

\section{Introduction}
Cyber-Physical System (CPS) structures can help in tackling modern and future technical and operational challenges in different domains such as transportation, healthcare, digital manufacturing, and renewable energy systems. 
% \footnote{PN: I am confused by this paragraph. Not sure if CPS addresses "societal" challenges; it is more commonly challenges that are technical, operational, etc. Also, more details on "what" improvements it has led to. And citation for the renewable energy example needed.} 
% In CPS, physical processes are controlled or monitored through computer-based algorithms, which commonly connect the internal network of an organization to the outside world through the Internet, and also involve human users' interaction with the cyber components through human-machine interfaces (HMI). 
However, the tight integration of the human, physical, and cyber components also increases the attack surface of these systems. %This tiding between Physical parts and Cyber parts make CPS more smart than traditional systems but also increases its vulnerability to cyber attacks which makes the security of such systems is crucial. 
%Several efforts have been made to analyze and mitigate such cyber attacks on CPSs. For instance, the industrial control systems' cyber emergency response team (ICS-CERT) generates a detailed yearly report (e.g. \cite{energyreport2016}) on the total number of cyber incidents, and the distribution of these incidents between different critical sectors. In this context, ICS-CERT provides incident response services to assess the extent of the cyber attack and assists the asset owner in developing strategies for improving ongoing cyber defenses.
%PN (06/09): I commented out this paragraph to save some space, and also as I wasn't sure if it was directly related to this paper. Mustafa, please uncomment if you believe it was better to keep it. 
% In \cite{energyreport2016},  out of the 796 reported cyber incidents between 2013-2015, 35\% affected the energy sector.
  %At the end of 2016, the (ICS-CERT) report \cite{energyreport2016} $ 20 \% $ of the total cyber incidents. These numbers show the huge vulnerability of the energy sector to such risks and attacks.
% \saurabh{We should not pitch smart grid so prominently. Our work applies to any interdependent CPS with multiple defenders. Smart grid security is considered a solved problem research-wise.}
% \saurabh{Mention that since many CPS have many legacy components, improving the security of all of them at the same time is impossible.}
Facing increasingly sophisticated attacks from external adversaries, CPS owners have to judiciously allocate their (often limited) security budgets so as to reduce security risks. Since a CPS typically has many legacy components, improving the security of {\em all} of them at the same time is infeasible.
This resource allocation problem is further complicated by the fact that a large-scale CPS consists of multiple interdependent subsystems managed by different operators, with each operator in charge of securing her own subsystem. Examples include the power grid (energy generators, utilities, domestic and industrial consumers, etc.) and a transportation network (federal and state transportation agencies, private toll road operators, vehicle owners, navigation software vendors, etc.). As a result, the security losses incurred by each operator will ultimately depend not only on her own security investments, but also on the decisions of other stakeholders in the CPS. 

Game theory has played a key role in reasoning about such security decision making problems, due to its ability to systematically capture incentives and optimal actions of defenders and attackers.
Specifically, existing work has modeled these scenarios as an {\em interdependent security game}~\cite{laszka2015survey,naghizadeh2016opting,miura2008security,chan2012interdependent,freudiger2009non,yan2012towards}.
% \footnote{\parinaz{additional citations from CSUR paper to be added}} 
% Mus(5/8/19): Added two related ccs papers from CSUR paper
These are a well-known class of strategic games where the security risk faced by a defender depends on her individual security investments, the security investments by other defenders she interacts with, and the attackers' optimal strategy in response to these investments.
% SB (5/29/18): Will this terminology (``interdependent security game'') be well known to the ACSAC reviewer? My guess is not. 
% Mus(5/29/18): defined
%
% In such large-scale interdependent systems, stepping-stone attacks are often used by external attackers to exploit vulnerabilities within the network in order to reach and compromise a particular valuable target within this network.  These stepping-stone attacks can be captured via the notion of {\it attack graphs} that represent all possible paths that attackers may take to reach their targets within the CPS \cite{homer2013aggregating}. On the other side, the responsibility of  each defender in such systems is defending some subset of the assets \cite{laszka2015survey,naghizadeh2016opting}. Usually, the defenders have restricted resources (i.e., security budget) that they can use to mitigate vulnerabilities in the network. 
%
However, existing work has relied on \emph{classical models} of decision making, where all defenders and attackers are assumed to make fully rational risk evaluations and security decisions~\cite{Alpcan:2010:NSD:1951874,hota2016optimal,modelo2008determining}.
%%\Saurabh{Add some more references to the classical models.}
% Mus{Done 10/14/18}

\begin{figure}[htbp]
\begin{minipage}[t]{0.48\linewidth}
  \includegraphics[width=\linewidth, keepaspectratio]{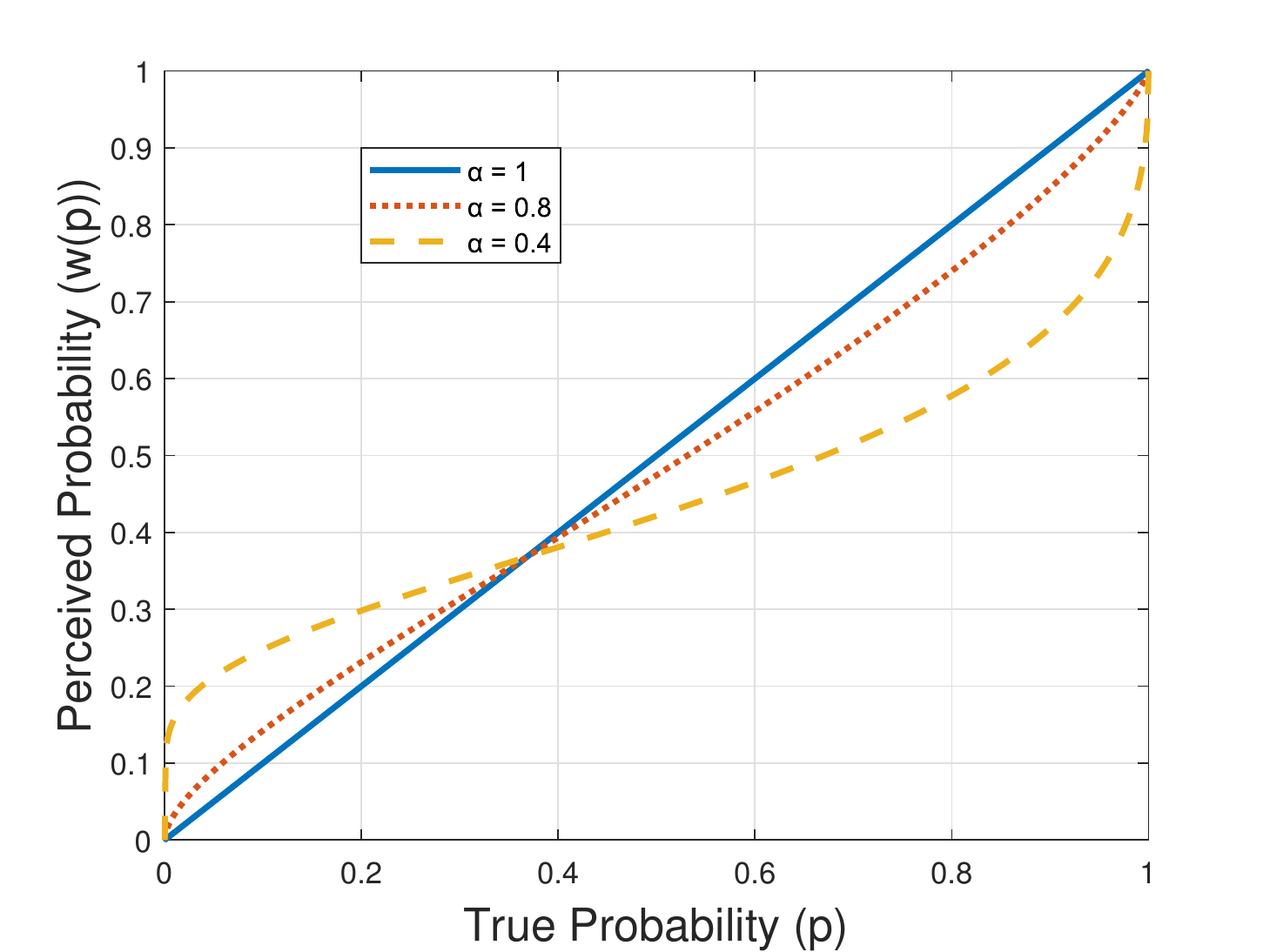}
  \caption{Prelec Probability weighting function which transforms true probabilities $p$  into perceived probabilities $w(p)$. The parameter $\alpha$ controls the extent of overweighting and underweighting, with $\alpha = 1$ indicating non-behavioral or rational decision making. The smaller the value of $\alpha$, the greater is the degree of misperception.}
  \label{fig:Prelec Probability weighting function}
\end{minipage}%
    \hfill%
\begin{minipage}[t]{0.48\linewidth}
   \includegraphics[width=\linewidth]{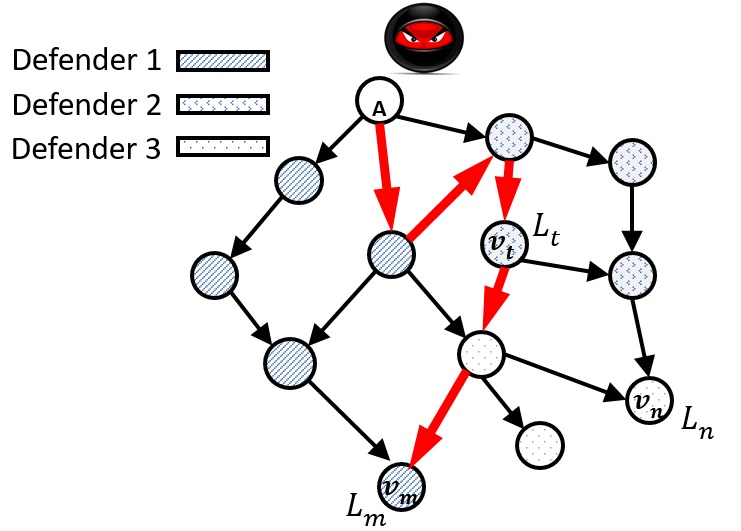}
  \caption{Overview of the interdependent security game framework. This CPS consists of three interdependent defenders. An attacker tries to compromise critical assets using stepping stone attacks starting from node $A$. Interdependencies between assets are represented by edges. The red attack path shows  example of how such interdependency affects the defenders. 
  %when defenders 2 and 3 invest on the edges leading to their own critical assets, they improve the security of defender 1's asset $v_m$ as well.
  }
  \label{fig:Overview attack Graph}
  \end{minipage}
 \vspace{-0.16in}
\end{figure}

Contrary to this focus on perfectly rational decision making for security games, behavioral economics and psychology have shown that humans consistently deviate from these classical models of decision-making. 
Most notably, research in
{\em behavioral economics},
% {\it prospect theory}, 
an important and emerging field for which Kahneman (in 2002) and Thaler (in 2017) won the Nobel Memorial prize in Economic Sciences, has shown that humans perceive gains, losses and probabilities in a skewed, nonlinear manner~\cite{kahneman1979prospect, mullainathan2000behavioral}.
% SB (10/13/18): I have clubbed Kahneman and Thaler together with a broader brush. Verify. 
% Mus(10/14/18): Makes more sense, Both of them have Nobel for behavioral economics. Thaler for "understanding the psychology of economics" and Kahneman for "areas of judgment and decision-making under uncertainty". 
In particular, humans typically overweight low probabilities and underweight high probabilities, where this weighting function has an inverse S-shape, as shown in Figure~\ref{fig:Prelec Probability weighting function}. Many empirical studies (e.g., \cite{gonzalez1999shape,kahneman1979prospect}) have provided evidence for this class of behavioral models, which form part of the foundation for prospect theory.
% SB (11/1/18): Mention in which domains the above empirical studies have been done, like: ``have confirmed this class of behavioral models in domains as varied as XXX, XXX, and XXX.''
% Mus: Done

While a rich literature on prospect theory exists in economics and psychology, most of the existing work studying the security of interdependent systems does not take into account the aforementioned human behavioral decision making effects. These effects are relevant for evaluating CPS security since decisions on implementing security controls are not made purely by automated algorithms, but rather through human decision making by plant managers, system operators, or CISOs \cite{dor2016model,pate2018cyber}.
% \footnote{\parinaz{not sure what kind of citations go here...?}}
% SB (5/9/19): Added
%To our knowledge, there is a relatively little investigation of the effect. 
%To our knowledge, there has been relatively little investigation of the effect of such behavioral decision-making on CPS security and robustness outside of the 
Three notable exceptions where behavioral decision making for security have been studied are \cite{sanjab2017prospect,7544460,9030279}. 
%In~\cite{hota2016fragility}, the authors used prospect theory %in a theoretic game framework 
%to study fragility of a common-pool resources
%(CPR) game to explore the effects of behavioral decision making on the utilization and fragility of common-pool resources, 
In \cite{9030279}, the authors studied the impact of probability weighting on the security investments of a single defender protecting isolated assets. In \cite{7544460}, the authors studied the impacts of probability weighting on certain specific classes of interdependent security games. 
Similarly, \cite{sanjab2017prospect} incorporated prospect theoretic models in the theoretical analysis of the security of drone-based delivery systems. 
However, there are limitations to all of these in the context of multi-defender interdependent CPS. First, in both of \cite{sanjab2017prospect,9030279}, the authors modeled a {\em single} defender CPS system while \cite{7544460} assumed that each node is managed by a different defender, neither of which is true in general for interdependent CPS. Moreover, prior work does not leverage human subject experiments to demonstrate the degree of bias of decision-makers in the behavioral models.

%In contrast, we base our model on human subject experiments, through which we find that the behavioral nature of the humans' security decisions is subject to multiple forms of bias, as opposed to the uni-dimensional models considered previously. 
%\saurabh{Make a high-level statement about what is missing in all these prior works (that we address in this paper). But what we address does not need to be spelled out here.}
%Mustafa: Done (10/14/18). Please check. 
% SB (11/1/18): Good job.

\noindent {\bf Our contributions}: \\
In this paper, we study the effects of human behavioral decision making %(nonlinear probability weighting, in particular) 
on the security of large-scale CPS \textit{with multiple defenders}. %in more general security game where each defender is responsible for defending a set of assets (i.e., subnetwork of the whole CPS network).
We design a reasoning and security investment decision making technique that we call \name (\textbf{B}eh\textbf{a}vioral \textbf{S}ecurity in \textbf{C}yber \textbf{P}hysical \textbf{S}ystems), pronounced as \textit{BASS-CPS}.
%\saurabh{Spell out the full form and say how it should be pronounced.}
% Mus(10/15/18): Done.
In such large-scale interdependent systems, stepping-stone attacks are often used by external attackers to exploit vulnerabilities within the network in order to reach and compromise critical targets \cite{hawrylak2012using,hota2016optimal,vu2014cybersage}. In stepping stone attacks, intruders compromise computing assets within a defended network by first gaining elevated privileges on an asset that is at the periphery of the network. From that, the attacker gains access to a connected asset and so on in a ``stepping stone" manner till some valued target deep inside the defended network is compromised. These stepping-stone attacks can be captured via {\it attack graphs}, representing the possible paths an attacker may take to reach targets within the CPS
\cite{modelo2008determining}. %\cite{homer2013aggregating}.
Through estimating which path(s) in the attack graph the current attack is taking, the defender can allocate the security resources appropriately. We propose a \emph{behavioral security game} model, consisting of multiple defenders and an attacker, in which the interdependencies between the defenders' assets are captured via an attack graph. 
Specifically, we consider two classes of defenders.
% SB (5/29/18): The above sentence does not appear to be true. The graph for the connection between the assets has nothing to do with behavioral security game model.
%Specifically, we study the effect of non-linear perceptions of attack probability by human defenders on the security of systems. We formulate the  best response of each behavioral defender in the security game as a convex optimization problem. In this context, we show that this behavioral game have Pure Nash Equilibrium (PNE). 
% \begin{quote}

\noindent\textit{\bf Behavioral defenders}: These defenders make security investment  decisions under two types of cognitive biases. First, following prospect-theoretic,  non-linear probability weighting models, they misperceive the probabilities of a successful attack on each edge of the attack graph. Second, they have a bias toward spreading their budget so that a minimum, non-zero investment is allocated to each edge of the attack graph. This second kind of bias is motivated by behavior that we observe in our human subject experiments. 
%These defenders make security investment  decisions based on prospect-theoretic, non-linear probability weighting. Specifically, they misperceive the probabilities of a successful attack on each edge of the attack graph, based on cognitive biases.
%,  This means their estimates of the probabilities on attack graph edges are not correct. % modeling attacks on CPS network.    

\noindent \textit{\bf Non-behavioral or rational defenders}: These defenders make security investment decisions based on the classical models of fully rational decision making. Specifically, they correctly perceive the risk on each edge within the attack graph of the CPS network.
%, and choose their investments accordingly.    
% \end{quote}

% SB (10/13/18): Would it be egregiously wrong and intolerable to label these as ``biased defenders'' and ``rational defenders''. Such terminology will mean more to a practical security reviewer, but if these would be flat out wrong, then let us stick to what we have.
% Mus(10/14/18): To my knowledge, I am not sure if biased defender has the same meaning of defender who has behavioral probability weighting. For non-behavioral, rational can be used instead. 

% Our proposed behavioral security game applies to any interdependent CPS with multiple defenders, and captures behavioral decision making under risk and its application in security defense investments.
% \saurabh{We jump too quickly into the results --- we need to say something more about our modeling and solution approach.}

We first analyze the security investments of the decision makers in the CPS, and compare the strategies of behavioral and non-behavioral defenders. We observe that behavioral defenders typically spread their budget throughout their subnetwork, while non-behavioral defenders concentrate their resources on protecting the more critical edges of their subnetwork. This distinction, which is a direct consequence of skewed risk perceptions, will result in higher overall system loss under behavioral probability weighting, due to underinvestment in the critical parts of the system. %lower security defenses in the most critical parts of the network for all defenders under behavioral probability weighting.
%\saurabh{More compelling would be to say that the overall system loss will be higher under behavioral probability weighting, due to underinvestment in the critical parts of the system.}
% Mus (10/14/18): done.
% In other words, we show that behavioral decision making leads to  suboptimal resource allocation  compared to non-behavioral decision making. 
% SB (9/30/19): Reviewed new text below. 
%There are domain-specific challenges to apply a behavioral model in the context of security of CPS.
% PN (4/1/20): re-wrote the above sentence; please feel free to switch back if earlier version is preferred. 
In conducting our analysis and obtaining these insights based on a behavioral model, we address several domain-specific challenges in the context of security of CPS.
These include augmenting the attack graph with certain parameters such as sensitivity of edges to security investments (Equation \eqref{eq:expon_prob_func}) and the estimation of baseline attack probabilities  (Table~\ref{tbl:cvss_cve_der_scada}). %and dependence of these probabilities across multiple edges (Section~\ref{sec:multi-hop-dependence}).
%and incomplete information sharing among the players (Section~\ref{sec:incomplete-information-games}).

% \footnote{PN(9/24/19): this statement is very unclear. Mustafa, could you please elaborate why this has been added (e.g., in response to a reviewer's comment, and if so, what the comment was)? 
% Mus(9/25/19): This was added in response to reviewer C:"yet it is not clear whether the authors build models by taking into account the unique challenges in security domains or conduct a  feasibility study to evaluate the model in a CPS system through previously explored models. Based on the discussion of the related work, I see that the main difference between this paper and several previous publications on game theory appears to lie in the adaptation of the target domain. Additionally,  there was nothing particularly challenging that needed to be done to apply the model in a security context (CPS)"}. 
One may wonder why we need to consider human cognitive biases in security decision making. Why can we not trust ruthlessly rational optimization algorithms which have been studied in the security context \cite{modelo2008determining,xie2010using} with such decisions? The kinds of decision making in our target application domain of interdependent CPS involves significant investments in security controls, security policies, or changes in the system architecture. Hence, the decision making is often done by system operators, plant managers, or security executives, albeit with help from threat assessment tools \cite{sheyner2002automated
,vu2014cybersage}.
%% Commented for Esorics
Also, at Security Operations Centers (SOCs), operators make near real-time decisions about prioritizing various security alerts.
%\footnote{\parinaz{citations to be added, perhpas the same tools we have used for benchmark comparison?}} 
There are many articles discussing the prevalence of human factors in security decision making, both in popular press~\cite{human-security-decision-making-1, human-security-decision-making-2} and in academic journals~\cite{dor2016model, fielder2016decision}, none of which however shed light on the impact of cognitive biases on the overall system security.
% \saurabh{I added above part based on MA's flagging a critique from ACSAC reviews. Verify.}
% Mus: Makes sense. But I think it may be better in discussion.
% SB (5/13/19): Added material in above. 
% \textcolor{red}{Issa suggests adding the following: Additionally SOC (Security Operation Center) operators have to take decisions on how to handle the many security alerts generated by the various security tools on different parts of the network. Behavioral decisions may lead to drop important alerts while spending precious time in further analysing false alarms.}

We perform a human subject study with 145 students to validate our behavioral model and to collect model parameters. We evaluate \name using two realistic interdependent CPS and attack paths through them. 
The first system is a distributed energy resource (DER) with attack scenarios developed by the US National Electric Sector Cybersecurity Organization Resource (NESCOR) working group~\cite{lee2013electric}. 
The second system is a SCADA industrial control system, modeled using NIST guidelines for ICS~\cite{stouffer2011guide}.
We do a benchmark comparison with a prior solution for optimal security controls with attack graphs~\cite{sheyner2002automated}, and quantify the level of underestimation of loss compared to the \name evaluation when defenders are behavioral.
In summary, this paper makes the following contributions: 
%\vspace{-1pt}
\begin{enumerate}[noitemsep,topsep=2pt,parsep=0pt,partopsep=2pt,leftmargin=*]
\item We propose a \textit{behavioral security game} 
  model for the study of security of multi-defender CPS where defenders' assets have mutual interdependencies.
  To the best of our knowledge, we are the first to bring in {\em behavioral} aspects of human decision-making to CPS security and we quantify the suboptimality of the security budget allocations due to behavioral decision making. %In this context, multiple behavioral defenders have interdependencies between their assets which is captured via an attack graph.

\item We illustrate the effects of a prospect theoretic model of decision making (specifically, nonlinear probability weighting) % and spreading behavior %, defined in our game model, with
through two interdependent real-world CPS. We also model the security-relevant aspects.

%including, incomplete information sharing among the actors and difficulty of securing some assets. 

\item We analyze the different parameters that affect the security of interdependent CPS under our behavioral model, such as the available security budget, budget distribution between defenders, types of defense mechanisms, degree of interdependency between defenders, and sensitivity of edges. Our insights in some cases are novel, while in other cases they run counter to prior work with purely rational defenders. 
% \textcolor{red}{We should add some text about study of incomplete information model.} 
% SB (10/1/19): Added. 
%, and the effect of suboptimal pattern of investments. % in behavioral security game on these parameters. 
\end{enumerate}

% SB (5/14/19): Chopped for space

The rest of the paper is organized as follows. Section \ref{sec:securitygames} presents the behavioral security game model and analyzes the differences in investment decisions between behavioral and non-behavioral defenders.  
In Section \ref{sec:human-experiments}, we present human experiments validation of our model. We evaluate the DER.1 and SCADA-based attack scenarios in Sections \ref{sec:DER.1} and \ref{sec:scada}, respectively. Section \ref{sec:Discussion} presents discussion of our approach and some limitations. We discuss related literature in Section \ref{sec:lit-review}. Section \ref{sec:conclusion} concludes the paper.

\section{Behavioral Security Games}\label{sec:securitygames}

In this section, we present our proposed model of behavioral security games, establishing a theoretical basis that can be used to model any multi-defender interdependent CPS. A simple example of our setup is shown in Figure \ref{fig:Overview attack Graph}, which represents a system consisting of 3 interdependent defenders. An external attacker,
%and therefore the loss in our experiments will be lower than theoretical prediction. On the contrary, if the defender assumes the attacker as behavioral and makes her investment decisions, this can lead to higher losses if the attacker is actually non-behavioral.}}
starting from node $A$, uses a stepping stone attack  (e.g., the path of bold edges in the attack graph) to reach the critical assets of the defenders.\footnote{We assume that defenders perceive the attacker as non-behavioral; in reality the attacker can be behavioral as well. Our assumption of a  non-behavioral attacker gives the worst case loss % for the system; 
as a behavioral attacker may not choose the path of true highest vulnerability due to probability misperceptions.} The critical assets are those that are associated with a financial loss when compromised (e.g., $v_{m}$ for defender 1 in the figure). Each defender aims to allocate her security budget on the network edges in a way that safeguards the attack paths reaching her critical assets. We formalize this scenario in this section.

% Mus: All of edges that on any attack path to my critical asset, I can put security investments on it. The three (middle) edges on red attack path have this property.  If the edges interconnects my assets, I will put larger amount on it but I can also put the budget on the other edges as it is on any attack path to my asset as mentioned above. Is it clear ?   

% PN: this was not still clear to me. I have modified the sentence I mentioned earlier, as that was implying individual defense, but seems that the example in the figure is based on joint defense. I have also commented out the last sentence to save space.}   

\subsection{Model and preliminaries}\label{sec:model}
% General_interdependent_network.pdf

% \saurabh{Use a small toy graph with 2 defenders and use that as a running example to explain the concepts.}

We study security games consisting of one attacker and multiple defenders interacting through an attack graph ${G}=({V}, \mathcal{E})$. The nodes ${V}$ of  the attack graph represent the assets in the CPS, while the edges $\mathcal{E}$ capture the attack progression between the assets. In particular, an edge from $v_i$ to $v_j$, $(v_i,v_j)\in \mathcal{E}$, indicates that if asset $v_i$ is compromised by the attacker, it can be used as a stepping stone to launch an attack on asset $v_j$. 
% For example, if an attacker gains the password required to access a power plant's control software ($v_i$), it can use it to attempt to alter the operation of a generator ($v_j$). 
The default probability that the attacker can successfully compromise $v_j$, having compromised $v_i$, is denoted by the edge weight $p^0_{i,j}\in [0,1]$. By ``default probability'' we mean the probability of successful compromise  without any security investment in protecting the assets. 
% \issa{Attacker's Skills: This is the worst case defense model which entails high cost. Do you assume that initial edge probabilities only depend on attacker skills, what about dependency strength of the head of the edge on its tail? Do not we assume that the attacker is completely rational irrespective of his skills? How do you define fairness in budget distribution?}
% \mustafa{ Yes, initial edge probabilities only depend on attacker skills. Yes, attacker is rational but has IT skills and domain knowledge mapped to the three levels (High, Medium and Low). Budget distribution is fair in our case studies if each defender has half of total budget as there is symmetry in critical assets. But generally, as you mentioned in another comment that it is according to its critical assets. Finally, I don't understand the sentence "what about dependency strength of the head of the edge on its tail?"}
Each defender $D_{k}\in D$ is in control of a subset of assets $V_k\subseteq V$, and can make security investments on a subset of edges $\mathcal{E}_k\subseteq \mathcal{E}$.
%\footnote{We will later differentiate between classes of defense mechanisms through the choice of the subsets $\mathcal{E}_k$. Specifically, the defenders employ \emph{individual defense} if $\mathcal{E}_k$ consists only of the edges incoming to $V_k$. In contrast, a choice of $\mathcal{E}_k=\mathcal{E}$ represents \emph{joint defense} in which defenders can spend their budget on any part of the CPS.}
This is motivated by the fact that a large CPS comprises a number of smaller subnetworks, each owned by an independent stakeholder. Among all the assets in the network, a subset $V_m\subseteq V$ are \emph{critical} assets, the compromise of which entails a financial loss for the corresponding defender. Specifically, if asset $v_m\in V_m$ is compromised by the attacker, any defender $D_{k}$ for whom $v_m\in V_k$ suffers a financial loss $L_m\in \mathbb{R}_+$.
%\footnote{For simplicity, we assume a loss of zero on the remaining assets $v\in V\backslash V_m$. Our insights will continue to hold if we consider losses on all assets.}.
%{\color{blue} Note to self: it could be useful to draw a non-critical between the variables in the case studies and the model's variables introduced here...}

% 
To protect the critical assets from being reached through stepping stone attacks, the defenders can choose to invest their resources in strengthening the security of the edges in the network. Specifically, let $x^k_{i,j}$ denote the investment of an eligible defender $D_k$ on edge $(v_i,v_j)\in \mathcal{E}_k$, and let $x_{i,j}=\sum_{D_k \in D} x^k_{i,j}$ be the total investment on that edge by all eligible defenders. 
In addition, let $s_{i,j}\in[1,\infty)$ denote the sensitivity of edge $(v_i,v_j)$ to the total investment $x_{i,j}$. 
%Taking the baseline $s_{i,j}$ to be 1, as the sensitivity of the edge to  security investments increases, 
For larger sensitivity values, the probability of successful attack on the edge decreases  faster with each additional unit of security investment on that edge. 
Then, the probability of successfully compromising $v_j$ starting from $v_i$ is given by,%  $p_{i,j}(x_{i,j},s_{i,j})$.  %(detailed in Section \ref{sec:attack-prob}). 
%
% We assume the following form for $p_{i,j}(x_{i,j},s_{i,j})$: %
\begin{equation}\label{eq:expon_prob_func}
p_{i,j}(x_{i,j})= p_{i,j}^0\exp{\Big(-  s_{i,j} \sum_{D_{k} \in {D} \text{ s.t. } (v_i,v_j)\in \mathcal{E}_k} { x^{k}_{i,j}}\Big)}.
\end{equation}
% \parinaz{why do we have the dependence of $p_{i,j}$ on $s_{i,j}$? these are not decision variables, right? They seem to be problem parameters, same as $p^0$s... If they are parameters, they should be dropped from the probability function's arguments.}
% Mus(5/12/19): You are correct. It is just a constant for each edge. Modified.
Under this assumption, the probability of successful attack on an edge $(v_i,v_j)$ decreases
exponentially with the sum of the investments on that edge by all defenders. This probability function falls within a class commonly considered in security economics (e.g., \cite{gordon2002economics,baryshnikov2012security}).
% SB (5/29/18): ``commonly considered'' needs more than one reference. 
% Mus (5/30/18): done
and further enables analytical tractability of the defenders' decision making problem. In this context, analytical tractability means guarantee of convexity of total loss functions (with the ability to solve and analyze the solutions), and existence of Pure Strategy Nash Equilibrium. Note that any log-convex function for the probability of successful attack enables such analytical tractability in these decision making problems~\cite{baryshnikov2012security}. Our proof in Appendix \ref{sec:proof-convexity} applies generally to any such log-convex function.
%We prove that any log-convex function for the probability of successful attack guarantee of convexity of total loss functions in our formulation (see the proof in Appendix \ref{sec:proof-convexity}).

The attacker initiates attacks on the network from a source node $v_s$ (or multiple possible source nodes),
%\footnote{\textcolor{red}{Our model captures the worst-case in which multiple attackers are colluding perfectly where each attacker has a different source node since it is a min-max model. This can be handled by adding a fake (virtual node) $v_f$ with edges with probabilities $1$ from these nodes to the fake node $v_f$.}}),
% SB (10/14/18): Can our formulation handle multiple possible source nodes? Mention it in the text.
%Mus (10/15/18): Done.
and attempts to reach a target node $v_m \in V_k$, i.e., a critical node for defender $D_k$. Let $P_m$ be the set of all attack paths from $v_s$ to $v_m$. We assume the worst-case scenario, i.e., the attacker exploits the most vulnerable path to each target\footnote{Our formulation also captures the case where each defender faces a different attacker who exploits the most vulnerable path from the source to that defender's assets.}. 
%and therefore the defenders assume the attacker uses the most vulnerable path to each critical asset. 
% SB (5/9/19): Why are we indexing by the path P_m?
% Mus(5/9/19): After our discussion, I added its meaning as set of all paths and added worst case explanation.
% SB (5/9/19): Resolved. 
% through some attack path $P_m$. We assume the worst-case scenario, i.e., the attacker's strategy is not known to the defenders. As a result, each defender chooses her investments to minimize the probability of a successful attack on the most vulnerable path to each of her critical assets. 
%Note that the attack success probability on an edge $(v_i,v_j)$ signifies the conditional probability that the attack on node $v_j$ succeeds given that node $v_i$ has been successfully attacked. We assume that these conditional probabilities across different edges are independent. We also study the dependence between edges in the same attack path in next section.
% SB (5/29/18): Y path later in this section.ou mean ``dges in the attackthe most vulnerable path to each of her critical assets''?
% Mus (5/30/18): exactly. modified.
% SB (11/1/18): Add in a sentence or two in discussion about how we can relax this assumption.
% Mus (11/1/18): Added final sentence in discussion. 
Mathematically, this can be captured via the following total loss function for $D_k$, 
\begin{equation}\label{eq:defender_utility}
\bar{C}_{k}(x_k,\mathbf{x}_{-k}) = \sum_{v_{m} \in V_{k}} L_{m} \hspace{0.3mm} \Big( \hspace{0.3mm} \underset{P \in P_{m}}{\text{max}}\prod_{(v_{i},v_{j}) \in P} p_{i,j}(x_{i,j}) \hspace{0.3mm} \Big)~.
\end{equation}
% SB (10/14/18): There is an assumption of independence of the attack steps that needs to be spelled out.
% Mus(10/15/18): Added sentence "We assume that the success of attack steps across different edges in the network are independent".

Each (rational) defender $D_k$ minimizes $\bar{C}_{k}(x_k,\mathbf{x}_{-k})$ in \eqref{eq:defender_utility}, which is the sum of losses of all of $D_k$'s critical assets, 
%For each $v_{m} \in V_{k}$, the total loss is due to the worst-case attack path $P$ to $v_{m}$. Each defender minimizes the above total loss function, 
subject to her total security investment budget $B_k$, i.e., $\sum_{(v_i,v_j)\in \mathcal{E}_k} x^k_{i,j} \leq B_k$. Note that $x_k$ and $ \mathbf{x}_{-k} $ are the vector of investments by defender $D_k$ and defenders other than $D_{k}$, respectively.
%A summary table of the %behavioral security game's variables is given in Appendix \ref{app:variables_summary_table}. 
% \parinaz{$s$ was missing in the table, I added it. Any other new notation that has been introduced?}
% Mus(5/12/19): Thanks a lot. I think no other notation introduced.
Our setup corresponds to a single-shot game where the defender moves first and spends all her security budget, after which the attacker moves. This investigation can serve as a foundation for future work on multi-shot games, where significant additional complexities will arise, including the need to create realistic models of human decision making in dynamic security games.
%Attack graph 
% PN (10/18/18) changed to security games instead of attack graph games. Let me know if you think attack graph games was a more appropriate name. 
% Mus(10/24/18): That's fine.
%Security games of this form have previously been studied in the case where all players behave rationally \cite{hota2016optimal}. 

% SB (10/14/18): Are we identical to the formulation in our previous paper? Spell out differences if any.
% Mus(10/15/18): Yes, but this paper assumes rationality in decision making.
% \subsection{Security Game Variables}
% We present our general security game framework variables that are used to model the security game. Table 1 summarizes all the variables introduced to describe the system variables of the main components of the main component of the security game which are the attack graph, defenders, losses, and
% attackers. These set of variables will be used on the rest of the paper frequently.
\subsection{Behavioral probability weighting}% function}
Behavioral economics and psychology literature has shown that humans consistently misperceive probabilities by overweighting low probabilities, and underweighting high probabilities \cite{kahneman1979prospect,prelec1998probability}.  More specifically, humans perceive a ``true'' probability $p$ as probability $w(p)$, where $w(\cdot)$ is known as a \emph{probability weighting function}.  A commonly studied functional form for this weighting function was formulated by Prelec in \cite{prelec1998probability}, and is given by
\begin{equation}\label{eq:prelec}
w(p) = \exp{\Big[-(-\log(p)\hspace{0.2mm})^{\alpha}\hspace{0.5mm} \Big]},  \hspace{3mm} p\in [0,1],
\end{equation}
% SB (10/14/18): We need to discuss in the ``Discusssion" section how tied we are to this specific weighting function, i.e., can we generalize to other functions?
% Mus(10/15/18): Added in Discussion.
where $\alpha \in (0,1]$ is a parameter that 
controls the extent of misperception. 
% We provide justification for use of this form of behavioral probability weighting through human subject experiments (Section~\ref{sec:human-experiments}). 
When $\alpha = 1$, we have $w(p) = p$ for all $p \in [0,1]$, which corresponds to the situation where probabilities are perceived correctly, i.e., a non-behavioral defender.  Smaller values of $\alpha$ lead to a greater amount of overweighting and underweighting, as illustrated in 
Figure \ref{fig:Prelec Probability weighting function}.  

%For the Prelec function, $p = \frac{1}{e}$ is the point of intersection between all of the curves.  
%In the next subsection, we incorporate this probability weighting function into the security game of Section \ref{sec:model}%., and define the behavioral security games that we will be studying.

%\subsection{Cost of Behavioral Game} 
We now incorporate this probability weighting function into the security game of Section \ref{sec:model}. 
%
%Behavioral Game is a game between different defenders in an interdependent network, 
In a \emph{behavioral security game}, each defender misperceives the attack success probability on {each edge} according to the probability weighting function in \eqref{eq:prelec}. 
\begin{comment}
Specifically, the perceived attack probability on an edge $(v_i,v_j)$ with total investment $x_{i,j}$ is given by, 
\begin{equation}\label{eq:perceived_edge_attack_probability}
  w(p_{i,j}(x_{i,j})) = \exp{\Big[-(-\log(p_{i,j}({x_{i,j}}))\hspace{0.2mm})^{\alpha}\hspace{0.5mm} \Big]}.
  % , \hspace{2mm} p_{i,j}(x)\in [0,1].
\end{equation}
\end{comment}
Substituting \eqref{eq:prelec} in the total loss function \eqref{eq:defender_utility}, a behavioral defender $D_{k}$ chooses her investments $x_k$ to minimize her \emph{perceived} total loss % via the utility function
\begin{equation}\label{eq:defender_utility_edge}
C_{k}(x_{k},\mathbf{x}_{-k}) = \sum_{v_{m} \in V_{k}} L_{m} \Big(  \underset{P \in P_{m}}{\text{max}}\prod_{(v_{i},v_{j}) \in P} w\left(p_{i,j}(x_{i,j}) \right) \Big),
\end{equation}
subject to her budget constraint $\sum x^k_{i,j} \leq B_k$. Note that $w\left(p_{i,j}(x_{i,j}) \right)$ is the perceived attack probability on the edge $(v_i,v_j)$ with total investment $x_{i,j}$. The total loss function $C_{k}(x_{k},\mathbf{x}_{-k})$ given by \eqref{eq:defender_utility_edge} is convex (see proof in Appendix~\ref{sec:proof-convexity}). 

A set of security investments by the defenders is said to be a Pure Strategy Nash Equilibrium (PNE) if no defender can improve her utility by unilaterally changing her investment \cite{mas1995microeconomic}. %\parinaz{Any preference for using Rosenthal's paper for the definition of NE, rather than a standard GT or microecon textbook?} 
The concept of PNE is widely used to determine the course of action among multiple players in %an adversarial 
a non-cooperative setting. % and is typically used to determine a course of action with perfectly rational players. 
In this paper, we study the security outcomes of the system at the PNE of the above game, and how those outcomes vary with the behavioral probability weighting parameter $\alpha$. 
% SB (9/30/19): Reviewed the text below.
Note that Nash equilibrium is also relevant for behavioral setups, since mistaken probability judgments of behavioral players do not necessarily imply limited strategic reasoning.

To find these Nash equilibria, we use the notion of {\it best response dynamics} \cite{hofbauer2003evolutionary}, where the investments of each defender $D_k$ are iteratively updated based on the investments of the other defenders. For our behavioral security game, in each iteration, the optimal investments for defender $D_k$ can be calculated by solving the convex optimization problem in \eqref{eq:defender_utility_edge}.
\begin{comment}
\begin{equation}\label{eq:minmax_formula_cost_opt}
\begin{aligned}
& \underset{x_k}{\text{minimize}}
& & \hspace{3mm} C_k(x_{k},\mathbf{x}_{-k}) \hspace{5mm}\;\\
& \text{subject to}
& & \sum_{(v_i,v_j)\in \mathcal{E}_k} x^k_{i,j} \leq B_k , \end{aligned}
\end{equation}
\end{comment}
%where $x_k$ is the vector of investments by defender $D_k$ and $ \mathbf{x}_{-k} $ is the vector of investments by defenders other than $ D_{k}$. % This methodology is consistent with the work of \cite{hota2016optimal}, which shows that the best response of each player can be calculated by solving a convex optimization problem although that work did not consider behavioral probability weighting. Thus,
%Throughout 
For our experiments, we run the best response dynamics until they converge to a Nash equilibrium\cite{hota2016optimal}, and then study the security outcomes at that equilibrium.

Our model can also capture the effects of learning by the human subjects as they proceed through multiple rounds of the game (in each round, each defender plays the single-shot game, allocating all her security budget). The model for defender $D_k$ in a multi-round defense game can be captured by our model in \eqref{eq:defender_utility_edge} with $\alpha_k(i)$ as the behavioral level in round $i$. While it may be expected that learning will move a subject toward a more rational model, i.e., $\alpha_k(j) > \alpha_k(i)$ for $j > i$, we find in practice that this is not the case for all subjects (Section~\ref{sec:human-experiments-A}). 
% For some, the learning goes toward more behavioral, for a second category $\alpha$ stays approximately constant, and the third category shows the expected behavior. 

\begin{figure*}[t]%t! 
\centering
%\begin{minipage}[t]{1.0\textwidth}
\begin{subfigure}[t]{.45\textwidth}
\centering
\begin{tikzpicture}[scale=0.5]

\tikzset{edge/.style = {->,> = latex'}};

\node[draw,shape=circle] (vs) at (-2,0) {$v_s$};
\node[draw,shape=circle] (v1) at (0,0) {$v_1$};
\node[draw,shape=circle] (v2) at (2,1) {$v_2$};
\node[draw,shape=circle] (v3) at (2,-1) {$v_3$};
\node[draw,shape=circle] (v4) at (4,0) {$v_4$};
\node[draw,shape=circle] (v5) at (6,0) {$v_5$};
\node[shape=circle] (L1) at (6,1) {$L_5=1$};

\draw[edge,thick] (vs) to (v1);
\draw[edge,thick] (v1) to (v2);
\draw[edge,thick] (v1) to (v3);
\draw[edge,thick] (v2) to (v4);
\draw[edge,thick] (v3) to (v4);
\draw[edge,thick] (v4) to (v5);
\end{tikzpicture}
\caption{An attack graph with a min-cut edge.} 
\label{fig:split_join_dependence_before}
\end{subfigure}%\label{fig:MPNE_example_both_behavioral_1}\hfill
\begin{subfigure}[t]{.45\textwidth}
\centering
\begin{tikzpicture}[scale=0.5]

\tikzset{edge/.style = {->,> = latex'}};

\node[draw,shape=circle] (vs) at (0,0) {$v_1$};
\node[draw,shape=circle] (v1) at (2,1) {$v_2$};
\node[draw,shape=circle] (v2) at (2,-1) {$v_3$};
\node[draw,shape=circle] (v3) at (4,0) {$v_4$};

\node[shape=circle] (L1) at (4,1) {$L_4=1$};

\draw[edge,thick] (vs) to (v1);
\draw[edge,thick] (vs) to (v2);
\draw[edge,thick] (v1) to (v2);
\draw[edge,thick] (v1) to (v3);
\draw[edge,thick] (v2) to (v3);
\end{tikzpicture}
\caption{An attack graph with a cross-over edge.} 
\label{fig:cross_over_edge_graph}
 \end{subfigure} 
%\vspace{0.1in}
\caption{The attack graph in (a) is used to illustrate the sub-optimal investment decisions of behavioral defenders. The attack graph in (b) is used in the human subject experiment to isolate the spreading effect.}
\label{fig:MPNE_main}
\end{figure*}
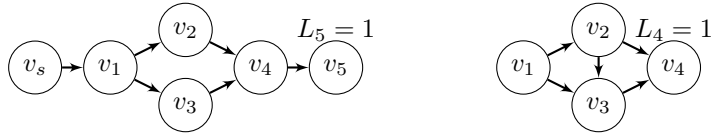

\vspace{-3mm}
\subsection{Motivational example}\label{ex:split_join}
We provide a simple example to illustrate the investment decisions by behavioral and non-behavioral defenders, and provide some intuition on why the optimal defense strategies under the two decision-making models differ.
In this section, we use the notion of a {\it min-cut} of the graph. Specifically, given two assets $s$ and $t$ in the graph, an edge-cut is a set of edges $\mathcal{E}_{c} \subset \mathcal{E}$ such that removing $\mathcal{E}_c$ from the graph also removes all paths from $s$ to $t$.  A min-cut is an edge-cut of smallest cardinality over all possible edge-cuts \cite{west2001introduction}. 
% The critical edges are those that are common in most of the attack paths from the attacker's source node to the defender's target assets (i.e., belongs to the min-cut set).
% SB (5/9/19): I do not think the above sentence is correct. We have also defined critical edges, so this is not really needed. 
% Mus(5/9/19): Removed.
As the example will show, the optimal investments by a non-behavioral defender (i.e., $\alpha = 1$) will generally concentrate the security investments on certain critical (i.e., min-cut) edges in the network.   
In contrast, behavioral defenders tend to spread their budgets throughout the network. 
% We will later use these insights to interpret how behavioral decision making affects the total losses in two case studies based on real-world interdependent CPS.  
%\begin{example}\label{ex:split_join}

Consider the attack graph shown in Figure \ref{fig:split_join_dependence_before}, with a single defender $D$ and a single target asset $v_5$ (with a loss of $L_5 = 1$ if successfully attacked).  Let the defender's budget be $B$, and let the probability of successful attack on each edge $(v_i, v_j)$ be given by $p_{i,j}(x_{i,j}) = e^{-x_{i,j}}$ (assuming $p_{i,j}^0 = 1$). This graph has two possible min-cuts, both of size $1$: the edge $(v_s, v_1)$, and the edge $(v_4, v_5)$.
The total loss function \eqref{eq:defender_utility} for the defender is given by
%\begin{small}
\begin{align*}
\bar{C}(\textit{x}) =  \max \left(e^{-(x_{s,1} + x_{1,2} + x_{2,4} + x_{4,5})}, e^{-(x_{s,1} + x_{1,3} + x_{3,4} + x_{4,5})}\right)
\end{align*}%
%\end{small}
which reflects the two paths from the source $v_s$ to the target $v_t$. Note that an optimal solution of a constrained convex optimization problem satisfies the KKT conditions \cite{hillier2012introduction}. One can verify (using KKT conditions \cite{hillier2012introduction}) that it is optimal for a non-behavioral defender to put all of her budget only on the min-cut edges, i.e., any solution satisfying $ x_{s,1} + x_{4,5} = B $ and $ x_{1,2}=x_{2,4}=x_{1,3}=x_{3,4}=0 $ is optimal.  The intuition of the above result is that for a non-behavioral defender, the probability of successful attack on any given path is a function of the sum of the security investments on the edges in that path. Thus, any set of investments on min-cut edges would be optimal since the sum of investments would be the whole security budget on each path of the graph. 
%The intuition of the above result is that for a non-behavioral defender, the probability of successful attack on any given path is a function of the sum of the security investments on the edges in that path. Thus, any set of investments that maintains the same total investment on each path of the graph will  be optimal. 
% SB (10/14/18): Expand on the above paragraph. A S&P reviewer will need more details to understand and believe this claim.
% Mus(10/16/18): Done.

Now consider a behavioral defender, i.e., a defender with $\alpha < 1$.  With the above expression for $p_{i,j}(x_{i,j})$ and using Prelec function \eqref{eq:prelec}, we have
$w(p_{i,j}(x_{i,j})) = e^{-x_{i,j}^{\alpha}}$. 
Thus, the total loss function \eqref{eq:defender_utility_edge} for a behavioral defender is 
\begin{align*}
%\begin{multline*}
C(x) = \max \left(e^{-x_{s,1}^{\alpha} - x_{1,2}^{\alpha} - x_{2,4}^{\alpha} - x_{4,5}^{\alpha}},~ % \right. \\
%\left.
e^{-x_{s,1}^{\alpha} - x_{1,3}^{\alpha} - x_{3,4}^{\alpha} - x_{4,5}^{\alpha}}\right),
%\end{multline*}
\end{align*}
which includes the two paths from the source $v_s$ to the target $v_5$. Again, one can verify (using the KKT conditions \cite{hillier2012introduction}) that the optimal investments are  
\begin{equation*}
\begin{aligned}
x_{1,2} &= x_{2,4} = x_{1,3} = x_{3,4} = 2^{\frac{1}{\alpha-1}} x_{s,1} .\\
x_{s,1} &= x_{4,5} = \tfrac{B-4x_{1,2}}{2}= \tfrac{B}{2+4( 2^{\frac{1}{\alpha-1}}) }. 
\end{aligned}
\end{equation*}
% SB (10/14/18): Provide the proof for the above in the Appendix (this is in addition to the main page budget). 
%Mus(10/20/18): Done.

Comparing these two cases, 
the optimal investments of the non-behavioral defender yield a total loss of $ e^{-B}$, whereas the investments of the behavioral defender yield a total loss of $e^{-2^\frac{\alpha}{\alpha-1}} e^{-\frac{B}{1+ 2^{\frac{\alpha}{\alpha-1}}}}$, which is larger than that of the non-behavioral defender. 

\textbf{Interpretation:} The reason for this discrepancy can be seen by examining the Prelec probability weighting function in Figure \ref{fig:Prelec Probability weighting function}.  Specifically, when considering an undefended edge (i.e., whose probability of successful attack is $1$), the marginal reduction of the attack probability on that edge as {\it perceived} by a behavioral defender is much larger than the marginal reduction of true attack probability on that edge. Thus the behavioral defender is incentivized to invest some non-zero amount on that edge. Therefore, a behavioral defender splits her investments among the two non-critical sub-paths in the attack path. Note that the same insight holds for different baseline probabilities, but 
% the amount of investments shifted to non-critical edges depends on the marginal reduction of the perceived attack probability on the non-critical edge (Figure \ref{fig:Prelec Probability weighting function}). This 
this shifting effect is greater when the slope of the behavioral probability weighting curve is higher (i.e., close to values of 1, 0, or where the cross-over happens between the behavioral curve and the diagonal).
% SB (5/9/19):  But what happens when the baseline probability on an edge is less than 1?
% Mus(5/9/19): Added the sentence "Note that..."
% SB (5/9/19): Resolved. 
%The magnitude of the perceived drop compensates for the fact that the 
% SB (10/14/18): I do not understand the previous sentence. 
% Mus(10/16/18): The last two sentences modified. Hope now more clear. 
A rational defender, on the other hand, correctly perceives the drop in probability, and thus prefers not to invest on the non-critical sub-paths%split her investments over the non-critical sub-paths
, instead placing her investment only on the critical edges $(v_s,v_1)$ or $(v_4,v_5)$ or both.
%(i.e., all on one and none on the other to any fractional investment on the two edges). Note that the  total loss function is exponential with $e^{-x_{s,1}} e^{-x_{4,5}} =  e^{-x_{s,1}-x_{4,5}}$.
%From a graph-theoretic standpoint, the critical edges are those that form the min-cuts; here there are two possible min-cuts.}
% SB (11/1/18): Verify above sentence.
% Mus(11/1/18): Correct and good intuition

%The third possibility where non-critical edges have higher sensitivity is covered in Experiment A.7 in which also behavioral defender has different investment decisions than non-behavioral defender. Specifically, the behavioral defender puts less investments on edges with higher sensitivity compared to non-behavioral defender.}

%In the above example, we assumed all edges have the same sensitivity to investments. We provide the analysis for different edges' sensitivities (i.e., where $s_{i,j}$ is the sensitivity of the edge $(v_i,v_j)$ to security investment) in Appendix~\ref{app:sensitivity}.

\subsection{Motivational Example with different sensitivities}
In the above example, we assumed all edges have the same sensitivity to investments. In cases where critical edges have equal or higher sensitivity than non-critical edges, the same insight as above holds.
Specifically, when edge $(v_i, v_j)$ has sensitivity $s_{i,j}$, one can verify (using KKT conditions) that the optimal investments by a behavioral defender are given by 
\begin{equation*}
\begin{aligned}
x_{1,2} &= x_{2,4} = x_{1,3} = x_{3,4} = 2^{\frac{1}{\alpha-1}} \left(\tfrac{s_{i,j}}{s_{s,1}}\right)^{\frac{\alpha}{1-\alpha}} x_{s,1} .\\
x_{s,1} &= \left(\tfrac{s_{s,1}}{s_{4,5}}\right)^{\frac{\alpha}{1-\alpha}}   x_{4,5};\hspace{1mm} x_{4,5} = B - \sum_{\forall (i,j)\neq (v_4,v_5) } x_{i,j}.
%&= \tfrac{B-4x_{1,2}}{2} . 
\end{aligned}
\end{equation*}
The insight here is that the investment decision has two dimensions: behavioral level and sensitivity ratio of non-critical edges to critical edges. Specifically, as the defender becomes more behavioral, she puts less investments on edges with higher sensitivity.

\begin{comment}

\end{comment}
\subsection{Spreading nature of security investments}

We augment our model with another aspect of behavioral decision making, which we call {\em spreading}. Such a behavioral defender spreads her defensive investments on 
% %multiple attack paths, 
all edges throughout the attack graph, even when some edges are highly unlikely to be exploited for attacks. 
Our use of spreading is inspired by {\em Na\"ive Diversification} 
% (or the {\em 1/n heuristic }) 
from behavioral economics~\cite{10.2307/2677899}, where humans have a tendency to split investments evenly over the available options.
% This refers to the fact that humans have a tendency to put defensive investments on 
% %multiple attack paths, 
% all edges throughout the attack graph,
% even when some of the edges are highly unlikely to be exploited for attacks. 
This phenomenon has not been reported earlier for security decision making, to the best of our knowledge, and we infer this behavior from our human subject study.
%in Section \ref{sec:human-experiments}. 
We capture this effect by adding another constraint to our model in (\ref{eq:defender_utility_edge}): for each defender $D_k$, we set $x^k_{i,j} \geq \eta_k$, where $\eta_k$ is the minimum investment the defender makes on any edge. The value $\eta_{k} = 0$ gives us the behavioral decision with no spreading, i.e., with only behavioral probability weighting.  %\parinaz{might we consider re-ordering 4.1 and 4.2?}

\begin{figure*}[t] 
%\centering
\begin{minipage}[t]{1.0\textwidth}
\begin{minipage}[t]{.47\textwidth}
\centering
   \includegraphics[width=\linewidth]{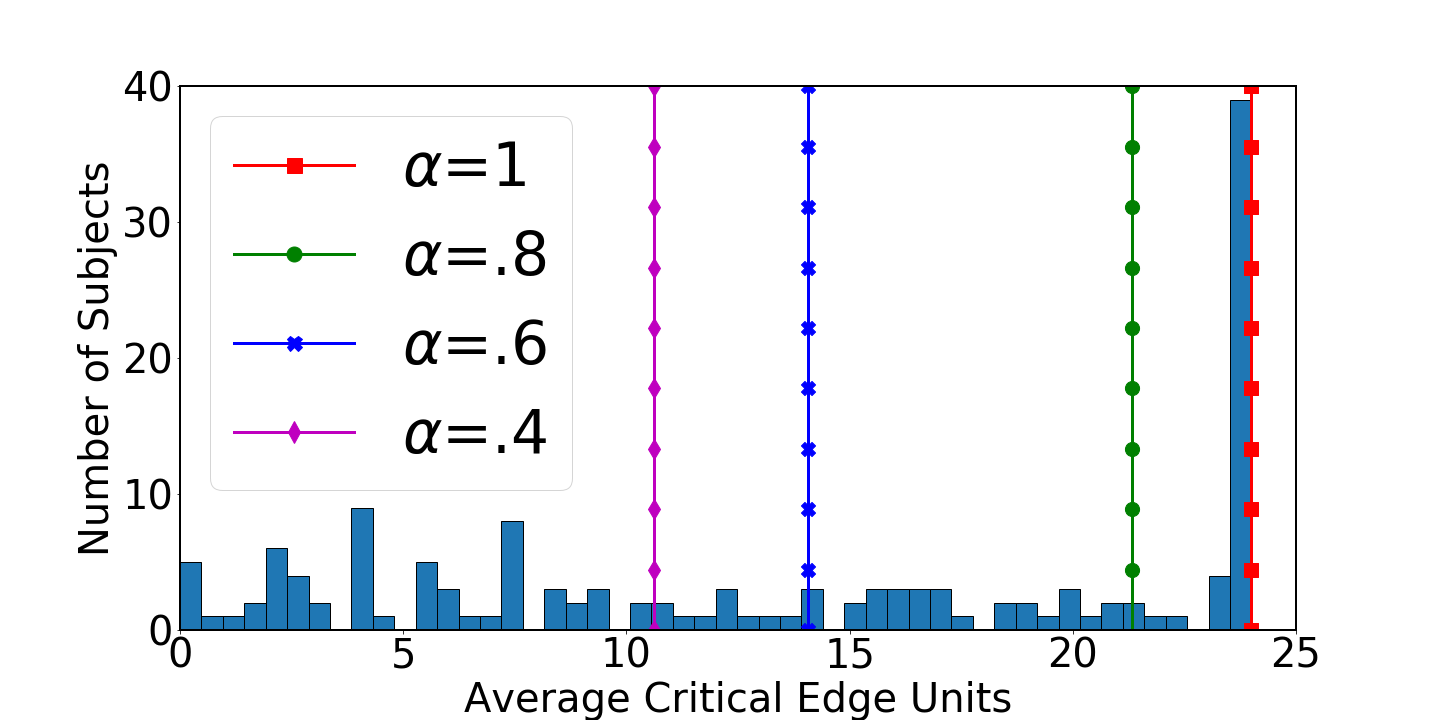}
  \caption{Histogram of human subjects' investments on the critical edge. The vertical red lines show the optimal allocations at specific behavioral levels ($\alpha$).}
  \label{fig:Avg_crit_red}
\end{minipage}\hfill
\begin{minipage}[t]{.47\textwidth}
\centering
   \includegraphics[width=\linewidth,height=55mm,keepaspectratio]{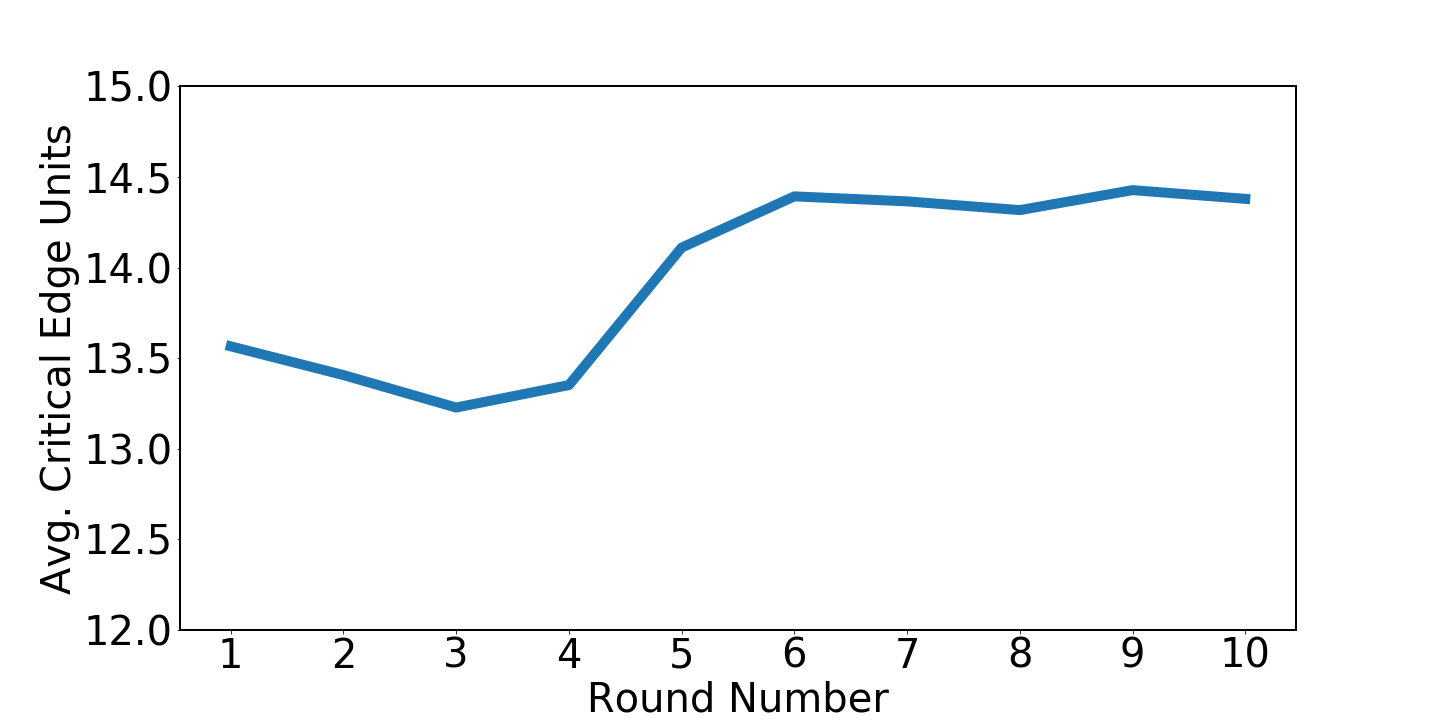}
  \caption{Average of all subjects' investments on the critical edge vs experiment rounds. The upward trend indicates that on average, subjects are learning.} % across rounds.} %We note that the human subjects learn across rounds (on average) and allocate more defense units to Critical Edge.}
  \label{fig:agg_multirounds}
  \end{minipage}%\hfill
  \end{minipage}
  \vspace{-5mm}
\end{figure*}

\begin{figure*}[t]
\begin{minipage}[t]{1.0\textwidth}
\begin{minipage}[t]{.47\textwidth}
\centering
   \includegraphics[width=\linewidth,height=55mm,keepaspectratio]{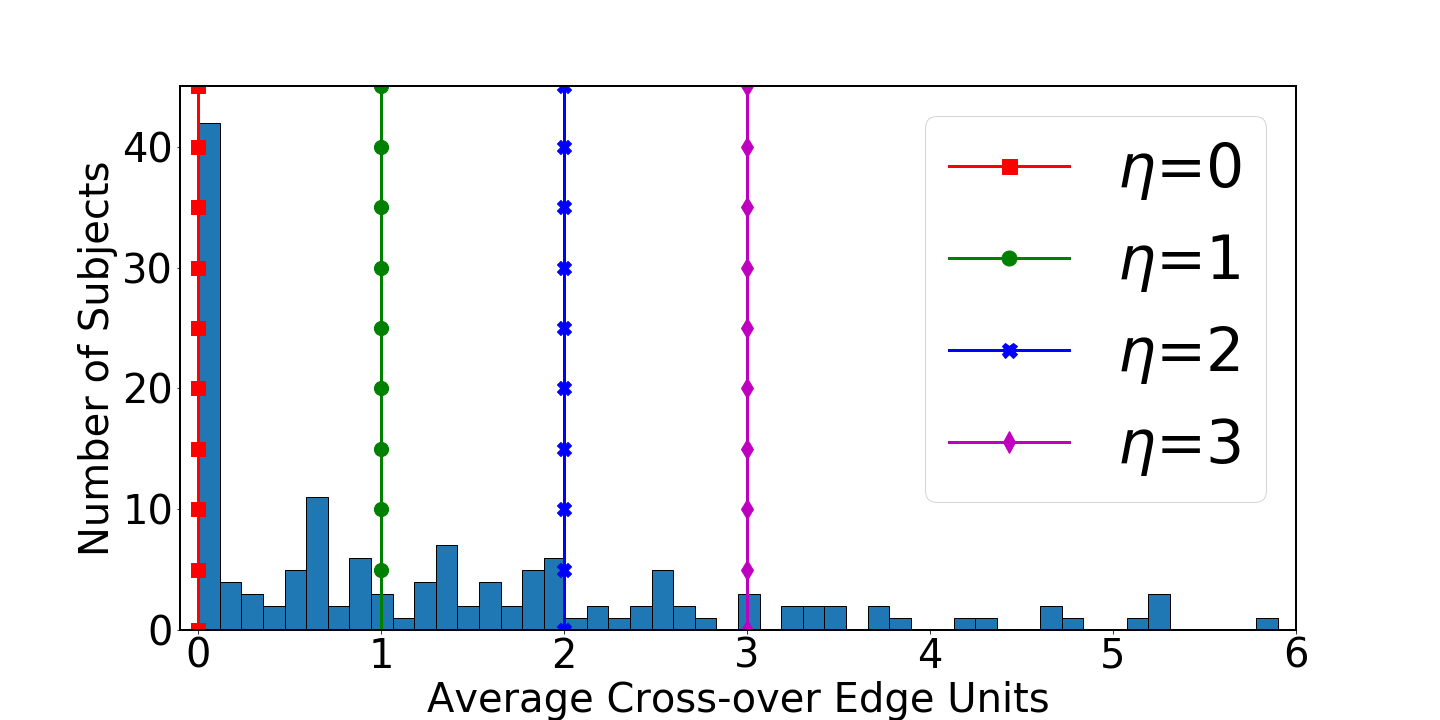}
  \caption{Histogram of human subjects' investments on the cross-over edge. The vertical red lines show the optimal allocations at specific spreading levels ($\eta$).}
  \label{fig:Avg_crit_blue}
\end{minipage} \hfill
\begin{minipage}[t]{.47\textwidth}
   \includegraphics[width=\linewidth,height=55mm,keepaspectratio]{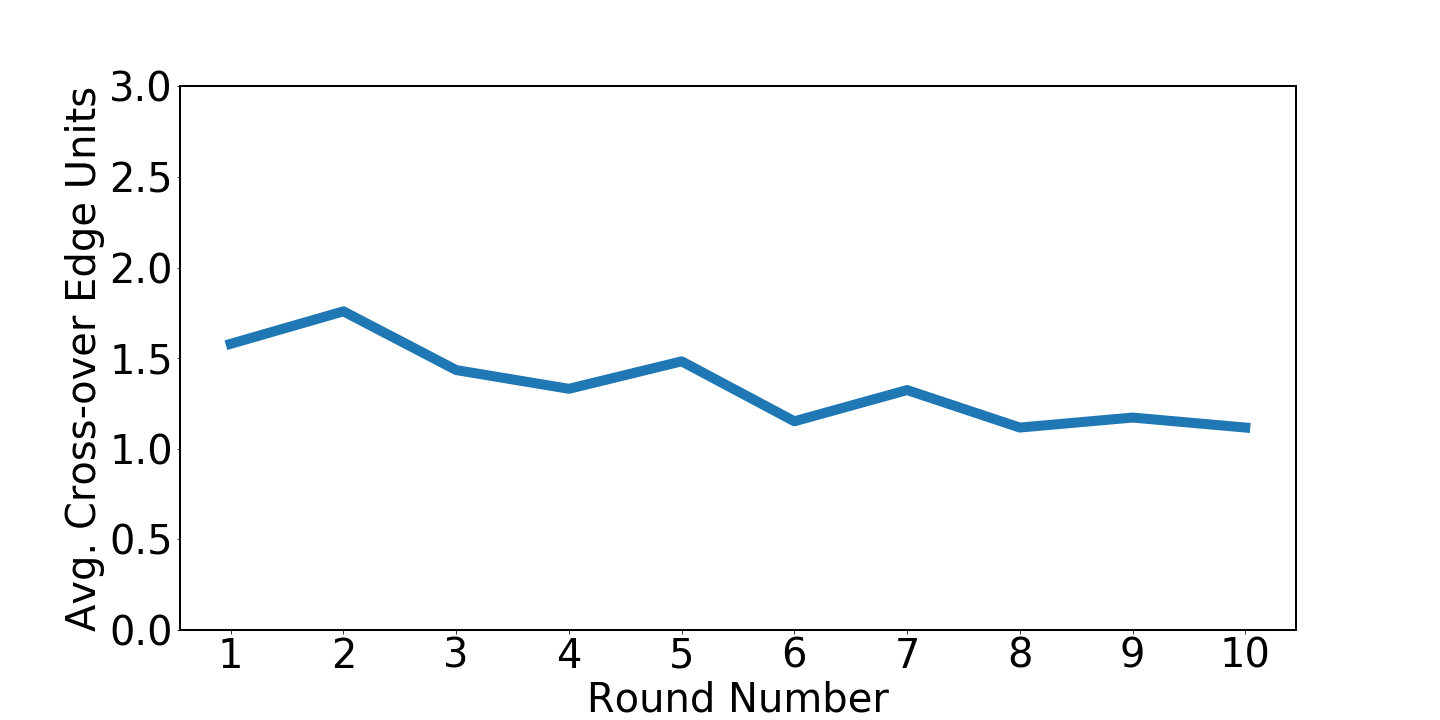}
  \caption{Average of all subjects' investments on the cross-over edge vs experiment rounds. There is only a weak  downward trend in spreading behavior.}%  across rounds.}
  \label{fig:agg_multirounds_blue}
\end{minipage}
\end{minipage}
\vspace{-3mm}
\end{figure*}

%\end{figure*}
%
%\begin{figure*}[t] 
%\centering
\section{Human Subject Experiments}\label{sec:human-experiments}
To validate our behavioral security model, incentivized experiments were conducted on 145 students in the Vernon Smith Experimental Economics Laboratory at Purdue University. Subject demographics are presented in Appendix~\ref{app: human-exp-extended}. Subjects participated in the role of a defender, and allocated 24 discrete defense units over edges in each network. 
% SB (5/15/19): Chopped because this part is down in the implementation weeds. 
% The following functional form was used to convert investment units into probabilities: $p_{i,j}(x_{i,j})=\exp (\frac{-x_{i,j}}{18.2})$, which was then rounded to 2 decimal places.\footnote{Dividing $x_{i,j}$ in the exponential is required to normalize the probability function for 24 units. The denominator of 18.2 is was chosen to target $1-\exp (\frac{-1}{18.2}) \approx 0.05$, a commonly over-weighted and elicited probability \cite{gonzalez1999shape}, to make a single unit of investment on a non-critical edge more attractive to behavioral subjects.} %\textcolor{red}{Issa: explain why you choose this function, what is the magic in 18.2?} Mus(5/13/19): Daniel added footnote.
Subjects made their decisions on a computerized interface, and faced 10 rounds for each network, receiving feedback after each round indicating whether the attack was successful or not (i.e., whether the valuable asset was compromised).
% SB (5/8/19): What feedback? If the valuable asset was compromised or not?
% Mus(5/8/19): Exactly. Changed.
% DW (2/29/20): If you want to be more explicit (given the constraints of space), you can replace after 'each network', with something like 'receiving feedback after each round on which path the attacker took, what edges were compromised, and whether the valuable asset was compromised or not.;
Subjects received comprehensive written instructions on the decision environment that explained how their investment allocation mapped into the probability of edge defense, and what was considered a successful defense. Subjects received a base payment of \$5.00 for their participation. In addition, we randomly selected one round from each network and if the subject successfully defended the critical node in that round she received an additional monetary payment.
  %Subjects were paid for 1 randomly selected round of the 10, and received an additional monetary payoff if they successfully defended the critical node in that period, and no additional monetary payoff otherwise.
% SB (5/8/19): I do not understand the above sentence. Is it the following? We chose one random round of the 10. If they had successfully defended the critical node in that round, then they get a monetary reward (how much?) and else they get nothing.
% Mus(5/8/19): I asked Daniel about this info.  Find the updated Sentence "We randomly selected one round of the 10.  If the subject successfully defended the critical node in that round they received $7.50, otherwise they received $0." Is $ as USD acceptable. I edited it a little bit.
 
\subsection{Network (A) with critical edge}
\label{sec:human-experiments-A}
This human experiment is on a network similar to Figure \ref{fig:split_join_dependence_before}, except that there is only one critical edge $(v_4,v_5)$ i.e., $v_s = v_1$. Figure \ref{fig:Avg_crit_red} shows the average investment allocation to the critical edge, based on 1450 investment decisions (i.e., 10 decisions %for each subject 
from each of the 145 subjects). It shows the proportion of subjects who are non-behavioral (those at the vertical red line of $\alpha = 1$, 27\%), as well as heterogeneity in $\alpha$, with observations further to the left being more behavioral. 
% The optimal allocations for specific levels of $\alpha$ when $\eta$ is non-binding are displayed as dashed vertical red lines. 
Subjects to the left of the $\alpha=0.4$ line (approximately 10 units allocated to the critical edge) are not necessarily exhibiting $\alpha < 0.4$. Those who allocate between 5 and 10 units to the critical edge could have a strong preference for spreading. Those who allocate less than 5 units to the critical edge cannot be explained by a strong preference for spreading, as there are 24 units in total to be spread over 5 edges. These subjects are probably using some other unidentified decision heuristic. %or just making fundamental mistakes.
% DW 2/29/20: In the economics paper, we (ex-post) identify `front-loading' and `back-loading' behavior (which we term as a preference for early or late revelation of the outcome) via cluster analysis.  Is that something we would want to briefly include here, as early revelation can describe less than 5 units on the critical edge?  One word of warning though - an explicit functional form for how all these biases interact with each other is quite a challenging task - as noted in the economics paper.  We may be inviting unwanted referee comments that we won't be able to address if we mention it.
 % Figure \ref{fig:crit_red} in Appendix~\ref{app: human-exp-extended} shows the distribution of allocation units for each individual subject between critical and non-critical edges. 
  Figure \ref{fig:agg_multirounds} shows the mean of subjects' investments in each round. After round 4, the average investment on the critical edge in each round is higher than the initial amount of investment in round 1. The average increase summed across the 10 rounds is one defense unit. This means that subjects become less behavioral on average through learning.
  % This set of multi-rounds human experiments shows that multiple rounds can be beneficial (on average) for behavioral defender to make better investment decisions.
  However, looking beyond the average, we note that individuals can be divided into three categories depending on their learning through rounds. The first category makes worse decisions in later rounds ($26.90\%$ of the subjects), the second category exhibits no learning ($40.69\%$), and the third category improves their investments  ($32.41\%$). 
  % SB (5/13/19): Removed as I do not think we can claim this nor is it needed. In effect, Issa's point need not be added because that is how all human subject experiments at univs are done. 
  % \textcolor{red}{Note that security experts likely belongs to the third category}. 
  % Mus(5/13/19): Got it. 
  
  %\textcolor{red}{Issa: A weakness in this experiment could be that the subjects are not security experts, therefore, the bad behavior. Security experts may most likely belong to the third category and with appropriate training would be already rational.}
 % Mus(5/13/19): Added the final sentence about 
 
\subsection{Network (B) with cross-over edge}
This experiment used the attack graph from Figure \ref{fig:cross_over_edge_graph}. This attack graph is suitable to separate the spreading behavioral bias from the behavioral probability weighting, since for any $0 < \alpha \leq 1$, the optimal decision is to put zero defense units on the cross-over edge ($v_2,v_3$). Figure \ref{fig:Avg_crit_blue} shows the average investment allocation on the cross-over edge based on 1450 investment decisions. We see that the proportion of subjects that are non-behavioral, i.e., invest nothing on the cross-over edge, is 29\%.  Figure \ref{fig:agg_multirounds_blue} shows the average of subjects' investments on the cross-over edge in each round, which shows a weak downward trend.
% Any downward trend in this sub-optimal allocation is weak. 
Taken together, these human experiments provide support for our behavioral model with the probability weighting and spreading factors.

\begin{comment}
\subsection{Spreading nature of security investments}

We augment our model with another aspect of behavioral decision making, which we call {\em spreading}. Such a behavioral defender spreads her defensive investments on 
% %multiple attack paths, 
all edges throughout the attack graph,
even when some edges are highly unlikely to be exploited for attacks. 
Our use of spreading is inspired by {\em Na\"ive Diversification} 
% (or the {\em 1/n heuristic }) 
from behavioral economics~\cite{10.2307/2677899}, where humans have a tendency to split investments evenly over the available options.
% This refers to the fact that humans have a tendency to put defensive investments on 
% %multiple attack paths, 
% all edges throughout the attack graph,
% even when some of the edges are highly unlikely to be exploited for attacks. 
This phenomenon has not been reported earlier for security decision making, to the best of our knowledge, and we infer this behavior from our human subject study.
%in Section \ref{sec:human-experiments}. 
We capture this effect by adding another constraint to our model in (\ref{eq:defender_utility_edge}): for each defender $D_k$, we set $x^k_{i,j} \geq \eta_k$, where $\eta_k$ is the minimum investment the defender makes on any edge. The value $\eta_{k} = 0$ gives us the behavioral decision with no spreading, i.e., with only behavioral probability weighting.  %\parinaz{might we consider re-ordering 4.1 and 4.2?}
\end{comment}

\section{CPS System A: Distributed energy system} \label{sec:DER.1}

\begin{figure}[t]
\centering
  \includegraphics[width=0.75\linewidth]{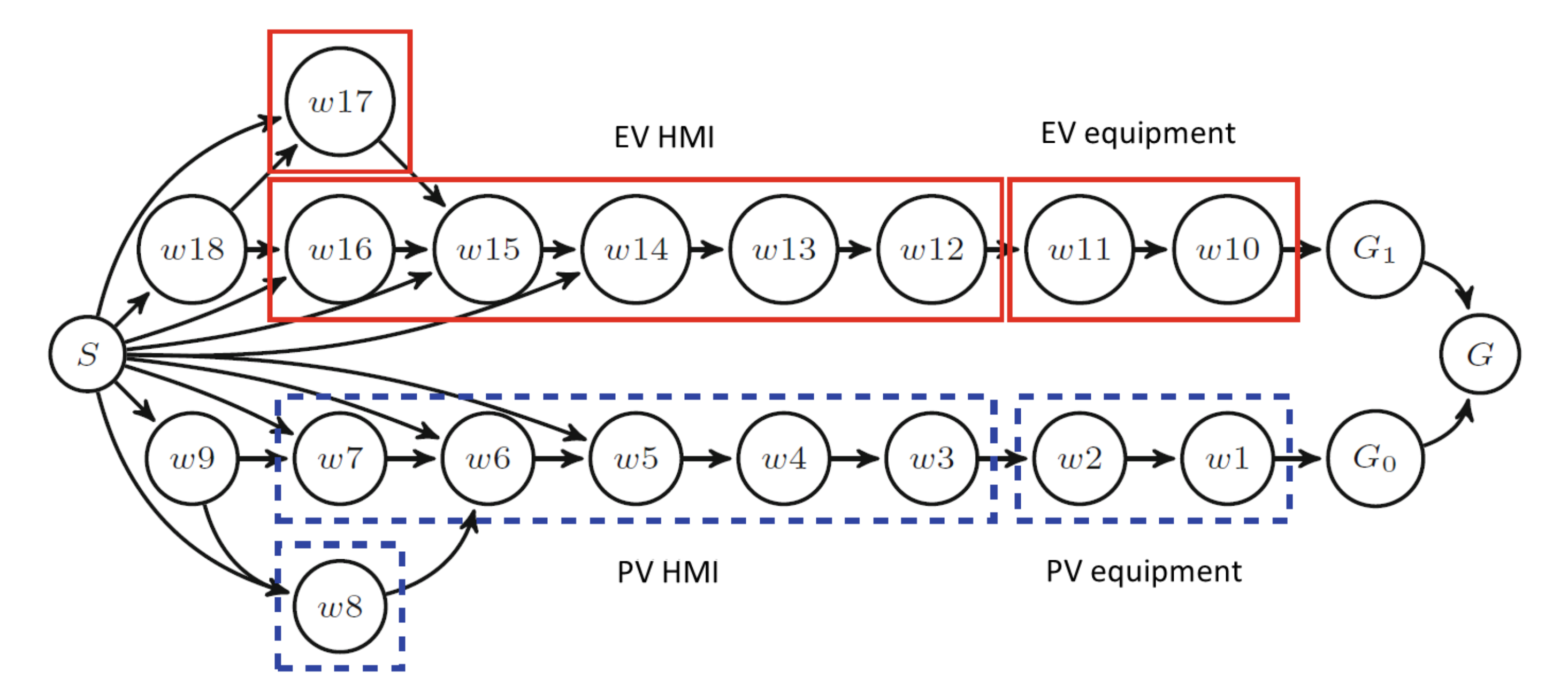}
  \caption{Attack graph of a DER.1 failure scenario adapted from
  \cite{hota2016optimal}. It shows stepping-stone attack steps that can lead to the compromise of PV (i.e., $G_{0}$) or EV (i.e., $G_{1}$). There are two defenders whose critical assets are $G_0$ and $G_1$, while $G$ is a shared critical asset.}
  \label{fig:nescor-attack-graph}
 %\vspace{0.25in}
 \vspace{-5mm}
\end{figure}

In this section, we use our proposed model to evaluate the security outcomes of a practical CPS. Specifically, we examine the effect of different system parameters on the degree of suboptimality of security outcomes due to behavioral decision making.

\subsection{DER.1 system description:}
 
The US National Electric Sector Cybersecurity Organization Resource (NESCOR) Technical Working Group has proposed a framework for evaluating the risks of cyber attacks on the electric grid \cite{lee2013electric}. 
A distributed energy resource (DER) is described as a cyber-physical system consisting of entities such as generators, storage devices, and electric vehicles, that are part of the energy distribution system \cite{lee2013electric}. 
  % This rank as the second top failure scenario is based on the risk score of this failure scenario. In this context, four teams assessed the risk score by dividing the impact of this failure scenario (i.e, $\mathbb{I}$) which has four levels from minor to significant over the cost (i.e., $\mathbb{C}$). The cost (i.e., $\mathbb{C}$) defines the amount of effort by the threat agent to carry out this failure scenario. For example, the least level , out of four levels, of $ C $ implies that the threat agent does not have almost any cost to trigger the attack scenario while the highest level means that it needs nation-state resources to successfully make the attack.
The DER.1 failure scenario has been identified as the riskiest failure scenario affecting distributed energy resources according to the NESCOR ranking. Here, there are two critical equipment assets: a PhotoVoltaic (PV) generator and an electric vehicle (EV) charging station. Each piece of equipment is accompanied by a Human Machine Interface (HMI), the only gateway through which the equipment can be controlled. 
% These assets are connected to other assets in a larger system (not shown in the figure) on the WAN or the LAN through routers and switches. 
% SB (10/25/18): It is not clear what are the different subnetworks. PV HMI and PV equipment are on one LAN and EV are on another?
% Mus(10/28/18): As shown in overview figure, PV and EV are on the same LAN . Other subnetworks here mean the other parts of the large system.
The DER.1 failure scenario is triggered when the attacker gets access to the HMI. The vulnerability of the system may arise due to various reasons, such as hacking of the HMI, or an insider attack.
% when the DER owner does not make the system password-protected, does not change the default password, or has weak authentication of this interface. 
Once the attacker gets access to the system, she changes the DER settings and gets physical access to the DER equipment so that they continue to provide power even during a power system fault.
Through this manipulation, the attacker can cause physical damage to the system, and can even lead to the electrocution of a utility field crew member. % which are very serious cases.
% SB (10/25/18): This is not clear and does not match with the detailed description that goes next. The DER equipment is PV and EV? What does G0 map to in this terminology? 
% Mus(10/28/18): PV and EV are DER equipments. G0 physical failure of PV equipment.
% SB (11/1/18): Resolved. 

% We will assume that the initial starting position of the attacker is within  this WAN. 

% {\color{red} P: I think there is a disconnect between the figure and the attack graph; the attack graph only represents only what goes on after the firewall...
% Mus: Exactly right. That figure should be cut from firewall. Is there a program to draw such things (HMI,...) instead of taking screen shot ?}
% {\color{red} P: just to make sure: did \cite{hota2016optimal} build their attack graph based on the model proposed by \cite{jauhar2015model}? Also, we should mention that the attack graph and its interpretation are taken from \cite{hota2016optimal} in the main text as well. Also, the boxes in the figure are not explained in the text. maybe we should add that too.

% Mus: Yes that's right. Boxes means that nodes (i.e., attack steps) are in same device.
% }

To analyze the above system within our behavioral security game model, we follow the model proposed by \cite{hota2016optimal}, which  maps the above high level system overview into an attack graph as shown in Figure \ref{fig:nescor-attack-graph}. %We followed the same procedure in to generate attack graph. Shortly, 
We generate the attack graph using the  CyberSage tool \cite{vu2014cybersage},
%\parinaz{We generate it using CyberSage, or \cite{hota2016optimal} has generated it using CyberSage?}
% Mus(5/13/19): I generated it also myself. So that is fine to claim that. I think so.
% PN(5/13/19): sounds good.
a Cyber Security Argument Graph Evaluation tool for CPS security assessment. 
% Cybersage can be used to generate %automatically 
% attack graphs given the workflow of a CPS, the security goals, and the attacker model. 
% SB (10/25/18): What is the relation between a security argument graph and an attack graph?
% Mus(10/28/18): Same thing. Changed it to attack graph.
In this attack graph, node labels starting with ``$w$'' are used to denote the non-critical assets/equipment used as part of the attack steps, and $ G_{0} $, $G_{1}$, and $ G $ represent the critical assets which are the attacker's goals. 
For the attacker's goals, $ G_{0} $ represents a physical failure of the PV system,  $ G_{1} $ represents a physical failure of the EV system, and $ G $ means that a failure of either type has occurred. The goal $ G $ may signify non-physical losses (e.g., reputation losses) for the DER operator as a result of a successful compromise.
The first defender is responsible for defending the critical asset $ G_{0} $, the second defender for defending $G_{1} $. % , and further, 
Both defenders share the common asset $ G $.

\subsection{Experimental setup:}
\noindent Each edge in the attack graph in Figure~\ref{fig:nescor-attack-graph} represents a real vulnerability. To create the baseline probability of attack on each edge (i.e., without any security investment), we first create a table of CVE-IDs (based on real vulnerabilities reported in the CVE database for 2000-2019), with each CVE-ID representing one  possible method for exploiting the vulnerability. %We score each CVE-ID using the Common Vulnerability Scoring System (CVSS) \cite{mell2006common}. %score for all of these scenario (
%which measures the potential impact of attacks % and environmental metrics (e.g., 
%based on integrity, confidentiality, and availability metrics. 
%In details, CVSS is composed of the following three access sub-metrics. 
\iffalse
\begin{itemize}
    \item Access Vector (AV), which measures whether or not the vulnerability is exploited locally or remotely.
    \item Attack Complexity (AC), which measures the complexity of attack required to exploit the vulnerability once an attacker has access to the target system.
    \item Authentication (AU), which measures whether or not an attacker needs to be authenticated to the target system in order to exploit the vulnerability successfully.
\end{itemize}
\fi    
We then followed previous works in \cite{homer2013aggregating,8005480} to convert the attack's metrics (i.e., attack vector, attack complexity, and need of authentication  \cite{mell2006common}) to a baseline probability of successful attack.
%We  and then dividing it by the highest possible CVSS score (which is 10).  %\cite{mell2006common}.\parinaz{Is the procedure described here our idea, or is it based on this reference? This type of citation make it seems like it is based on the reference...}
% Mus(5/13/19): Our idea as these scores out of 10.
Table \ref{tbl:cvss_cve_der_scada} illustrates the process.
%\vspace{-16mm}
% \saurabh{In this table the top part is for DER.1 and the bottom SCADA? If so, demarcate with a double line and mention the application in a row.}
% Mus(5/14/19): Done. 

\begin{table}[t]
\vspace{0.2in}
\caption {Baseline probability of successful attack for vulnerabilities in DER.1 and SCADA failure scenarios.}
\label{tbl:cvss_cve_der_scada}
\centering
\resizebox{\columnwidth}{!}
{%
\begin{tabular}{|l|l|l|l|}
\hline
\multicolumn{1}{|l|}{\text{\bf Vulnerability (CVE-ID) }}
& \multicolumn{1}{l|}{\bf Edge(s)}
& \multicolumn{1}{l|}{\bf Attack Vector}
%& \multicolumn{1}{l|}{\bf Attack Complexity}
& \multicolumn{1}{l|}{\bf Score} \\
\cline{1-4}
\hline
\multicolumn{4}{|l|}{\bf DER.1 application} \\ 
\hline
Physical access (CVE-2017-10125) & ($w_{9},w_{7}$),($w_{18},w_{16}$) & Physical & 0.71 \\  %High & High & High &    
\hline
Network access (CVE-2019-2413) & ($w_{9},w_{8}$),($w_{18},w_{17}$) & Network & 0.61 \\ %& Low & Low & None 
\hline
Software access (CVE-2018-2791) & ($w_{7},w_{6}$),($w_{8},w_{6}$) & Network  & 0.82 \\ %& High & Low & None 
\hline
Sending cmd (CVE-2018-1000093) & ($w_{6},w_{5}$),($w_{15},w_{14}$) & Network &  0.88 \\ %& High & High & No
\hline
\hline
\multicolumn{4}{|l|}{\bf SCADA application} \\ 
\hline
Control Unit (CVE-2018-5313) & (Vendor,Control1),(Vendor,Control2) & Local & 0.78 \\
\hline
Remote authentication (CVE-2010-4732) & (S, Vendor) & Network & 0.9 \\  
\hline
Remote cmd injection (CVE-2011-1566) &  (Control,RTU1),(Control,RTU2) & Network  & 1.0\\
\hline
Authentication bypassing (CVE-2019-6519) & (Corp,DMZ1),(Corp,DMZ2) & Network & 0.75\\
\hline
\end{tabular}%
}
%\vspace{-0.1in}
\end{table}
% SB (5/13/19): Chopped for space. 

We next explore the effects of behavioral decision making on the system's security. Specifically, we measure %the system's security by 
the total system loss, given by the sum of the losses experienced by the two defenders in the system. Loss only accrues if $ G_{0} $, $ G_{1} $, or $ G $ are compromised. 
We assume that the total budget available at the defenders' organization is $B$, and that an amount $ BT $ of this budget is set aside for security investments. We refer to $ BT  <  0.2 B$, $ 0.2 B < BT < 0.6 B $, and $ BT > 0.6 B $, as low, medium, and high security budgets, respectively. For our numerical simulations in this failure scenario, we set $B = \$20$.
%, so that $BT = \$1$ corresponds to low security budget, $\$5$ and $\$10$ to moderate security budgets and  $\$20$ to  high security budget.  
%Note that we sometimes use other intermediate values (e.g. $BT = 5$) corresponding to moderate security budget.
%We assume that the total budget of all defenders $ B =  \sum B_{k} $. In practical situations, the defenders can spend amount $ BT $ of this budget in security investments. We refer to low percentages investments (i.e., $ BT  <  0.2 B$ ) as low security budget. Also, we refer to medium percentages investments (i.e., $ 0.2 B < BT < 0.7 B $) as moderate security budget. Finally, high security investments (i.e., $ BT > 0.7 B $) is a high security budget. For our numerical simulations in this failure scenario, we use an example of $B = \$20$ to illustrate our general setup. For example, $BT = \$1$ is a low security budget, $BT = \$10$ is moderate security budget and $BT = \$20$ is a high security budget. 
For most of our simulations, we vary $\alpha \in [0.4,1]$ which is consistent with the range of behavioral parameters from prior experimental studies (e.g., \cite{gonzalez1999shape, abdellaoui2000parameter}) as well as from our user study. We observe that the same insights hold for $\alpha \in (0,0.4)$ (e.g., in the baseline comparison in Figure~\ref{fig:comparison_baseline_s&p_DER}). This is expected since the form of the probability weighting function does not change as $\alpha$ decreases.

\subsection{Results and insights}
\begin{figure}[htbp]
%\vspace{-0.2in}
\begin{minipage}[t]{0.48\linewidth}
    \includegraphics[width=\linewidth,keepaspectratio]{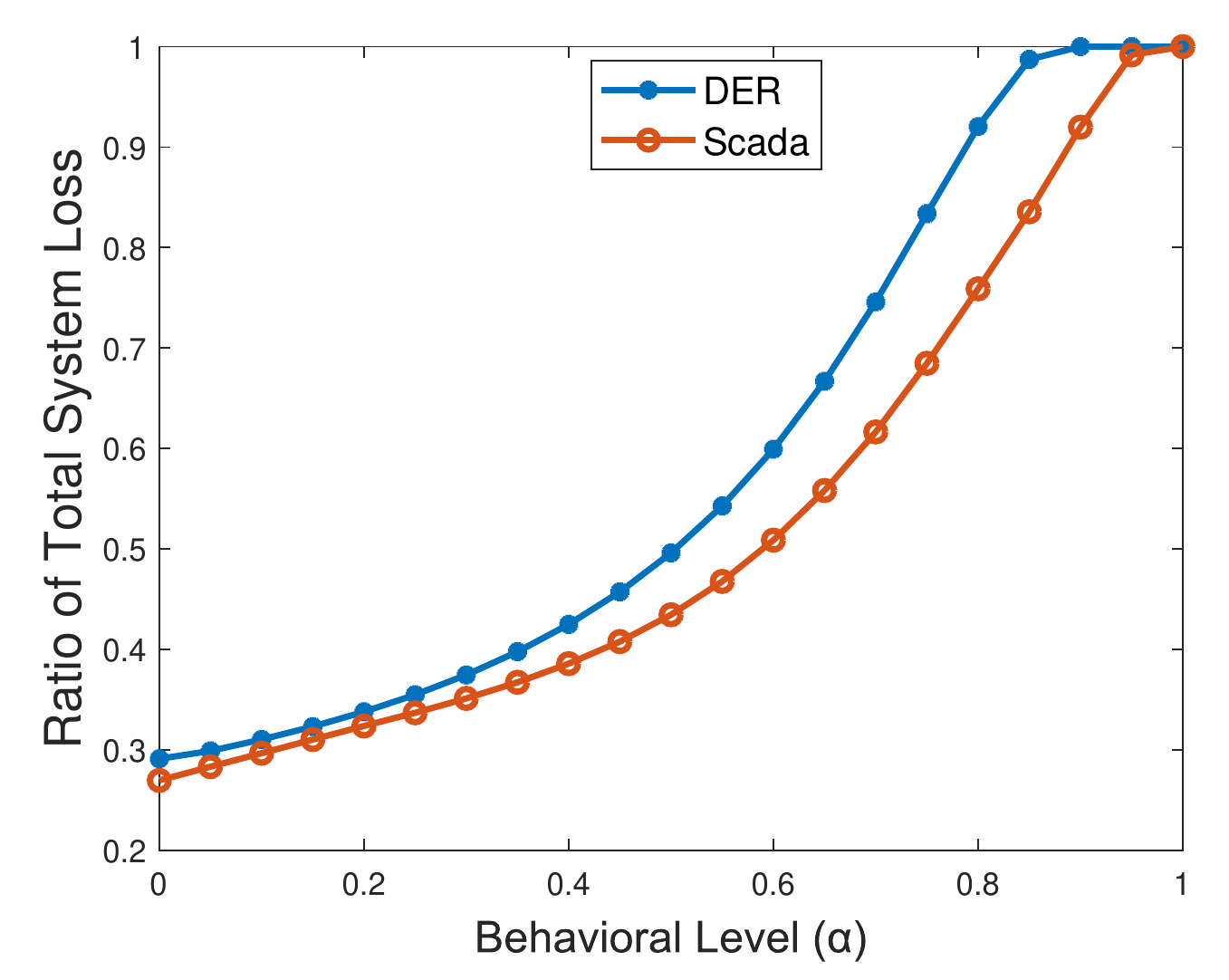}
  \caption{The ratio of loss estimated by \cite{sheyner2002automated} to the (true) loss estimated by \name for different behavioral levels, with $\eta = 0$.}
  \label{fig:comparison_baseline_s&p_DER}
\end{minipage}%
    \hfill%
\begin{minipage}[t]{0.48\linewidth}
        \includegraphics[width=\linewidth,keepaspectratio]{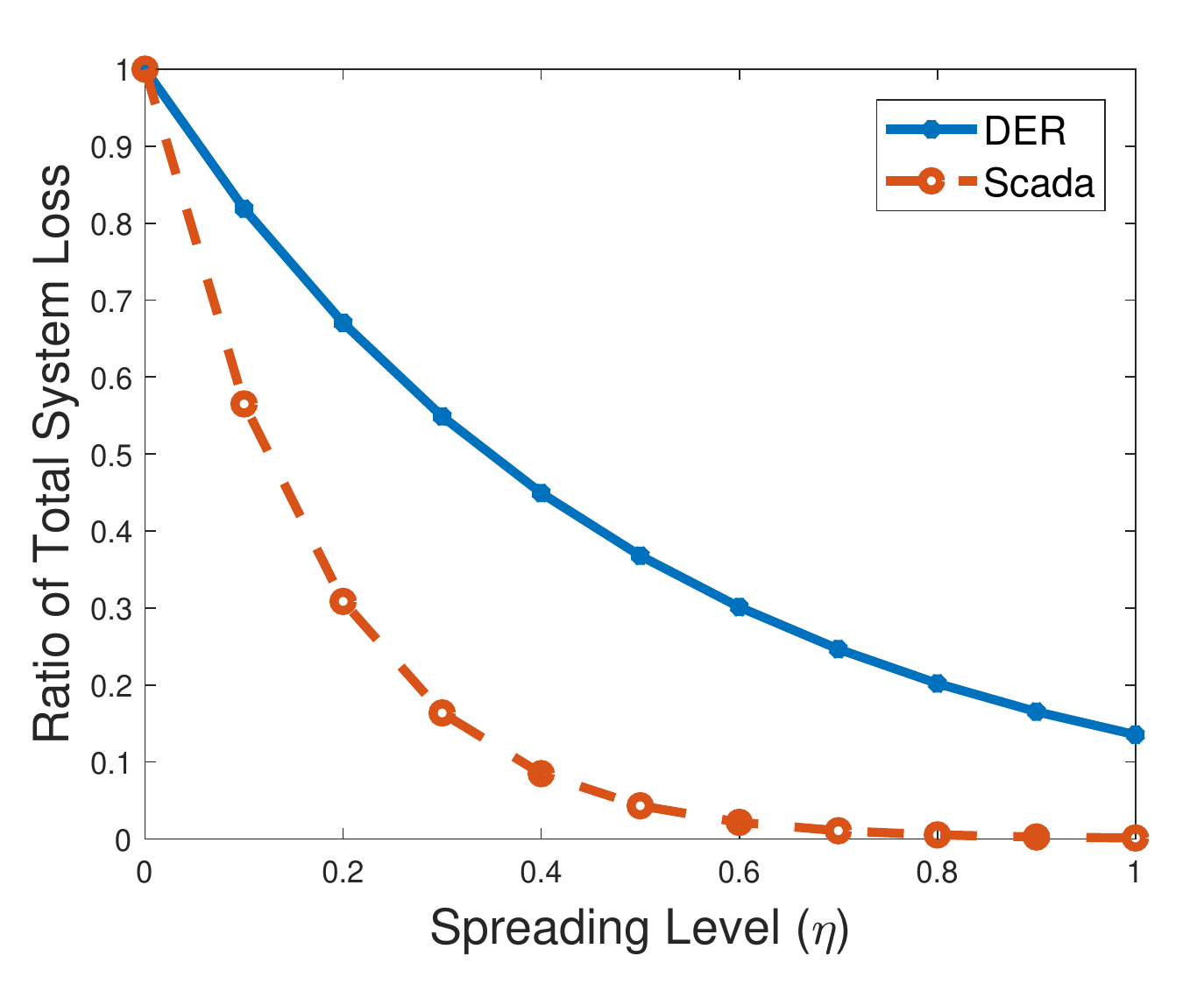}
  \caption{The ratio of loss estimated by \cite{sheyner2002automated} to the (true) loss estimated by \name for different spreading levels, with $\alpha = 1$. }
  \label{fig:comparison_baseline_s&p_DER_spreading}
\end{minipage} 
\vspace{-0.1in}
\end{figure}

\begin{figure*}[t] 
%\centering
\begin{minipage}[t]{1.0\textwidth}
\begin{minipage}[t]{.32\textwidth}
\centering
  \includegraphics[width=\linewidth]{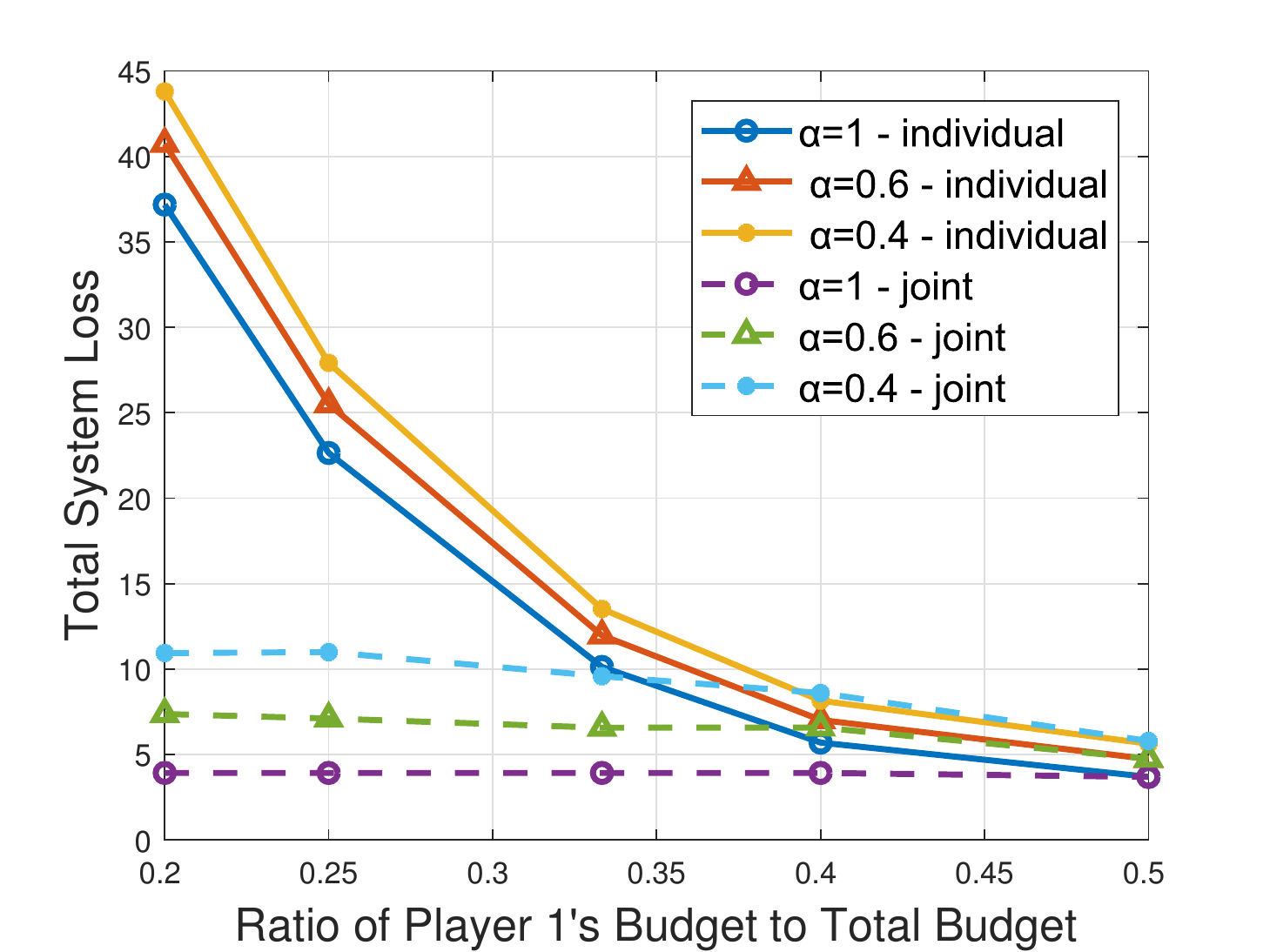}
  \caption{Comparison between individual defense and joint defense investment mechanisms. We observe that the joint defense mechanism can outperform the individual defense scenario, particularly at higher budget asymmetry.}% Total security budget  }
  \label{fig:Defense_mechanisms}
\end{minipage}\hfill
\begin{minipage}[t]{.32\textwidth}
\centering
 \includegraphics[width=\linewidth]{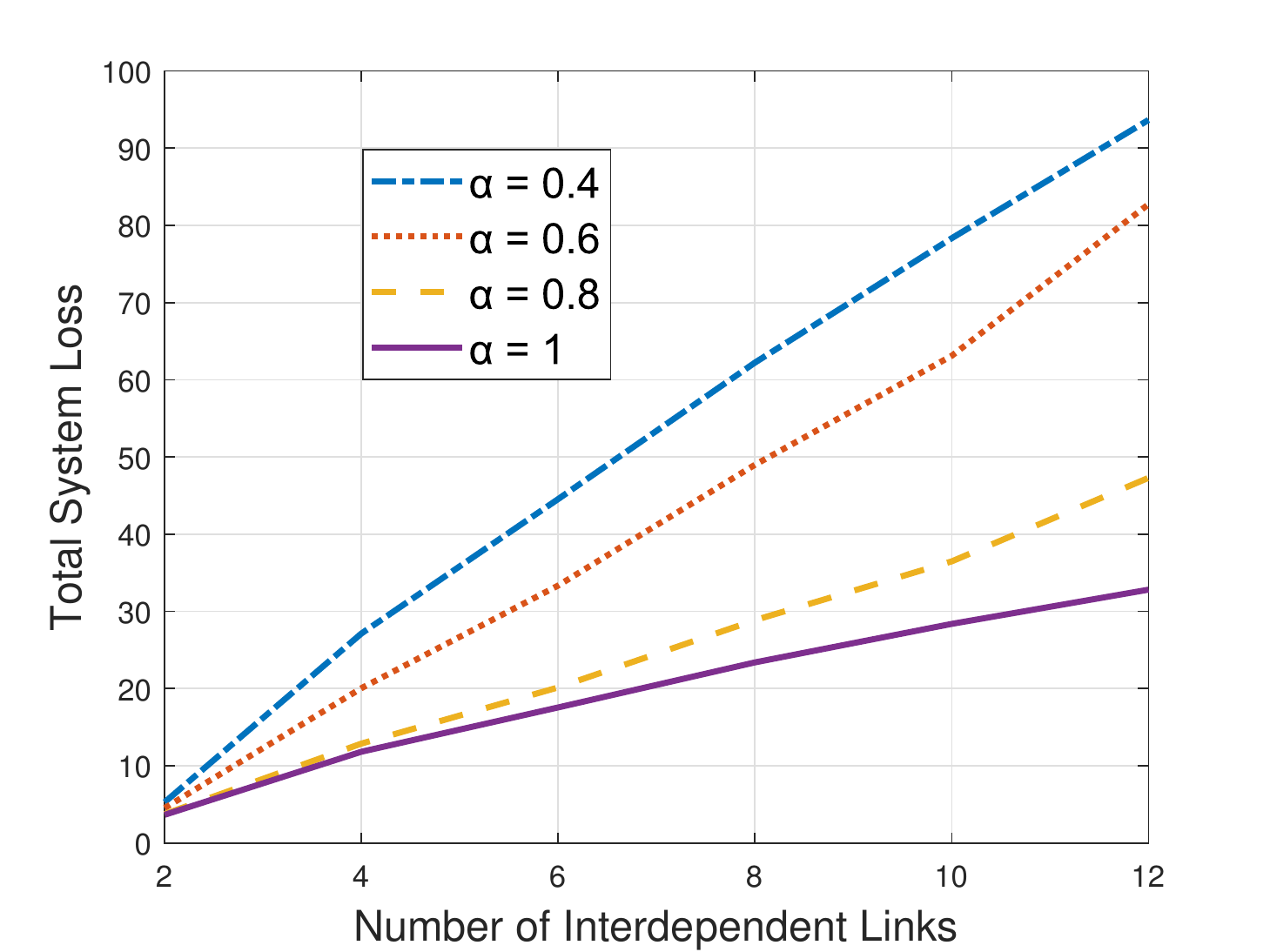}
  \caption{Total loss as a function of the number of interdependent links between defenders. We observe that the effect of increasing the degree of interdependency is more pronounced when the degree of behavioral decision making is higher.}% as the suboptimal investment decisions affects increases with extra non-critical edges from interdependent links. We consider medium budget configuration.}
  \label{fig:Interdependency_Effect_Nescor}
\end{minipage}\hfill
\begin{minipage}[t]{.32\textwidth}
\centering
  \includegraphics[width=\linewidth]{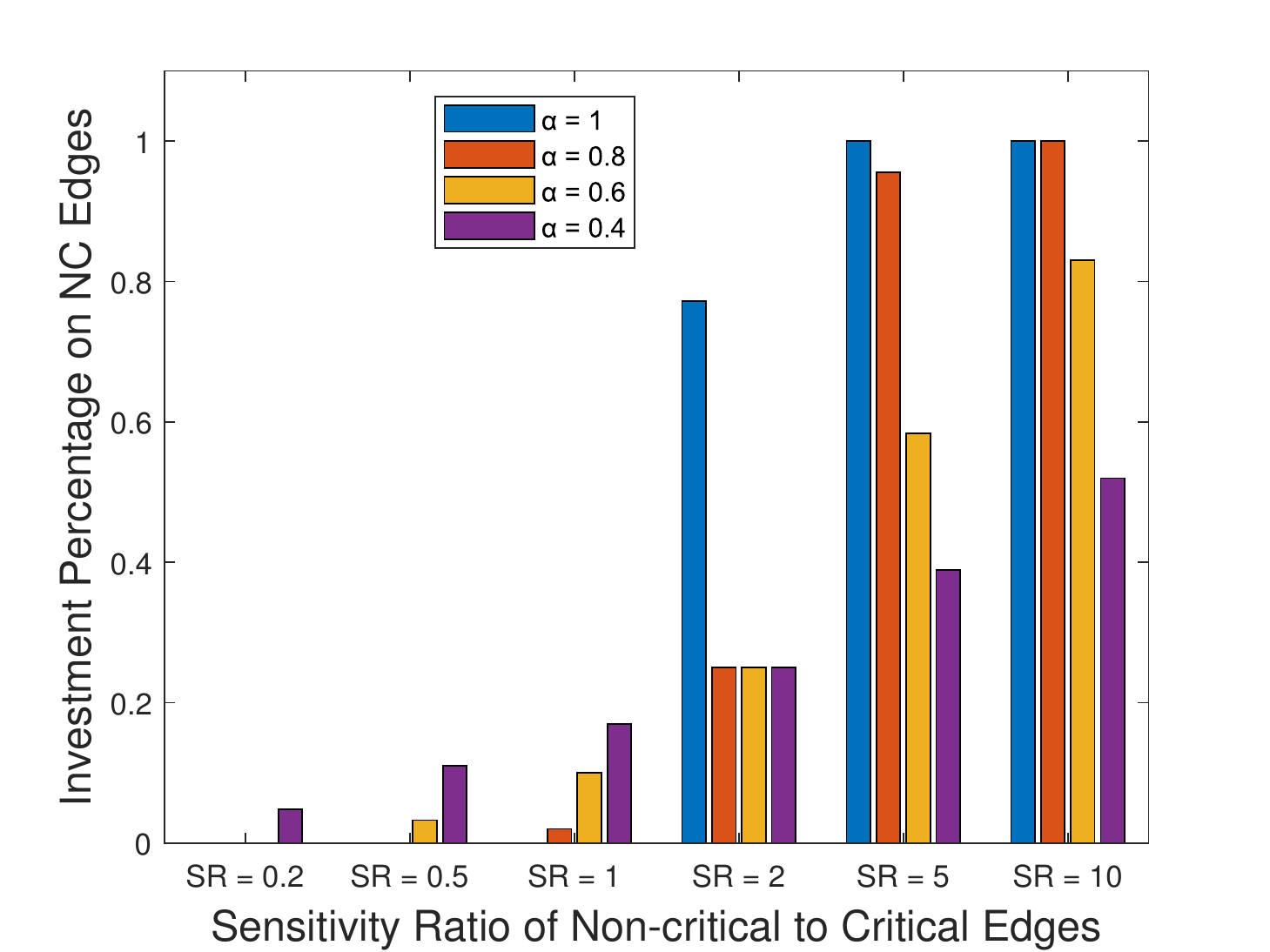}
  \caption{Percentage of investments on non-critical (NC) edges vs the sensitivity ratio (SR) between non-critical to critical edges. As SR increases, the increase in investments on NC edges is slower  for behavioral defenders.}
  \label{fig:sensitivity-exp-DER1}
\end{minipage}
\end{minipage}
%\vspace{-0.1in}
\end{figure*}

\noindent {\em\textbf{Experiment A.1: Baseline comparison}:} 
We begin by comparing \name with the seminal work of \cite{sheyner2002automated} on attack graph generation and investment decision analysis.
In \cite{sheyner2002automated} the defense mechanism is to select the minimal set $C$ of edges that, if removed from the attacker's arsenal, will prevent her from reaching the target asset (there can be multiple sets in case of non-uniqueness). This is equivalent to our min edge-cut. More recent approaches (e.g., \cite{gonzales2015cloud,zhang2016network}) conceptually follow the same strategy for security investment with attack graphs. % in  \name. 
%\parinaz{this is somewhat vague; is this the same as min-edge cut? A type of 'attack' can be used on multiple edges in the graph, and so removing an 'attack' the way it is written now may imply that one can remove all such edges.}
% Mus(5/13/19): Replaced attacks by edges. Exactly, it is the min-edge cut in our technique.
%Here we compare the ratio of the total system loss due to defense placement by \cite{sheyner2002automated} which depends on determining minimal sets $C$ of attacks that if removed from attacker's arsenal, It cannot reach her goal. 
We compare~\cite{sheyner2002automated} with loss under both behavioral and non-behavioral defense placements in \name. % For \cite{sheyner2002automated}, after generating the attack graph, we generate the possible minimal sets in which defense would be allocated.
% \parinaz{again, this is somewhat confusing. do we mean that we find the min edge-cut of the attack graph and consider that to be the decision that \cite{sheyner2002automated} would make?} \textcolor{blue}{Mus: This seminal paper for minimization analysis gives you the min edge-cut (they call minimal set) but you choose how to allocate defense}. 
% \saurabh{Any allocation over the edges in the minimal set will be optimal.}
% SB (5/14/19): Should we mention the above sentence ``Any allocation ...''?
% Mus(5/14/19): I think we mention that BASCPS non-behavioral is identical to them. Thus, no need for it.
We compare the two methods in Figure \ref{fig:comparison_baseline_s&p_DER} by calculating the ratio of total system loss estimated under the method of \cite{sheyner2002automated} over that estimated by \name. %Note that in our attack graph in Figure \ref{fig:System_overview_DER1}, we have multiple minimal sets of same size. 
% 
% \parinaz{This to me seems an inaccurate statement. \name does not protect the system, it just helps us understand how a human would behave, and also how bad things can get. Mustafa, please let me know if you agree. I think we need to be careful with what we claim \name to be: not a defense tool, but rather a tool to understand inefficiencies due to human behavior.} 
% \saurabh{\name can also be used as a defense allocation tool should we want to make automated decisions. Set $\alpha = 1$ and run our algorithm. It will tell you how much to invest on which edge.}
Note that the defense investments given by \name for non-behavioral defenders is identical to that determined by \cite{sheyner2002automated}. 
% The defense investments under behavioral decision making on the other hand % (i.e., bounded rationality) are suboptimal compared to \cite{sheyner2002automated}. % Therefore, leading to higher losses. 
However, if indeed investment decisions are made by subjects with human decision making biases, prior work will underestimate the loss in the system. The degree of underestimation can be as high as 3.34X in DER.1 and 3.57X in SCADA
% \parinaz{we have not introduced SCADA yet.}
% Mus(5/13): Removed.
and that degree is inversely related to the value of $\alpha$. 
Figure \ref{fig:comparison_baseline_s&p_DER_spreading} 
%compares the comparison between the total system loss of the DER system when protected by \name for different spreading levels.
compares the system loss for rational defenders vs that determined by \name under different spreading levels by behavioral defenders. The under-estimation is as high as 8.33X for DER.1 and 826X for SCADA and that degree is directly related to the value of $\eta$.

%\vspace{0.05in}
%\subsubsection*
\noindent {\em \textbf {Experiment A.2: Choice of defense mechanism.}} 
Next, we compare the security outcomes under two potential defense mechanisms. 
%, now returning to the two-defender network. 
The first defense mechanism is individual defense, in which each defender can spend her security budget only on the edges connecting assets inside her subnetwork. In the previous experiment, we have only considered these types of individual investments. An alternative to this defense mechanism is joint defense,  where each defender can choose to spend her security budget on any edge in the network (i.e., she can choose to help defend the subnetworks of other defenders).
% In general, joint defense can provide  advantages over individual defense; especially we show that it can play a key role in mitigation of suboptimal security decisions due to budget asymmetries. % over subnetworks discussed above. 
%For example, in DER system, the government may give more EV vendor rather than PV or vice versa. In this scenario, joint defense mitigate such asymmetry in budget distribution.
Figure \ref{fig:Defense_mechanisms} illustrates that a joint defense mechanism always outperforms the individual defense mechanism.
The advantage of joint defense is higher under asymmetric budget allocation among the defenders.
% The symmetric case is the only case where both individual and joint defense mechanisms yield the same total loss.
Further, the improvement under joint defense holds with both behavioral and non-behavioral defenders. A good example of this is that if both defenders are non-behavioral, and defender 1 has only $ 20\% $ of the total security budget, the joint defense mechanism decreases the total loss by $ 88.5\% $ over the individual defense mechanism. Also, under the same budget allocation, if both defenders are behavioral with $\alpha_1 = \alpha_2 = 0.6$, the joint defense mechanism decreases the total loss by $81.2\% $ over individual defense. % mechanism.

We also note that two behavioral defenders who cooperate can, despite their suboptimal decisions, achieve a {\em lower} total loss compared to two non-behavioral defenders making individual defense decisions in the asymmetric budget setup. For instance, from Figure \ref{fig:Defense_mechanisms}, we observe that two defenders with a high degree of behavioral decision making (i.e., $ \alpha_1 = \alpha_2 = 0.6$) will attain a {\em lower} total loss than two non-behavioral defenders with selfish decisions when the budget asymmetry is 40:60 or greater. 
% Begin unclear
This is explained by the fact that enabling cooperative defense allows the defender with higher budget to put part of her excess security budget as security investment on the other defender's subnetwork. It therefore has considerable potential in mitigating the effects of suboptimal behavioral decision making. 
% End unclear
%Mus: Done

\noindent {\em \textbf {Experiment A.3: Interdependency among different defenders.}} 
We next study the effects of the degree of interdependence among the defenders on the system's security. In the DER.1 failure scenario, the degree of interdependency increases if the HMI of the PV and the EV are communicating. Therefore, making any progress in the attack steps towards one device affects the other device as well (e.g., having software access to the PV enables sending commands to either the PV or the EV). This is represented in Figure \ref{fig:nescor-attack-graph} by introducing additional edges interconnecting the ``$w$'' nodes from the two different sides of the network. %In particular, we show that as the interdependency between the different defenders (i.e., number of interdependent links) is an important parameter that shows the consequences of the suboptimal security decisions on the total system loss. 
%% Esorics: To add that we used BT=10 far above 
%We consider a moderate security budget of $BT = 10$ for this experiment.

Figure \ref{fig:Interdependency_Effect_Nescor} illustrates that as the number of interdependent edges between the two defenders increases, the total system loss increases in both non-behavioral and behavioral security games.
% A good example of this phenomenon is that if both defenders are non-behavioral and there are $12$ interdependent links among the two defenders (i.e., the PV and EV HMIs are communicating at all levels), the total system loss increases $ 500\% $ over the case of $2$ interdependent links (i.e., when the interdependence is only between the final target asset $G$.)% through $G_{0}$ and $G_{1}$).
Further, we observe that the degradation is more pronounced  under behavioral decision making. % in behavioral security games. 
For instance, suppose there are $12$ interdependent links among the two defenders (i.e., the PV and EV HMIs are communicating at all levels). If both defenders are behavioral with $\alpha_1 = \alpha_2 = 0.6$, the total system loss increases by $1230\% $ over the case of $2$ interdependent links. This increase is smaller (500\%) when the defenders are non-behavioral. %This can also be observed by noting  that the slope of the loss becomes larger as $\alpha$ decreases. 
The insight behind these differences is that as the number of interdependent links increases, the number of paths between the source and target nodes increases as well.
% , due to the increased number of criss-crossing edges between the two attack paths. 
% SB (10/25/18): I changed from ``the increased number of parallel edges'' to ``the increased number of criss-crossing edges''. Verify. 
% Mus(10/28/18): Makes more sense.
In this case, behavioral defenders will increasingly spend their budget suboptimally on the criss-crossing edges % where most of these edges are on different paths 
instead of optimally investing on the critical edges, leading to higher losses.
% SB (10/25/18): Degree of interdependency is usually a factor people worry about in this community. So I suggest to keep this. But I do not understand this last sentence. Are we saying that the correct decision is not to invest in these criss-crossing edges. If so, give the intuition why. 

%%%%%%%%%%%%%%%%%%%%%%%%%%%%%%%%%%%%%%%%%%%%%%%%%%%
\noindent {\em \textbf {Experiment A.4: Sensitivity of edges to investments.}} 
%In this experiment, we consider two scenarios where the sensitivity of edges to investments are different. In other words, the successful attack probability on some edge with an investment amount may come down a lot while for some other edge, with the same investment, it comes down only a little. 
We next consider the effects of different sensitivities of edges to security investments. Recall that higher sensitivity edges are those for which the probability of successful attack decreases faster with each unit of security investment. We show the result in Figure \ref{fig:sensitivity-exp-DER1} by using as the independent variable the ratio of sensitivity of non-critical to critical edges. 
First, assume critical edges correspond to mature systems that are already highly secure and difficult to secure further. 
% In practice, this may capture scenarios in which hardware components are less sensitive to defense investments compared to software changes (on non-critical edges in Figure \ref{fig:nescor-attack-graph}). 
For our model, this translates to  high (resp. low) $s_{i,j}$ for non-critical (resp. critical) edges. % while $s_{i,j}$ is low for critical edges.
We observe that as the sensitivity ratio increases, all defenders put more investments on the non-critical edges, but the increase is slower in behavioral defenders. 
% \parinaz{description of comparison in words}
% \saurabh{addressed}
%The second scenario that the sensitivity of the critical edges (near to hardware components) are higher 
However, lower sensitivity ratio will result in investing almost all budget on these critical edges, even for behavioral defenders.

\section{CPS System B: SCADA Control System}\label{sec:scada}

We next evaluate \name on the SCADA system in Figure \ref{fig:Scada_High_level_overview}. We provide a subset of the evaluations here; the remainder, which provide identical insights to DER.1 are in Appendix \ref{app: scada_extended}.

\subsection{SCADA system description:}
\begin{figure}[t]
\centering
  \includegraphics[width=\linewidth,keepaspectratio]{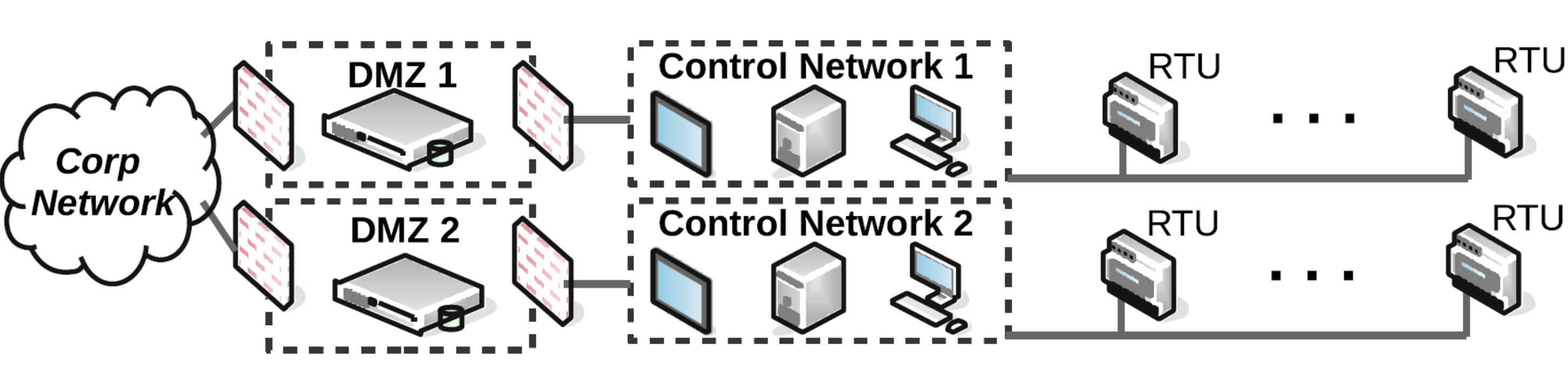}
  \caption{A high level network overview of a SCADA-based system, adapted from \cite{hota2016optimal} consisting of two control subnetworks. These two  subnetworks are interdependent due to sharing a common vendor for their control network and RTUs, and through their common connection to the corporate network. % where each control network is directly connected to a multiple RTUs.
  }
  \label{fig:Scada_High_level_overview}
\end{figure}

The SCADA system (in Figure \ref{fig:Scada_High_level_overview}) is composed of two control subsystems, where each incorporates a number of cyber components, such as control subnetworks and remote terminal units (RTUs), and physical components, such as, valves controlled by the RTUs. 
This system is architected following the NIST guidelines for industrial control systems \cite{stouffer2011guide}. For example, each subsystem is separated from external networks through a demilitarized zone (DMZ). 
% The purpose of a DMZ is to add an additional layer of security between the local area networks of each control subsystem and the external/corporate networks, from where external attackers may attempt to compromise the system. 
The system implements firewalls both between the DMZ and the external networks, as well as between the DMZ and its control subnetwork. Therefore, an adversary must bypass two different levels of security to gain access to the control subnetworks.

\begin{figure}[t]
\vspace{-3mm}
\begin{center}
  \includegraphics[width=0.9\linewidth]{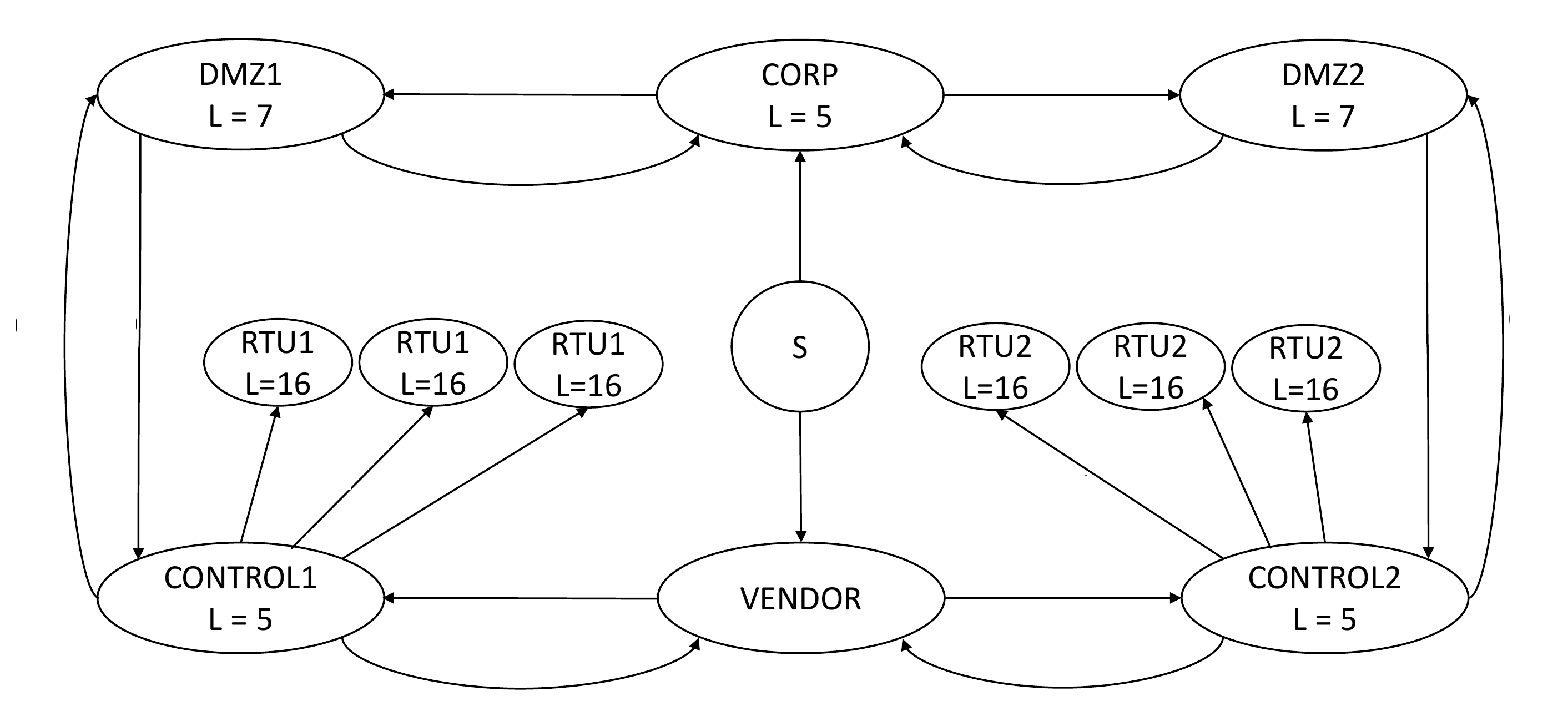}%Attack_Graph_Scada.pdf
  \caption{The attack graph for a SCADA-based control network, adapted from \cite{hota2016optimal}. The attacker's starting node is $S$. Each target node has an associated loss (denoted as L within the node).}%The initial attack success probabilities are shown on the edges.
  \label{fig:Scada_attack_Graph}
  \end{center}
  \vspace{-7mm}
\end{figure}

Mapping this system to our proposed security game model, each control subnetwork is owned by a different defender. These two subsystems are interdependent via the shared corporate network, as well as due to having a common vendor for their control equipment. % remote terminal units (RTUs). %This interconnection captures interdependencies between the two defenders. 
The resulting interdependencies map to the attack graph shown in Figure \ref{fig:Scada_attack_Graph}. The ``Corp'' and the ``Vendor'' nodes connect the two subnetworks belonging to the two different defenders and can be used as jump points to spread an attack from one control subsystem to the other. A critical node has the loss amount denoted within the node (``L = X").
%We consider \emph{external} attackers from either the corporate network or through the vendor network which is more practical than \cite{hota2016optimal}.
% The authors in \cite{hota2016optimal} studied a similar attack graph in face of \emph{internal} attackers who can directly access the RTUs.  %which is not a practical assumption. 
We consider \emph{external} attackers who attempt to gain access to the RTUs through attacks initiated  from either the corporate network or the vendor network. 
% This assumption captures the means available to cyber attackers, and is essential for a practical evaluation of risks in cyber-physical systems. Plus, we are considering the human biases (i.e., behavioral model) on the part of the defenders. 
% In this attack graph, the numbers in the name of each node indicates the defender who owns the asset. Each node (``L = '') also indicates the amount of loss incurred by its owner if the asset is successfully compromised. 
The corporate network ``CORP'' is owned by both defenders. The compromise of a control network ``CONTROL $i$'' leads to loss of control of all 3 connected RTUs.% in that network; 
% to capture this, the corresponding edges from the control units to the RTUs have an initial attack success probability of 1.
%, and are assumed to be indefensible.
% SB (10/31/18): ** Initial ** attack success probability of 1 or ** always ** this probability is 1, i.e., these edges are not defensible.
% Mus(10/31/18): Not fixed in all experiments so i removed "are assumed to be indefensible".

Now, we present our simulations of failure scenario of Scada system with studying some important parameters in the behavioral security game. 

\subsection{Experimental Setup:}
% We present simulations of the above failure scenario and discuss the effects of defenders' behavioral decision making. 
Similar to the DER.1 system, the choice of baseline probability of attack is also based on CVE vulnerabilities (Table \ref{tbl:cvss_cve_der_scada}). The choice of the budget levels has a similar intuition as the DER.1 failure scenario, but differs in values due to the increased number of critical assets in this attack graph. Specifically, in this section, $ BT = \$10$ and $\$20 $ reflect low and moderate budgets, respectively, and $ BT \geq \$30$ reflects high budgets.

\subsection{Results and insights}

%PN(10/28/18): The graph looks great! Thanks for revising Mustafa. 
\begin{figure*}[t] 
%\centering
\begin{minipage}[t]{1.0\textwidth}
\begin{minipage}[t]{.32\textwidth}
\centering
\includegraphics[width=\linewidth]{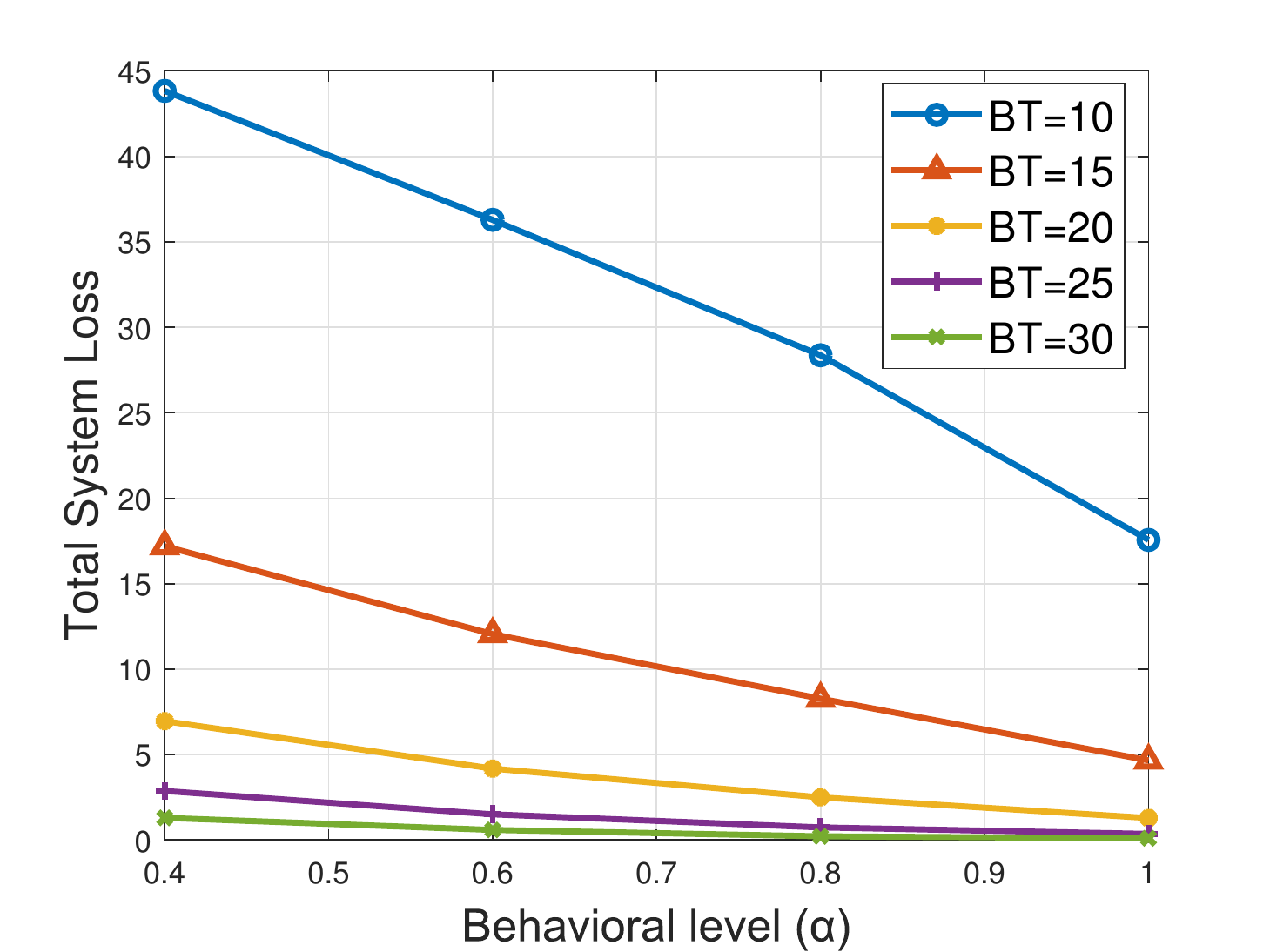}
\caption{Total loss of the SCADA system as a function of budget. The effects of human behavior decision making are higher for this attack graph compared to the DER.1 scenario, due to the increase in the degree of interdependency and the number of critical assets.}% $ BT $ is the total available security budget for all defenders.}
\label{fig:Effect_BT_Behavioral_Scada}
\end{minipage}\hfill
\begin{minipage}[t]{.32\textwidth}
\centering
\includegraphics[width=\linewidth]{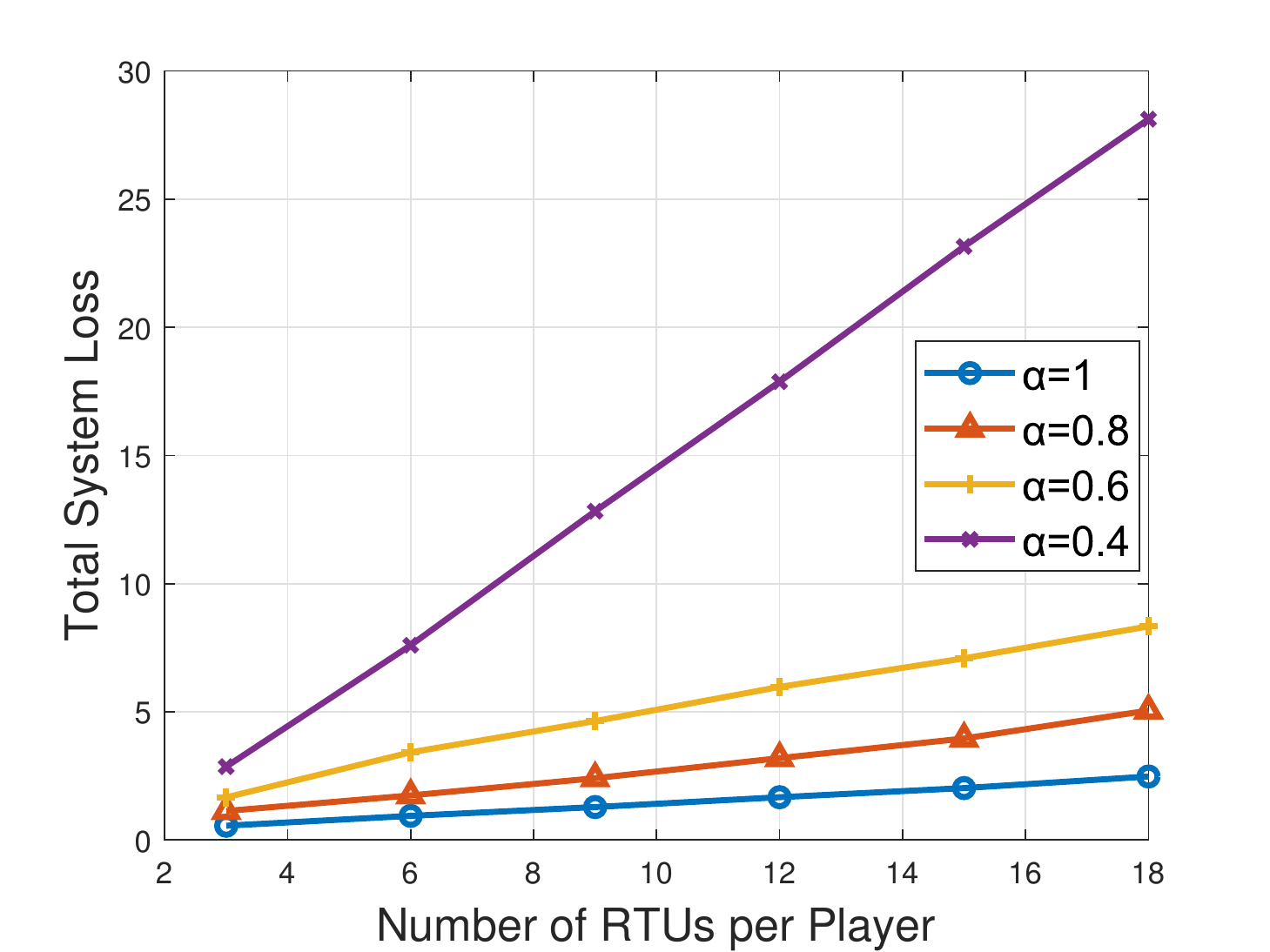}
\caption{Total loss as a function of the number of RTUs per defender. %where $ BT = \$25$. 
We observe that the effect of increasing the number of critical assets is more pronounced when the degree of behavioral decision making is higher %(i.e., $\alpha$ decreases ) 
as the suboptimal decisions affects all critical assets.% (RTUs) of the player's subnetwork.
}
\label{fig:Scada_Effect_RTUs}
\end{minipage}\hfill
\begin{minipage}[t]{.32\textwidth}
\centering
  \includegraphics[width=\linewidth]{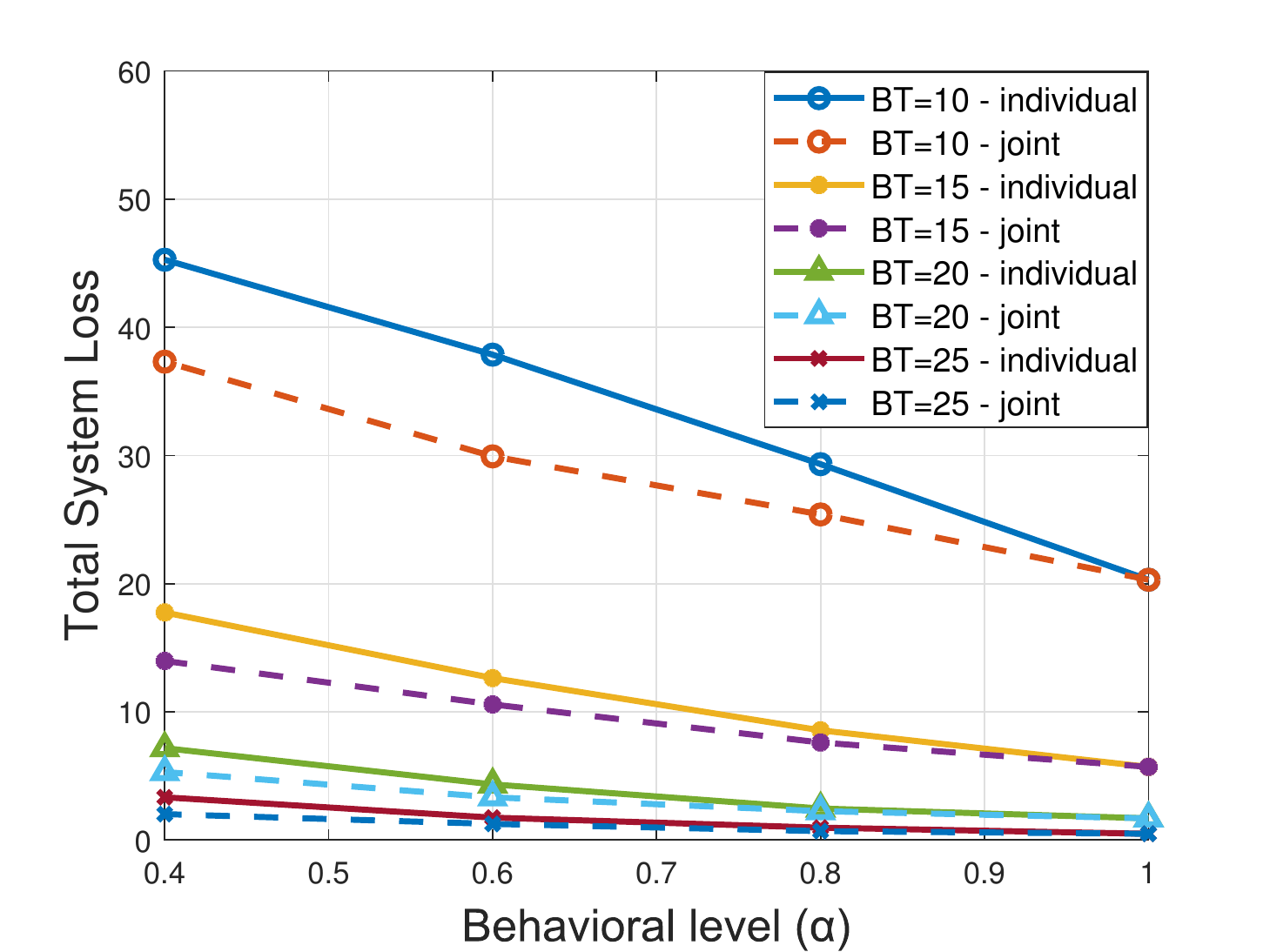}
  \caption{ Comparison between individual and joint defense investment mechanisms. We observe that joint defense can outperform individual defense (i.e., lower losses), particularly when behavioral probability weighting is higher and under asymmetric budget distribution. %The security budget distribution is $\frac{1}{3}$ and $\frac{2}{3}$ for defenders 1 and 2 respectively.
  } 
\label{fig:Effect_Joint_Behavioral_Scada}
\end{minipage}
%  \label{figure:Synthetic_Errors_Detection}
\end{minipage}
%\vspace{-0.1in}
\end{figure*}

\vspace{0.01in}
\noindent {\em\textbf{Experiment B.1: Comparison with the DER.1 attack graph.}} Figure \ref{fig:Effect_BT_Behavioral_Scada} shows the effect of behavioral probability weighting on the total system loss with two behavioral defenders. Comparing Figure \ref{fig:Effect_BT_Behavioral_Scada} with Figure \ref{fig:Effect_SecurityBudget} shows that the relative difference in total loss between non-behavioral and behavioral defenders is much higher in the SCADA network compared to the DER network. This is due to more interdependencies between the two defenders in the SCADA attack graph-- 4 edges per defender in SCADA compared to 1 in DER. Also, there are more  critical assets per defender in SCADA, as all assets (firewalls, control units, and RTUs) are critical. In general, tighter interdependencies among defenders imply that behavioral decision making is going to be more damaging.

%  {\color{red} P: to be fair this is not really the effect of the security budget, right? is a comparison between two attack graphs. We should either change the subheading or make them two separate ones.}
%here that mis-perception of risk for very rich defenders doesn't affect a lot but this case isn't the general case as in most of practical cases restricted security budget is the common scenario. Therefore, the effect of behavioral attitude plays a role here.

% \begin{figure}
%   \includegraphics[width=\linewidth]{Figures/Scada_BT_Behavioral_Effect.eps}
%   \caption{
%   %The effect of behavioral probability weighting on the total real loss of Scada system with different levels of security (i.e., changing security budget). 
%   Total expected loss of the SCADA-based system as a function of budget availability. 
%   The effects of human behavior decision making are higher for this attack graph compared to the DER attack scenario, due to the increase in the degree of interdependency and the number of critical assets. $ BT $ is the total available security budget for all defenders. 
%   %much than DER case as there are multiple RTUs (i.e., target nodes) with higher degree of interconnections.
%   }
%   \label{fig:Effect_BT_Behavioral_Scada}
% \end{figure}

%\subsubsection*
\vspace{0.01in}
\noindent {\em\textbf{Experiment B.2: Number of critical assets. }} Here we consider scaling up the SCADA system, with each defender owning more physical assets and correspondingly having a larger number of RTUs. We make replicas of these RTUs for each defender, and assume that the defender has the same security budget $BT=25$, done to measure the effect of suboptimal decisions as the CPS size grows. Again, we notice that as the number of RTUs increases, the difference between system loss with behavioral players versus non-behavioral players is magnified as shown in Figure \ref{fig:Scada_Effect_RTUs}. For instance, we observe that when $ \alpha= 1 $, a change from 3 RTUs per player to 18 RTUs per player yields a relative increase of $ 344\% $ in system loss, while the same change with $ \alpha = 0.4 $ results in a higher increase of $ 882\%$. As the size of the CPS grows, the budget stays constant and hence the magnitude of the loss increases.
As the decision becomes behavioral, at large CPS sizes, the scarce defense budget is improperly allocated and this magnifies the loss. 
The impact of behavioral decisions is more damaging here than in the DER case because of the highly critical edges between the control network and the RTUs. A non-behavioral defender protects these critical edges, but a behavioral one mistakenly distributes her budget between these and other edges.
\vspace{0.01in}
\noindent {\em\textbf{Experiment B.3: Choice of defense mechanism.}}
As in the DER system, we observe the merits of cooperation (i.e., joint defense) in decreasing the total loss to the defenders (Figure \ref{fig:Effect_Joint_Behavioral_Scada}). The effect is more pronounced for a higher degree of behavioral bias of the defenders. For example, at moderate budget ($BT = 20$), the relative decrease in total system loss due to joint defense at $\alpha_1 = \alpha_2 = 0.4$ is 25\% while $\alpha_1 = \alpha_2 = 0.8$, the decrease is lower (10\%). Thus, as the defenders exhibit higher degree of cognitive bias, it is more advantageous to adopt joint defense mechanisms. 

% SB (10/31/18): Add an example quantitative number to prove the above claim.
% Mus(10/31/18): Added sentence beginning with "For example,"
% SB (11/1/18): Resolved.

% \begin{figure}
%   \includegraphics[width=\linewidth]{Figures/Joint_indiv_comparison}
%   \caption{%The effect of asymmetry of budget distributions between the two defenders of Scada. 
%  Comparison between individual defense and joint
% defense investment mechanisms. We observe that the joint defense mechanism can outperform (i.e. lower losses) compared to the individual defense scenario, particularly when behavioral level is higher (i.e., lower $\alpha$ ). $ BT $ is the total available security budget for all defenders. We fix that budget distribution at $\frac{1}{3}$ and $\frac{2}{3}$ for defenders 1 and 2 respectively.  %This effect is shown for different levels of behavioral levels with fixing total security budget at $\$10$. 
%   %The effect of human behavior increases such asymmetry effect and its reflection on the total real expected loss of the network.}
%  }
%   \label{fig:Effect_Joint_Behavioral_Scada}
% \end{figure}

\vspace{0.01in}
\noindent {\em\textbf{Experiment B.4: Centralized defense merits.}}
Next, we compare the outcomes of decentralized decision-making by two defenders with a centralized defense method where a central planner owns both subnetworks and aims to minimize the system loss % (i.e., the sum of expected losses of all critical units) of all defenders) 
by allocating the budget appropriately. %minimization in which a central planner (e.g., official authority) takes care of minimizing th 
%We note that in centralized defense the game is just a single player (i.e., official authority ) that takes the responsibility of distributing the budget on the whole huge network. 
We observe that when the defenders have a symmetric budget, the outcomes of central and distributed planning coincide. As the budget asymmetry increases, the difference between the two defense methods increases as well. Figure \ref{fig:Central_Behavioral_Scada} (in Appendix~\ref{app: scada_extended}) shows this trend for two budget choices, $ BT = 10$ and $ BT = 20$. We observe that with the lower budget, the suboptimality due to selfish decision making is more pronounced. 
Further, we observe that interventions by a social planner have a higher benefit %of the system 
when the defenders are behavioral. % decision makers. 
In particular, when $ BT = 10 $, the relative difference between social and selfish for $ \alpha = 1 $ is $5.7\%$, while it is higher (14\%) for $ \alpha=0.6 $.
\section{Limitations and  Discussion} \label{sec:Discussion}

%\hspace{4mm} 
\textbf{Guiding security decision-makers:} Here we have studied the impacts of human behavioral decision making
%, represented by non-linear weighting of actual probabilities of attacks to determine human-perceived probabilities. We have seen through the various system configurations that different levels of loss may arise from behavioral decision making, 
on the security of an interdependent system.
% and for individual stakeholders. 
We believe this opens up a new dimension of {\em intervention} in securing interdependent systems. How does one guide security decision makers to the appropriate levels of behavioral weighting for a given system? How does one incentivize multiple stakeholders to cooperate to achieve greater system-level security? 
What is the role of a central regulatory agency in incentivizing the security-beneficial investments by individual stakeholders? In this context, the insights gained from our analysis would be useful for configuring real-world systems with optimal parameter choices. For example, following the result of Fig.~\ref{fig:Defense_mechanisms}, we can start to quantify the benefit of joint defense relative to individual defense by each defender.

We can also quantitatively show the decision maker the improvement in system security (under various systems conditions) when moving from her current (sub-optimal) investments to that given by a (rational) algorithm (e.g., \name with $\alpha = 1$ or ~\cite{sheyner2002automated}). In contrast to the line of work that studied placing response actions solely based on the system admin's intuition (e.g., \cite{lemay2011model,sheyner2003tools}), \name solves a rigorous convex optimization problem \eqref{eq:defender_utility_edge} to determine how to distribute the mitigation actions across the critical edges.

\noindent \textbf{Human subject experiments:}
%Note that using students  to simulate trained security defenders can be considered a limitation of our work. T
The use of student subjects as opposed to security professionals is for practical reasons, as it is difficult to collect enough number of security professionals at one time and place for a controlled experiment and importantly, it would be very expensive to incentivize such professionals with monetary payments. For reference, the average payment a subject earned for the two networks described plus the participation fee was \$10.93.   
Although, we think that students (with instructions) are proxying for general ``human'' performance, it should be noted that they are (most likely) not the best case performers. On the other hand, the empirical evidence of differing behavior of professional (of a relevant field) and student subject pools is mixed. There are cases where behavior differs in an expected way, with professionals being less biased than students. However, more generally the biases found in student subject pools do exist in professional subject pools.\footnote{See Part IV, in Chapter 17, 
\cite{frechette2015handbook} for a survey and discussion on this topic.}
In our environment, we anticipate that security professionals exhibit higher levels of rational behavior relative to our subject experiment. However, even small deviations from rationality can result in sub-optimal security investments that are empirically important due to the magnitude of losses associated with compromised `real-world' assets.

\noindent \textbf{Loss aversion:} In addition to misperceptions of probabilities, empirical evidence shows that humans perceive utilities and losses differently than simple expected values. In particular, humans avoid uncertain outcomes and overweight losses compared to gains (loss aversion). A richer behavioral model, {\em cumulative prospect theory} \cite{kahneman1991anomalies}, incorporates these aspects. However, in our setting, this richer model does not significantly change the total loss function of the defenders.
% DW: 2/29/20 : perhaps 'does not change the total loss function of the defenders in any meaningful way'.
Specifically, the attack on an asset is either successful or it is not. If the reference total loss is zero for each asset (i.e., without a successful attack), then the index of loss aversion only scales the loss $L_m$ by a scalar constant without changing the optimal decision on the investments. 
% DW: 2/29/20 : I believe that loss aversion, in and of itself, is not dependent on the number of outcomes.  Rather it is the observation from Prospect Theory that people are risk-averse over gains and risk-seeking over losses that is irrelevant if there are only two outcomes, as this is a matter of risk aversion (or more specifically utility curvature), which is irrelevant with 2 outcomes.  I guess it depends on what exactly you are calling 'loss aversion'.

\noindent \textbf {Uncertainty in estimation of probability of successful attacks:} In our evaluation, we have assumed that each defender has a correct probability assessment (i.e., $p^{0}_{i,j}$ is estimated correctly). However, there are practical scenarios in which this assumption does not hold. In this context, we should note that the behavioral decision making affects the problem of resource allocations even under such uncertainty. We provide a brief discussion and analysis of such scenario in Appendix~\ref{app:uncertainity_attack_probability}.

\noindent \textbf{Multi-hop dependence:}   In several cybersecurity scenarios, the ease of an attacker in achieving an attack goal depends not just on the immediate prior attack step but on steps farther back. In such scenarios, the simpler formulation of using probabilities on each edge and assuming independence of the events of traversing the different edges can lead to inaccurate estimates. However, we follow several prior works (e.g.,~\cite{modelo2008determining, xie2010using}) that leveraged the property that in most cases, a node has the highest dependence on the previous node, in order to build computationally tractable analysis tools. Moreover, to handle this issue in our model, we introduce the notion of {\em k-hop dependence} whereby the probability of reaching a particular node depends not just on the previous hop node, but nodes up to $k$ hops away. We provide a brief discussion of such setup in Appendix~\ref{app:multi-hop-dependence}.

\section{Related Work}\label{sec:lit-review}

\noindent {\bf Security in interdependent systems}: 
The problem of securing systems with interdependent assets has been handled in several prior works~\cite{modelo2008determining, xie2010using}. The common theme is that a successful attack to one asset may be used to compromise a dependent asset. The notion of attack graphs \cite{homer2013aggregating} is a popular abstraction for capturing the security interdependencies. The specific works differ in what the assets are (physical or virtual, resource-constrained nodes, networking assets, etc.), the level of observability into the states of the assets, and the probabilistic reasoning engine used. Our work here differs from these works 
%the prior work on this topic 
in that the prior work creates algorithms to make the security control decisions, while we are considering humans with cognitive biases making these decisions. 

\noindent{\bf Game-theoretic modeling of security}: 
% SB (5/29/18): Editorial - Casing of the section/sub-section headings is inconsistent. One suggestion is to capitalize each word except for supporting words (and, to, the, ...).
% Mus (5/30/18): done
 Game theory has been used to describe the interactions between attackers and defenders and their effects on system security. % in security of system such as modeling attacks and defenses 
A commonly used model in this context is that of two-player games, where a single attacker attempts to compromise a system controlled by a single defender (e.g.,  \cite{roy2010survey,nguyen2009security,you2003kind,lye2005game,wang2010network}). Game theoretic models have been further used in  \cite{yan2012towards,bedi2011game} to study the interaction between one defender and (multiple) attackers attempting Distributed Denial of Service (DDoS) attacks. 
%\textcolor{red}{The work in \cite{varian2004system} considers multiple defenders model in which the security level is a function of the sum of investments by the two defenders}\footnote{PN(9/24/19): it looks odd that we only give more detail on this one paper, and put the others together. Any particular reason for this? Mus(9-25): I moved it into models in which security level increases in sum of investments. This is in response to the following comment of Reviewer D "the reference you are seeking for the legitimacy of treating security as a situation in which the increase in security is a result of the sum of investments is Hal Varian, "System reliability and free riding""}. 
Our work differs from both lines of literature in that we consider the interdependencies between {multiple} defenders in the network.

Game theoretic models for the study of CPS security have been proposed in~\cite{zhu2011stackelberg,perelman2014network}. %Specifically, the
The authors in \cite{zhu2011stackelberg} proposed
a Stackelberg game model, in which a defender attempts to maintain high performance in a SCADA control system against cyber attacks launched by fully rational jammers. %correlated rational decision making jammers that try to disrupt the communications between different sensors that exchange state measurements. 
%The problem of security resource allocation for smart city infrastructures was studied in \cite{ferdowsi2017colonel} as a Colonel Blotto game. 
The work in \cite{perelman2014network} proposed a single-defender game-theoretic approach to minimize loss due to attacks in water distribution networks.  
%however, this work does not take into account the interdependencies between the infrastructure components. 
However, these works have not taken into account the interdependencies between multiple defenders.

%The authors in \cite{zhu2011stackelberg} proposed a Stackelberg game model, in which a defender attempts to maintain high performance in a SCADA control system against cyber attacks launched by fully rational jammers. 
%The work in \cite{perelman2014network} proposed a single-defender game-theoretic approach to minimize loss due to attacks in water distribution networks.  
%however, this work does not take into account the interdependencies between the infrastructure components. 

% \footnote{PN (06/09): \cite{ahmed2017model} is in the first sentence. We should describe it and compare with that as well.}

% \footnote{PN: this is somewhat of a vague description. Can we have more details on the control system, type of CPS, type of attack, and difference from this work?} 

% Also, \cite{ahmed2017model} uses a state space model to detect attacks on the water distribution network,  where they take into account the multistage processing in CPS systems. % where each stage is dependent on the previous stage.
% \footnote{PN: what is the main  difference between our work and \cite{ahmed2017model}?} 

Lastly, the major difference of our work with all aforementioned literature is that existing work has focused on classical game-theoretic models of rational decision making, while we analyze behavioral models of decision making. 
% These behavioral models are meant to capture the imperfections in the human defenders' risk perceptions and reasoning. 
Notable departures from classical economic models within the security and privacy literature are  \cite{acquisti2009nudging,adams1999users}, which identify the effects of behavioral decision making on individual's personal privacy choices using human subject experiments without exploring rigorous mathematical models of players’ behavior. The importance of considering similar models in the study of system security has been recognized in the literature \cite{cranor2008framework}. To the best of our knowledge, the only study that provides a theoretical treatment of behavioral decision making in certain specific classes of interdependent security games is \cite{7544460}. That research, however, is theoretical in scope and does not consider the more realistic attack scenarios that we consider here. %\textcolor{red}{Note that our work does not focus on privacy decision making}. PN(9/24/19): commented this out, it seemed out of place here. 

\noindent \textbf{Human behavior in security:}
% SB (9/30/19): Done a pass on this new material.
%\textcolor{red}{
A large body of literature has considered models from behavioral economics in the context of security applications, such as internet security and information security, via psychological studies; see  \cite{anderson2007information} for survey or   
through human subject experiments~\cite{redmiles2018dancing,baddeley2011information}.
% \footnote{PN(9/24/19): I'm somewhat confused about the distinction we are making here between empirical studies, human subject studies, and psychological studies. could you please elaborate? Also, are any of these works theoretical in nature?
% Mus(9-25): All of them is either surveys from psychological point of view or human subject experiments. To my understanding, empirical studies are similar to human subject experiments.}
Our work differs from these in that we explore a rigorous mathematical model of defenders' behavior, model the  interaction between multiple defenders (in contrast to the study of only one defender for all of these studies), and consider interdependent assets (in contrast to these studies which reason about binary decisions on isolated assets). 
% We also %\footnote{PN(9/24/19): by saying most of these studies, do we mean some of them are multi-defender? In that case, IMO this statement is unclear, we should either specify except for which works, or not state this as our contribution. 
% Mus(9-25): Done}
% model security failure scenarios through attack graph representation, and consider the resource allocation of security resources on the CPS.
%}
%For example, \cite{verendel2008prospect} has studied the effect of prospect theory\footnote{PN(9/24/19): prospect theory is a theoretical model, right? My understanding of this statement is that this paper is supposed to be a human subject study, so saying that is studies a theoretical model seems inconsistent. }
% Mus(9-25): Removed this reference. You are correct. I meant by this a theoritical study but for simple buy or not. What do you think? where to add it if needed?} when a decision-maker is faced with a binary decision between two prospects for buying protection against security threats.} 

% Our work presents the effects of behavioral decision making on the security outcomes of interdependent CPS.

% \footnote{PN (06/09): Please verify that it is ok to claim this and that it is not a hyperbole, especially as the submitted CDC paper technically would be the first paper doing that?}

\section{Conclusion}\label{sec:conclusion}
We studied \emph{behavioral security games} to evaluate the effects of human behavioral decision making on the security of large-scale CPS with multiple defenders. % where  we model stepping-stone attacks by the notion of {\it attack graphs}. 
%We have proposed a \emph{behavioral security game} model in which the interdependencies between the defenders' assets is also captured via the attack graph. 
We compared the strategies of behavioral and rational defenders, where behavioral defenders exhibit nonlinear probability weighting and a tendency to spread security investments across assets.  %,
We observed that behavioral decision makers tend to allocate their budget across the network, while non-behavioral decision makers concentrate their budget on critical edges. 
We presented two real case studies of interdependent CPS: a distributed energy resources system and a SCADA-based control network. We studied the effects of several game parameters including the defense budget availability and distribution, and collaborative defense mechanisms, on the security risks of the system in the presence of behavioral decision making. 

We find that the suboptimal pattern of resource allocation by humans exhibiting behavioral decision making characteristics can considerably  increase the security risks of interdependent CPS,  %facing cyber attacks, 
with the suboptimality becoming more pronounced under moderate security budgets. Using better security risk evaluations and expert input (which can alter the patterns of behavioral decision makers toward rational decision making) can ultimately lead to improvements of CPS security. A complementary approach is to enable collaborative defense strategies, or centralized planning, which can lead to a reduction in security risks even with behavioral decision making.

 \bibliographystyle{splncs04}
 {\footnotesize
\bibliography{refs}}

\section*{Appendix}
\appendix

%%% Appendix: Convexity of the Total Loss Function
\section{Convexity of the Total Loss Function}\label{sec:proof-convexity}
\begin{lemma}
Let the probability successful function $p_{i,j}(x_{i,j})$ be twice-differentiable and log-convex. Then, the total loss function in \eqref{eq:defender_utility_edge} is convex.
\end{lemma}
% \begin{proof}
\emph{Proof :} 
\begin{comment}
Let  the feasible defense strategy for defender $D_k$ be $$X_{k} \triangleq \lbrace{x_{i,j}^{k} \in \mathbb{R_{\geq \texttt{0}}} , (v_{i}, v_{j}) \in \mathcal{E}_{k} \!: \!\!\sum_{(v_{i}, v_{j}) \in \mathcal{E}_{k}} \!\!x_{i,j}^{k} \leq B_{k}  \rbrace}.$$
\end{comment}
 We drop the subscript $i,j$ in the first part of this analysis for better readability. Now, beginning with the probability weighting function defined in \eqref{eq:prelec}, we have
\begin{align*}
 w(p(x)) &= \exp\Big[ -(-\log(p(x)))^\alpha\Big] = (g \circ h)(x),
\end{align*}
where $ g(x) = \exp(-x) $ and  $ h(x) = (-\log(p(x)))^\alpha$. 
%Next, it is easy to show that $h''(x)$
we prove that $h(x)$ is concave:
\begin{align*}
h'(x) &= -\alpha (-\log(p(x)))^{\alpha-1} \frac{p'(x)}{p(x)} \\
h''(x) &= \alpha (\alpha-1) (-\log(p(x)))^{\alpha-2} \frac{(p'(x))^2}{(p(x))^2}\\
& + \alpha (-\log(p(x)))^{\alpha-1} \left[ \frac{(p'(x))^2 - (p(x))(p''(x))}{(p(x))^2} \right].
\end{align*}
Since $0 \leq p(x)\leq 1$, we have $0 \leq -\log(p(x)) \leq \infty $ for all $x$. Moreover, $0 < \alpha \leq 1$ and thus the first term in the R.H.S. of $h''(x)$ is negative. Also, since $p(x)$ is twice-differentiable and log-convex, $(p'(x))^2 < (p(x))(p''(x))$ \cite{boyd2004convex}, which ensures that the second term is also negative. Therefore, $h(x)$ is concave.
% (defined in \eqref{eq:defense_strategy_space})
Since $ g(x) $ is convex and non-increasing while $ h(x) $ is concave, $ w(p(x)) $ is convex. 

Now, with the total loss function defined in \eqref{eq:defender_utility_edge},
$ w(p_{i,j}(x_{i,j})) $ is convex (as shown above). Since $w(p_{i,j}(x_{i,j}))$ is monotone, thus $ \displaystyle \prod_{(v_{i},v_{j}) \in P} w(p_{i,j}(x_{i,j})) $ is convex \cite{boyd2004convex}. Moreover, the maximum of  a set of convex functions is also convex \cite{boyd2004convex}. Finally, since the total loss function $ C_{k}(x_k,\mathbf{x}_{-k}) $ is a linear combination of convex functions, the total loss function defined in \eqref{eq:defender_utility_edge} is convex. That concludes the proof.
%\end{proof}
% \ref{ex:split_join}
%%%%%%%%%%%%%%%%%%%%%%%%%%%%%%%%%%%%%%%%%%%%%%%%%%%%%%%%%%
\iffalse
% Appendix: Motivational Example with different sensitivities
\section{Motivational Example with different sensitivities}\label{app:sensitivity}
In the motivational example in Section~\ref{sec:model}, we assumed all edges have the same sensitivity to investments. In cases where critical edges have equal or higher sensitivity than non-critical edges, the same insight as above holds.
Specifically, when edge $(v_i, v_j)$ has sensitivity $s_{i,j}$, one can verify (using the KKT conditions) that the optimal investments by a behavioral defender are given by 
\begin{equation*}
\begin{aligned}
x_{1,2} &= x_{2,4} = x_{1,3} = x_{3,4} = 2^{\frac{1}{\alpha-1}} \left(\tfrac{s_{i,j}}{s_{s,1}}\right)^{\frac{\alpha}{1-\alpha}} x_{s,1} .\\
x_{s,1} &= \left(\tfrac{s_{s,1}}{s_{4,5}}\right)^{\frac{\alpha}{1-\alpha}}   x_{4,5};\hspace{1mm} x_{4,5} = B - \sum_{\forall (i,j)\neq (v_4,v_5) } x_{i,j}.
%&= \tfrac{B-4x_{1,2}}{2} . 
\end{aligned}
\end{equation*}
The insight here is that the investment decision has two dimensions: behavioral level and sensitivity ratio of non-critical edges to critical edges. Specifically, as the defender becomes more behavioral, she puts less investments on edges with higher sensitivity.
\fi
%%%%%%%%%%%%%%%%%%%%%%%%%%%%%%%%%%%%%%%%%%%%%%%%%%%%%%%%%%
%% Appendix: Human Experiments Details
\section{Human Experiments Details}\label{app: human-exp-extended}

\textbf{Human Experiments Demographics:}
The 145 human subjects in our experiment are comprised of 78 males (53.79\%) and 67 females (46.21\%). They belong to various majors on campus, with the three largest being Management/Business (24.8\%), Engineering (24.2\%), and Science (23.5\%). Regarding year in college, 6.9\% are 1st year, 13.1\% are 2nd year, 21.38\% are 3rd year, 35.86\% are 4th year, and 22.76\% are graduate students. Regarding the GPA distribution, 44.83\% have GPAs between 3.5 and 4, 35.17\% between 3 and 3.5, and 17.93\% between 2.5 and 3.

\iffalse
\textbf{Individual's decisions}: Here, we show the distribution of allocation units for each individual subject in our human experiments. Specifically, Figure \ref{fig:crit_red} shows the distribution of allocation units for each individual subject between  critical and non-critical edges. Also, Figure \ref{fig:crit_blue} shows the distribution of allocation units for each individual subject between the cross-over edge and other edges.

\begin{figure}%[t]
\begin{minipage}[t]{0.48\linewidth}
 \includegraphics[width=\linewidth]{Figures/ind_red2.pdf}
  \caption{Sorted subjects' allocations on critical edge vs non-critical edges. Only 26\% of the subjects are considered non-behavioral (subject numbers 1-14 above).}
  \label{fig:crit_red}
 \end{minipage}%
    \hfill%
\begin{minipage}[t]{0.48\linewidth}
  \includegraphics[width=\linewidth]{Figures/ind_blue.pdf}
  \caption{Sorted subject's investments on cross-over edge vs other edges. Only 18\% of the subjects are considered  non-spreaders (subject numbers 45-54 above).}
  \label{fig:crit_blue}
\end{minipage} 
\end{figure}
\fi

%%%%%%%%%%%%%%%%%%%%%%%%%%%%%%%%%%%%%%%%%%%%%%%%%%%%%%%%%%
%% Appendix: Extended results DER.1

\begin{figure*}[t] 
%\centering
\begin{minipage}[t]{1.0\textwidth}
\begin{minipage}[t]{.47\textwidth}
\centering
  \includegraphics[width=\linewidth]{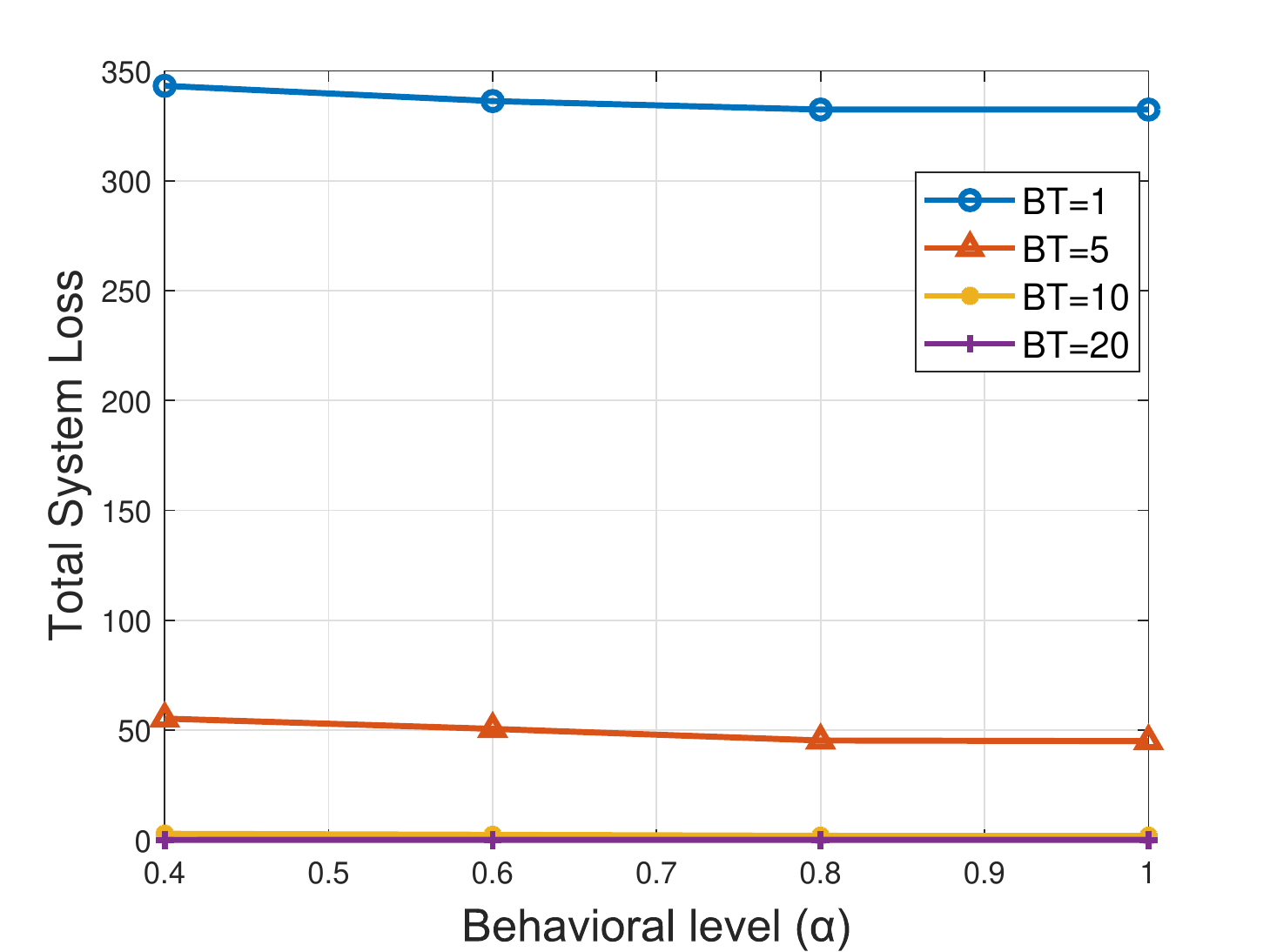}
  \caption{Total loss as a function of the security budget. The adverse effects of behavioral decision making are most severe with intermediate budgets. At either high or low budgets, the amount of the budget, rather than its allocation, determines security.}% is crucial in determining security. } % so that the effects of behavioral decision making become secondary.
  \label{fig:Effect_SecurityBudget}
\end{minipage}\hfill
\begin{minipage}[t]{.47\textwidth}
\centering
  \includegraphics[width=\linewidth]{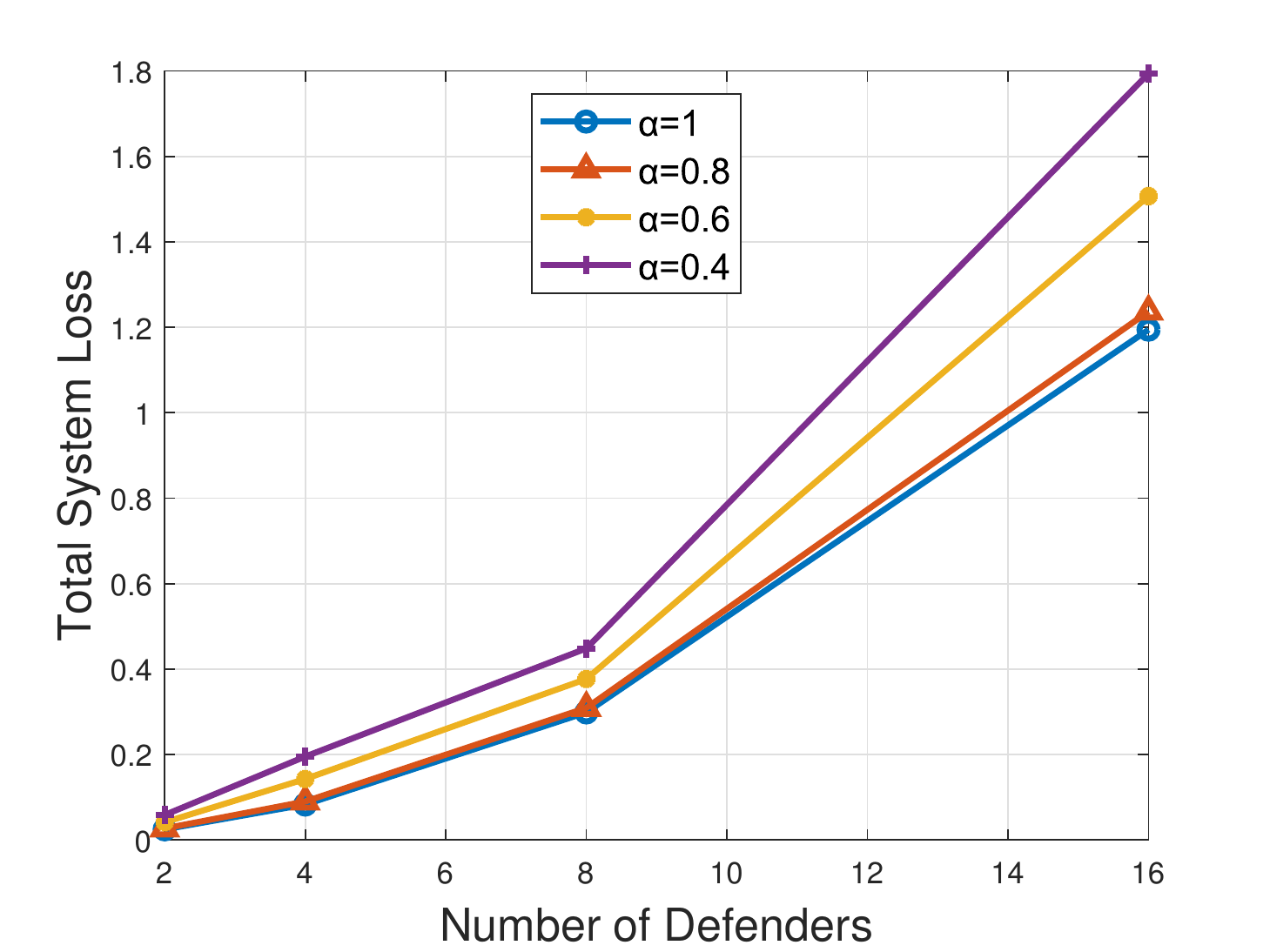}
  \caption{Total loss as a function of the number of defenders. % where each defender has security budget of \$10. 
  We observe that the loss increases super-linearly (i.e., the per-defender loss is increasing as the CPS grows). This is due to the increased risks resulting from interdependencies in the defenders' critical assets.}
  \label{fig:Effect_Defenders}
\end{minipage}
\end{minipage}
%\vspace{-0.1in}
\end{figure*}

\section{DER.1 Extended Results}\label{app:DER-results-extended}
\vspace{0.05in}
\vspace{0.05in}
\noindent {\em \textbf{Experiment A.5: Amount of security budget.}}
We next consider the system defense, and vary the available total security budget for defending the DER system, under symmetric budget distribution between the two defenders. Figure \ref{fig:Effect_SecurityBudget} shows the effect of behavioral decision making on the security level of the DER system as a function of the total budget. We observe that the effect of behavioral decision making with intermediate security budgets (i.e., $ BT= 5 $ and $ BT=10 $) is more severe than both very low budget (i.e., $ BT = 1 $) and  very high budget (i.e., $ BT = 20 $). For instance, with $ BT=1 $, the total loss when both defenders are behavioral with $ \alpha_1 = \alpha_2  = 0.6 $ has a relative increase of $ 1.11\% $ over a system in which both defenders are non-behavioral, while at $ BT = 5$ and $ BT = 10$, a similar change in behavioral levels yields a higher increase of $ 12.41\% $ and $ 28.24\% $ in the total loss, respectively. 

%\issa{I do not see the curve of BT=10 on the figure. It would be great to have another figure with X-axis is BT and Y-axes is the loss for two curves $\alpha=1$ and $\alpha=0.6$. Such curve helps identify the budget needed to get losses below a certain value for each defender scenario.}
%Mus(10/17/18): Done. Added detailed values on a new figure with added explanation.

This effect can be intuitively interpreted by noting that with a low security budget, all defenders suffer mainly due to the scarcity of protection resources, and are less affected by how their budget is distributed. At very high budgets on the other hand, the total loss experienced by both behavioral and non-behavioral decision makers is very close to zero (since the probabilities of successful attack on each edge decrease exponentially with the amount of investment on that edge), and thus suboptimal investment decisions do not have a considerable impact on overall security.  %This is because as the budget spent on each edge increases, the loss probabilities decrease, exponentially, towards negligible values, so that suboptimal investment decisions do not have a considerable impact on the system's security.
In other words, we observe that judicious (non-behavioral) investments are most crucial in determining the security of the system when the \emph{allocation}, rather than the amount, of the security budget is the deciding factor in determining risks.  

\noindent {\em \textbf{Experiment A.6: Number of defenders.}}
% So far, our experiments consisted of only two critical assets, PV and EV, representing a two-defender game.
We create a network with multiple defenders where we make replicas of these two subnetworks, and assume that each new installed equipment corresponds to a new defender's subnetwork. We consider a symmetric distribution of the security budget over all defenders, with each defender having a security budget of \$10. % to defend her assets. 
We notice in Figure \ref{fig:Effect_Defenders} that as the number of defenders increases, the difference between total losses between non-behavioral and behavioral games increases in a super-linear manner. 
For instance, we observe that when the number of defenders is 4, a change from non-behavioral to  behavioral defenders ($\alpha = 0.6$) increases the loss by 8.65\%%. For 16 defenders, the corresponding increase in loss is higher (26.17\%). 
, while the same change in $ \alpha $ in the larger network with 16 defenders results in a higher increase of $ 26.17\%$. This phenomenon is due to the interdependencies between the subnetworks. For instance, if there are two defenders, each will incur a loss in two cases: when either her target asset is successfully compromised or the other defender's target asset is successfully compromised (as it can lead to the compromise of their common goal $G$). On the other hand, if there are 16 defenders, for each defender, there are 16 possible paths through which she suffers a loss. %{\color{red} does this explanation justify the non-linearity too?
%Mus: I don't get your point ?.
%}
% PN(10/23/18): so I think this explanation so far to me only would justify an increase in total loss. But why is there an *exponential* increase? Why is it not a linear increase for example?
% Mus(10/24/18): As the last paragraph mentions, my loss will be due to the effect of each target asset plus extra loss due to non-behavioral sub-optimal investments. For non-behavioral, it is almost linear increase in number of assets. 
% PN(10/28/18): Ok, I see. I now agree that this tells me it is super-linear growth. I see that we have changed it to super-linear and I agree with this, as exponential growth has a very concrete definition of how super-linear it is.
This also explains why the total loss in the system increases with increasing number of defenders---the individual budget of each defender stays the same but the number of ways in which her asset can be compromised increases linearly. 
% SB (10/25/18): When # defenders goes up, why does total system loss go up since there is higher security budget now?
% Mus(10/28/18): I gave this explanation below. Is it convincing? "if there are two defenders, each one will incur a loss in two cases: when either her  target asset is successfully compromised or the other defender's target asset is successfully compromised (as it can lead to the compromise of their common goal). On the other hand, if there are 16 defenders, for each defender, there are 16 possible outcomes in which she suffers a loss."
% SB (11/1/18): Added material and resolved.

\noindent {\em\textbf{Experiment A.7: Asymmetry in budget distributions.}}
We next analyze the effect of asymmetric budget distribution in the two-defender network facing an attacker. The total security budget is \$10. %game for both non-behavioral (i.e., $ \alpha = 1$) and behavioral (i.e, $ 0 < \alpha < 1$) games. 
Figure \ref{fig:Defense_mechanisms} illustrates the total loss as a function of the fraction of defender 1's budget.  %in the budget distribution as the relative difference between total real loss when both defenders are behavioral (i.e., having 0 < $\alpha$ < 1) and non-behavioral defenders (i.e., $\alpha = 1$). 
For the individual-defense loss (solid lines), we observe that the suboptimality of behavioral decision making is more pronounced with higher budget asymmetries. For example, if defender 1 has $ 20\% $ of the total budget, the relative increase in total loss %of security game 
from $\alpha = 1$ to $\alpha=0.4$ is $ 25\% $. In contrast, the same change of $ \alpha $ when the budget is symmetric results in only a $ 6\% $ relative increase in the total loss. This observation can be explained by two facts. First, with suboptimal behavioral allocation, the poorer defender wastes even her constrained budget on non-critical edges. Second, the richer player also allocates her resources suboptimally. This leads to this magnified relative increase in losses under budget asymmetry.

\begin{figure*}[t] 
%\centering
\begin{minipage}[t]{1.0\textwidth}
\begin{minipage}[t]{.47\textwidth}
\centering
\includegraphics[width=\linewidth]{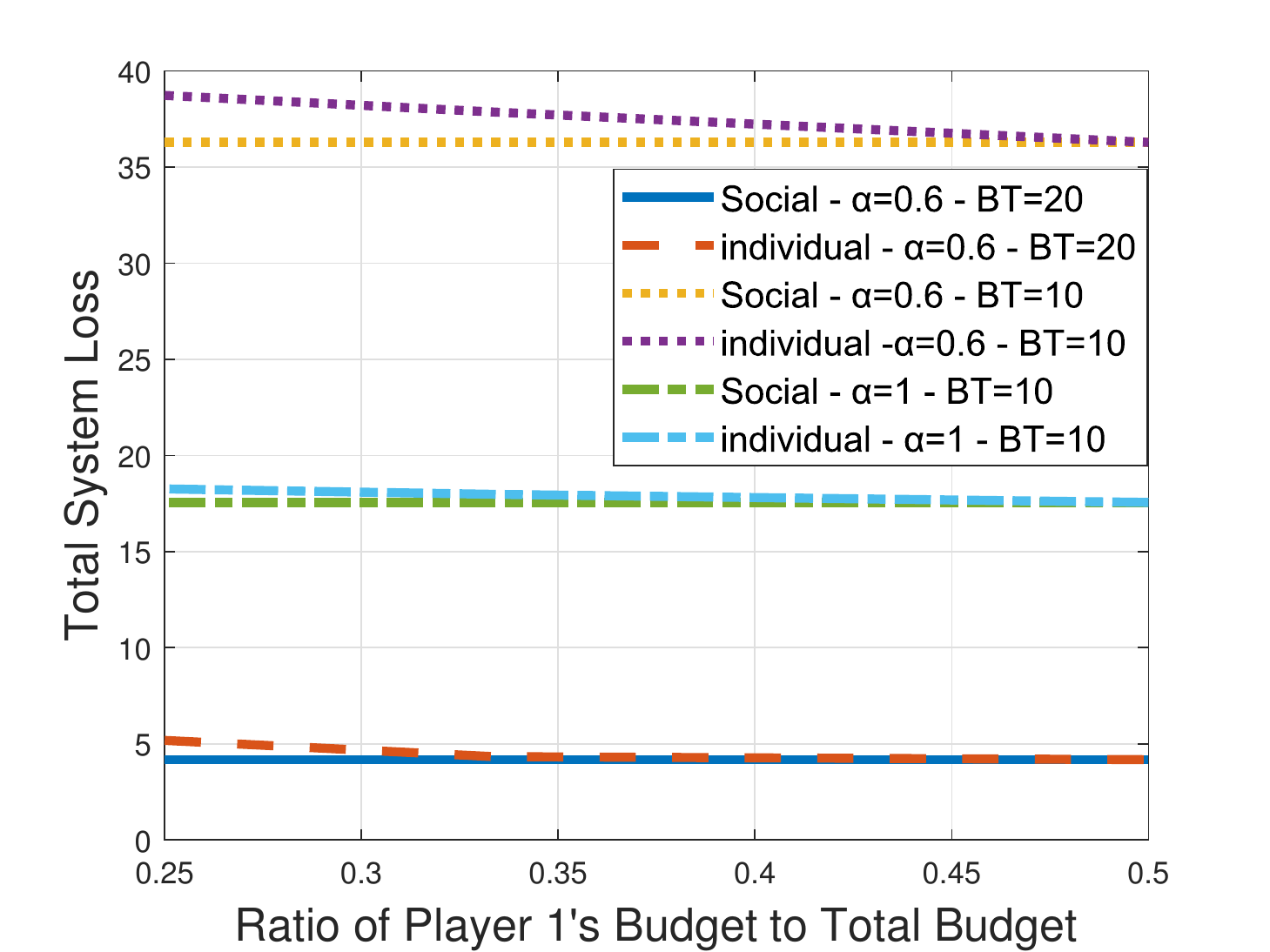}
  \caption{A comparison of expected losses between centralized and distributed decision making. Social planning is most effective under moderate total budget and high budget asymmetry between the defenders, and when the defenders are behavioral.}% decision makers.}
  \label{fig:Central_Behavioral_Scada} 
\end{minipage}\hfill
 \begin{minipage}[t]{0.47\linewidth}
   \includegraphics[width=\linewidth]{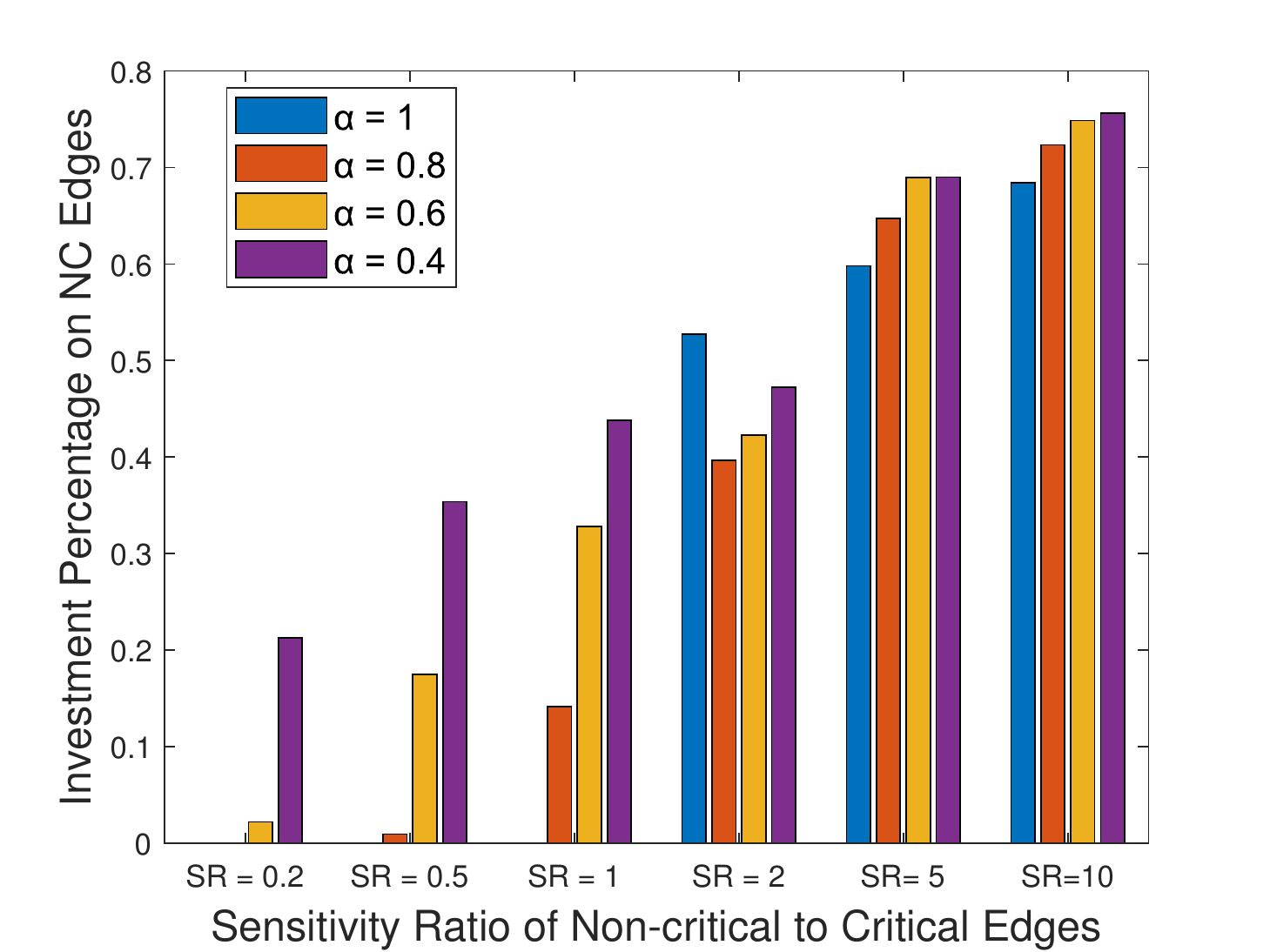}
  \caption{ The percentage of investments on non-critical (NC) edges vs the sensitivity ratio (SR) between non-critical to critical edges, for different $\alpha$ in the SCADA attack graph. As SR increases, the increase in investments on NC edges is slower for behavioral defenders}
  \label{fig:sensitivity-exp-Scada}
 \end{minipage}%
\end{minipage}
\end{figure*}

\section{Scada Extended Results} \label{app: scada_extended}
\vspace{0.01in}
\noindent {\em\textbf{Experiment B.5: Amount of security budget.}} 
%
%We study the effect of security budget change to see how does it mitigate the effect of the behavioral attitude. 
Figure \ref{fig:Effect_BT_Behavioral_Scada} shows the effects of budget availability on system security. We observe that for moderate %security 
budgets, behavioral decision making increases the total loss up to $ 60 \% $ over non-behavioral decision making. On the other hand, with high budgets 
%availability 
($ BT \geq 30 $ in our simulations), behavioral decision making has a negligible effect on the total loss, which is very close to zero. The intuition is similar to that in Section \ref{sec:DER.1}: the negative effect of behavioral decision making is more pronounced with moderate budgets as the allocation of the security budget for protecting the different assets is most critical.  

\begin{figure*}[t]%{.47\textwidth}
\begin{minipage}[t]{1.0\linewidth}
\begin{minipage}[t]{0.47\linewidth}
\includegraphics[width=\linewidth]{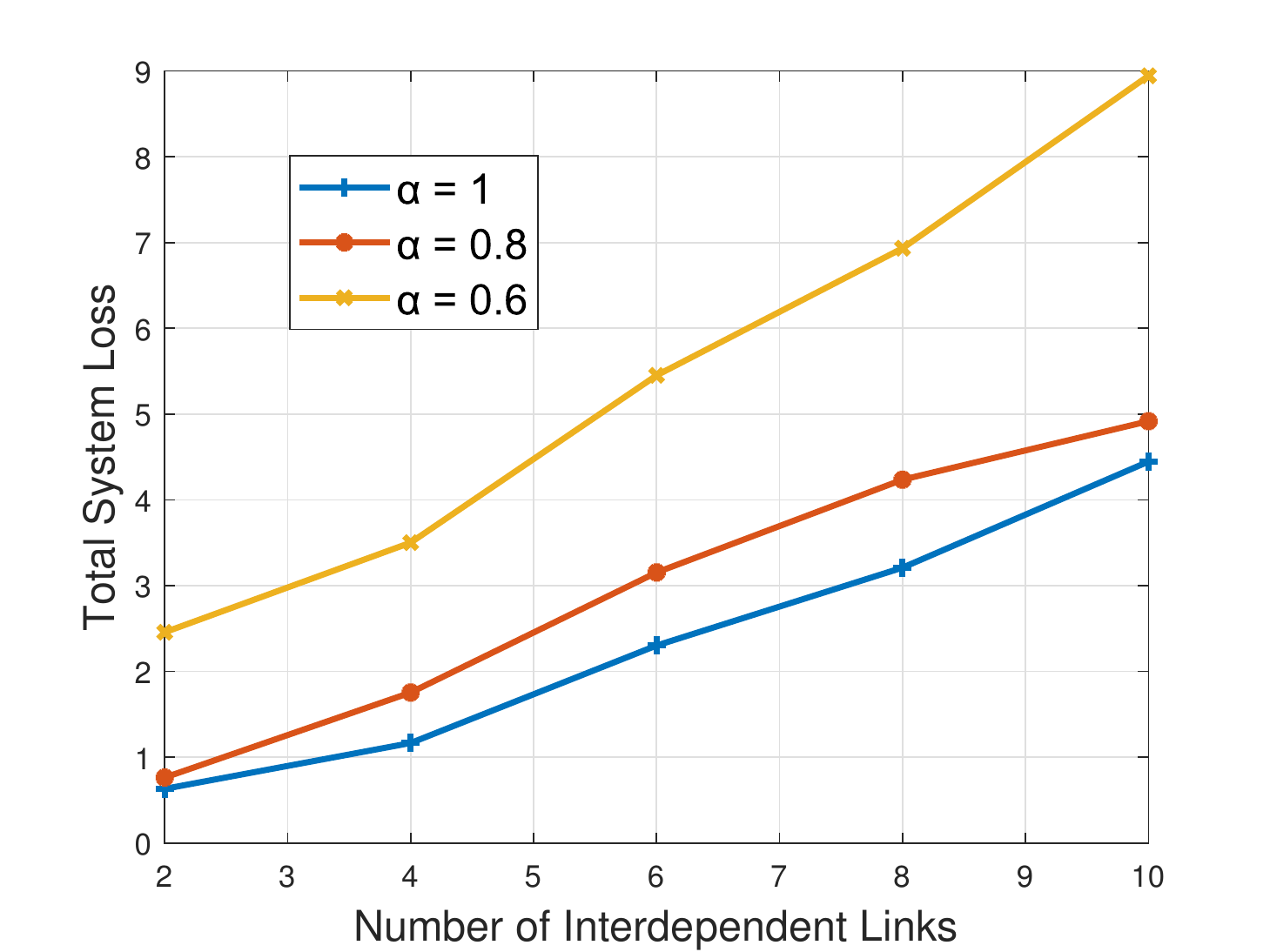}
  \caption{The effect of increasing the degree of interdependency on the total system loss (effect is more pronounced when the degree of behavioral decision making is higher).}
  \label{fig:Interdependency_Effect_Scada}
  \end{minipage}\hfill 
  \begin{minipage}[t]{0.47\linewidth}
  \includegraphics[width=0.9\linewidth]{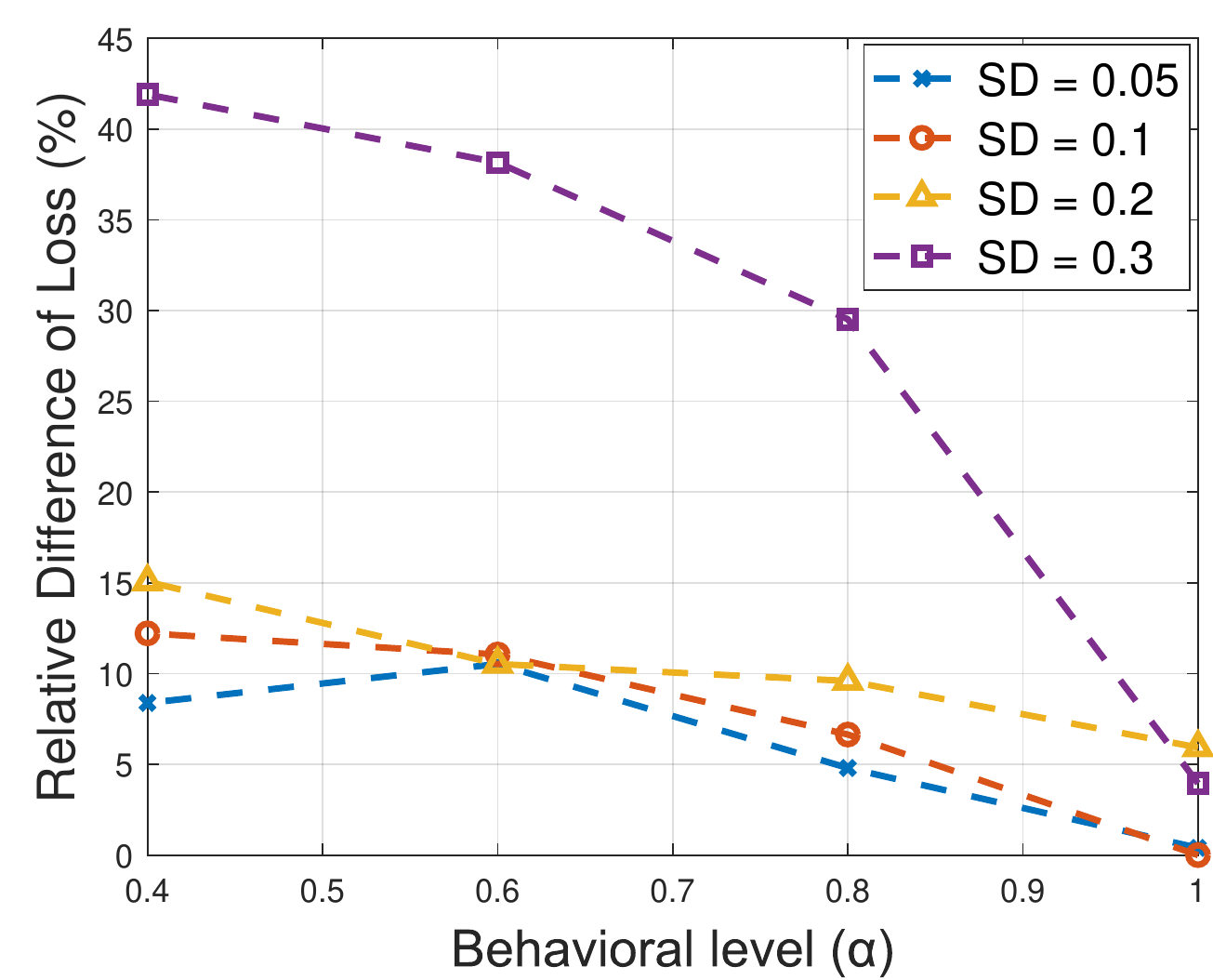}
  \caption{Percentage of relative difference in total system loss due to uncertainty for DER.1 failure scenario.}
  \label{fig:prob_perturbation_der}
  \end{minipage}
% \vspace{-0.1in}
 \end{minipage}
\end{figure*}

\noindent {\em \textbf {Experiment B.6: Interdependency among different defenders.}} 
As in the DER system, we again observe effect of interdependency between defenders on the security of the SCADA system. We consider medium budget choice (i.e., $BT = 25$) for this experiment. In the SCADA system, the degree of interdependency increases if assets from one subnetwork can access assets in the other, without going through the Corporate or Vendor nodes. 
% main control units (e.g., control units or firewalls) are communicating.
For example, if the attacker gets access to Control unit $1$, this enables her to compromise RTU2 as well, in addition to RTU1. Figure \ref{fig:Interdependency_Effect_Scada} illustrates this effect---as the number of interdependent edges between the two defenders increases, the total system loss increases in both non-behavioral and behavioral security games. The highest level of interdependency is when there are two edges between DMZ1 and DMZ2, between Control1 and Control2, and the controller to the 3 RTUs of the other defender. An example of this phenomenon is that if both defenders are non-behavioral and the level of interdependency is the highest, the total system loss is higher by $ 462\% $ over the case of the lowest level of interdependency ($2$ interdependent links).
We also see that as the interdependency between the different defenders increases, the suboptimal security decisions have greater adverse impact on the total system loss. 

\noindent {\em \textbf {Experiment B.7: Sensitivity of Edges to Investments.}} 
In this experiment, we repeat the sensitivity sweeping analysis of Experiment A.7 for the SCADA attack graph. The investments under different sensitivity ratios are shown in Figure \ref{fig:sensitivity-exp-Scada}, with insights similar to those explained for the DER.1 attack scenario.

%%%%%%%%%%%%%%%%%%%%%%%%%%%%%%%%%%%%%%%%
 \section{Uncertainty in Attack Probability} \label{app:uncertainity_attack_probability}
Recall that in our evaluation, we have assumed that each defender has a correct probability assessment (i.e., $p^{0}_{i,j}$ is estimated correctly). However, there are practical scenarios in which this assumption does not hold (see Section~\ref{sec:Discussion}).
Here, we analyze the effectiveness of behavioral decision making when uncertainty is taken into account. In other words, when probabilities in the attack graph cannot be simply assumed correct. We model such uncertainty by replacing $p^{0}_{i,j}$ in \eqref{eq:expon_prob_func} with $q^{0}_{i,j}$ where $q^{0}_{i,j} \sim 
\mathcal{N}(p^{0}_{i,j},\,\sigma)$ where $\sigma$ is the standard deviation (uncertainty) of the probability of attack. The interesting question that we answer here is: How well does behavioral decision
making compare to rational decision making when
true attack graph edge values are uncertain? In our experiments, we varied $\sigma$ from 0 to 0.3 with the restriction of $0 < q^{0}_{i,j} \leq 1$ since it is a probability. Figure~\ref{fig:prob_perturbation_der} shows the percentage of relative difference in total system loss due to uncertainty for DER.1 failure scenario for the different uncertainty levels.
%\footnote{TC(9/30/19)Again, considering a lower limit of alpha=0.4 makes more sense. This would change the discussion, obviously. Mus(9/30/19): Addressed} 
Note that rational defenders have much lower relative loss since they put most investments on critical edges. On the other hand, the effect is higher for moderate and highly behavioral defenders (i.e., $0.4 \leq \alpha \leq 0.6$) since such uncertainty shifts the probability to different regions in Figure~\ref{fig:Prelec Probability weighting function}. The same insights holds for SCADA system. We remove the details of SCADA uncertainty experiments in the interest of space. This shows that even if the probability of successful attack is not estimated perfectly, the rational behavior (which is the security investments placed by $\alpha = 1$) gives more accurate estimation of the  expected system loss compared to the decisions with behavioral biases. 
%\fi

\section{Multi-hop dependence}
\label{app:multi-hop-dependence}

To handle the multi-hop dependence (mentioned earlier in Section~\ref{sec:Discussion}) in our model, we introduce the notion of {\em k-hop dependence}, whereby the probability of reaching a particular node depends not just on the previous hop node, but nodes up to $k$ hops away. The value of $k$ is a specification provided by the security admin, either for the entire graph or for individual nodes. 
% as a global $k$ value for the entire network, or as a vector of size equal to the number of nodes in the attack graph, in which element $k_i = \theta$ means node $i$ has $\theta$-hop dependence; in practice, most elements have value one, which has been leveraged to build computationally tractable analysis tools~\cite{modelo2008determining, xie2010using, wang2013exploring}. 

\begin{figure*}[t]%t! 
\centering
%\begin{minipage}[t]{1.0\textwidth}
\begin{subfigure}[t]{.48\textwidth}
\centering
\begin{tikzpicture}[scale=0.5]

\tikzset{edge/.style = {->,> = latex'}};

\node[draw,shape=circle] (vs) at (-2,0) {$v_s$};
\node[draw,shape=circle] (v1) at (0,0) {$v_1$};
\node[draw,shape=circle] (v2) at (2,1) {$v_2$};
\node[draw,shape=circle] (v3) at (2,-1) {$v_3$};
\node[draw,shape=circle] (v4) at (4,0) {$v_4$};
\node[draw,shape=circle] (v5) at (6,0) {$v_5$};
\node[shape=circle] (L1) at (6,1) {$L_5=1$};

\draw[edge,thick] (vs) to (v1);
\draw[edge,thick] (v1) to (v2);
\draw[edge,thick] (v1) to (v3);
\draw[edge,thick] (v2) to (v4);
\draw[edge,thick] (v3) to (v4);
\draw[edge,thick] (v4) to (v5);
\end{tikzpicture}
\caption{} 
\label{fig:split_join_dependence_before_app}
\end{subfigure}%\label{fig:MPNE_example_both_behavioral_1}\hfill
\hfill
\begin{subfigure}[t]{.48\textwidth}
\centering
\begin{tikzpicture}[scale=0.5]

\tikzset{edge/.style = {->,> = latex'}};

\node[draw,shape=circle] (vs) at (-2,0) {$v_s$};
\node[draw,shape=circle] (v1) at (0,0) {$v_1$};
\node[draw,shape=circle] (v2) at (2,1) {$v_2$};
\node[draw,shape=circle] (v3) at (2,-1) {$v_3$};
\node[draw,shape=circle] (v4) at (4,1) {$v^{a}_4$};
\node[draw,shape=circle] (v6) at (4,-1) {$v^{b}_4$};
\node[draw,shape=circle] (v5) at (6,0) {$v_5$};

\node[shape=circle] (L1) at (6,1) {$L_5=1$};

\draw[edge,thick] (vs) to (v1);
\draw[edge,thick] (v1) to (v2);
\draw[edge,thick] (v1) to (v3);
\draw[edge,thick] (v2) to (v4);
\draw[edge,thick] (v3) to (v6);
\draw[edge,thick] (v4) to (v5);
\draw[edge,thick] (v6) to (v5);

\end{tikzpicture}
\caption{} 
\label{fig:split_join_dependence_after}
 \end{subfigure} 
\caption{The attack graph in (a) is converted to the attack graph (b) to consider 2-hop dependency whereby the probability of reaching $v_5$ depends on which path is used.}
\label{fig:MPNE_main_app}
\end{figure*}
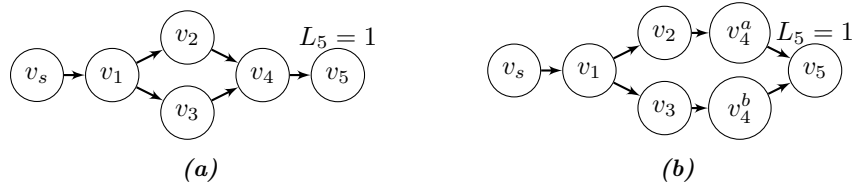

\name handles this type of dependencies by considering all paths of length $k_i$ ending at node $i$. Specifically, let this set of paths be $P_i$. Then the original attack graph is converted to one where node $i$ has $|P_i|$ incident edges with a splitting of the previous node into multiple virtual nodes if needed. The different probabilities on the different edges capture the relative ease of an attacker to reach $i$ through the different paths. We show an example of this in Figure~\ref{fig:split_join_dependence_after} which is created from attack graph in (a) in which node $v_5$ has 2-hop dependence, say because if the attacker has come to $v_4$ through $v_2$, then her job is easier than if she has come through $v_3$. There are 2 paths; correspondingly, the previous node $v_4$ is split into two virtual nodes $v_4^a$ and $v_4^b$. For the example case of the upper path being easier, $p_{v_4^a,v_5} > p_{v_4^b,v_5}$. This same approach can be used when the dependence is not on contiguous nodes, but on a decision taken early on in the attack path, by splitting into multiple paths. One subtle consideration is that when there is a security investment on an edge of the original attack graph, that should be mirrored on all the edges that have been derived from that edge. One potential problem with our approach is that it blows up the size of the attack graphs. The helpful factor here is that most nodes have only 1-hop dependence in practice and this property has been leveraged in the past to build computationally tractable analysis tools~\cite{modelo2008determining, xie2010using, wang2013exploring}.

\end{document}